\documentclass{article}
\usepackage{arxiv}
\usepackage[utf8]{inputenc} 
\usepackage[T1]{fontenc}    
\usepackage{hyperref}       
\usepackage{url}            
\usepackage{booktabs}       
\usepackage{amsfonts}       
\usepackage{nicefrac}       
\usepackage{microtype}      
\usepackage{lipsum}		
\usepackage{makeidx}
\usepackage{amssymb,amsfonts,amsthm,amsmath}
\usepackage{mathtools}
\usepackage{bbm}
\usepackage{chicago}
\usepackage[space]{grffile} 
\usepackage[section]{placeins}
\usepackage{float}
\usepackage[boxed]{algorithm2e}
\usepackage{bbding}
\usepackage{chngcntr}
\usepackage[toc,page]{appendix}
\usepackage{pgfplots}
\usepackage{dsfont}

\usepackage{natbib}
\usepackage{xcolor}
\usepackage{graphicx}
\usepackage{subcaption}
\usepackage{mwe}
\usepackage{makecell}
\usepackage[titles]{tocloft}

\usepackage{array, booktabs, caption}
\pgfplotsset{compat=1.16}

\def\HiLi{\leavevmode\rlap{\hbox to \hsize{\color{gray!30}\leaders\hrule height .8\baselineskip depth .5ex\hfill}}}

\title{Reinforced Deep Markov Models With Applications in Automatic Trading}

\author{
  Tadeu A. Ferreira \\
  Department of Statistical Sciences\\
  University of Toronto\\
  \texttt{tadeu.ferreira@mail.utoronto.ca}
}

\begin{document}

\maketitle
\setlength{\algomargin}{0.2in}

\begin{abstract}
Inspired by the developments in deep generative models, we propose a model-based RL approach, coined \textbf{Reinforced Deep Markov Model (RDMM)}, designed to integrate desirable properties of a reinforcement learning algorithm acting as an automatic trading system. The network architecture allows for the possibility that market dynamics are partially visible and are potentially modified by the agent’s actions. The RDMM filters incomplete and noisy data, to create better behaved input data for RL planning.  The policy search optimisation also properly accounts for state uncertainty. Due to the complexity of the RKDF model architecture, we performed ablation studies to understand the contributions of individual components of the approach better. To test the financial performance of the RDMM we implement policies using variants of Q-Learning, DynaQ-ARIMA and DynaQ-LSTM algorithms. The experiments show that the RDMM is data efficient and provides financial gains compared to the benchmarks in the optimal execution problem. The performance improvement becomes more pronounced when price dynamics are more complex, and this has been demonstrated using real data sets from the limit order book of Facebook, Intel, Vodafone and Microsoft. 		
\end{abstract}


\section{Introduction and Related Work}	
\label{Background}

\normalsize
Many developments in reinforcement learning have been introduced to improve the shortcomings of value iteration algorithms like Q-learning as in \citet{Watkins1989,Watkins1992}. Due to slow convergence and difficulties with large state-action spaces, classical algorithms in reinforcement learning exhibit poor performance in extracting information from the environment. In Q-learning, the changes in a state $x_{t}$ do not directly back-propagate to previous states $x_{t-1},x_{t-2},..$ \citep{Wiering2011}. This issue is known as \textit{exploration overhead}\index{exploration overhead}. Model-based reinforcement learning methods were introduced to model the dynamics of the transitions/rewards to expedite convergence and make the algorithm more data-efficient. However, the inability to reproduce the environment accurately due to  model bias\index{model bias},  has been reported as a common issue in model-based RL, which can hinder the benefits of learning the environment dynamics. Incorporating different sources of uncertainty into planning has shown good results in many recent studies aimed at mitigating model bias, improving model robustness and data efficiency as in \citet{Deisen2011}, and \cite{McAllister2016}.  \citet{Deisen2011} employ Gaussian processes into their probabilistic inference for learning control (PILCO) algorithm to model the Markov decision process dynamics which allows them to incorporate the uncertainty of the model parameters into planning and control. \citet{McAllister2016}  took a step further, including also the uncertainty regarding the agent's belief distribution. 

A model's ability to handle noisy data is also important; \citet{McAllister2016} points out that noise produces a substantial variation in the controller output, causing instability in the cart-pole problem. To cope with noisy observations, they extended PILCO \citep{Deisen2011} into a partially observable Markov decision process (POMDP) framework by filtering the observations and using the predictions with respect to the filtering process. The policy is formed using this smoother version of the observed values.

Our proposed architecture combines these successful approaches with advances in deep generative models such as variational auto-encoders (VAEs) by \citet{Rezende2014} and \citet{Kingma2014}, and deep Markov Models (DMM) by \citet{Krishnan2015,Krishnan2016}.

We modify and build upon the model structure of the DMM and add a gradient-based policy learning framework aiming to obtain the desirable properties mentioned before, such as fast convergence, data-efficiency, and the ability to handle noisy and incomplete observations. To achieve this, we need to extend the basic DMM structure presented in \citet{Krishnan2015,Krishnan2016} into a more appropriate state-space model (SSM)\index{state space model}\index{SSM} that accounts for actions and rewards, and reframe it into a model-based reinforcement learning framework inspired by some of the ideas in \citet{McAllister2016}, where the policy is optimised with respect to the filtering process instead of the observations themselves. State uncertainty is incorporated into the policy search process, which helps with data-efficiency. The policy $\pi$ is optimised to maximise the expected return for a finite number of steps. We refer to our approach hereafter as a \textbf{reinforced deep Markov model (RDMM)}\index{RDMM}. 

We begin with a description of POMDP and the intuition of our model architecture in Section \ref{RL_math}. After explaining the RDMM motivation, we give the complete model specifications in Section \ref{modelespecs}. In Section \ref{modeloptimisation}, the objective functions are defined along with the parameters that need to be optimised. The implementation of the learning algorithm is outlined in Section \ref{RDMMlearning}. A description of the optimal liquidation, the problem that we are trying to solve, is exposed in Section \ref{liquidation}. We conduct tests in a simulated environment using synthetic price dynamics and the results are presented in Section \ref{simulations_and_results}. The environment's complexity is raised by replacing the artificial price dynamics with real prices taken from the limit order book of Intel, Facebook, Microsoft and Vodafone in Section \ref{realex}. Finally, we end this article with a discussion and conclusions about the results in Section \ref{conclusions}.

\section{Mathematical formulation}
\label{RL_math}

In a POMDP\index{POMDP} with a finite and discrete number of states $Z=(z_1,...,z_N)$, the agent only perceives a set of observable states $\Omega=(o_1,...,o_N)$ instead of observing subsets of $Z$ directly. An observation function $O$ is defined in the context of an imperfect sensor, which assigns a probability to each 3-tuple $O(s',a,o)$, representing the probability of observing $o$ at state $s'$ after executing $a$. The idea of introducing limitations on the agent with respect to its perception of the environment is more complete and reflective of reality compared to the classical MDP. 

Using the POMDP framework, we assume a discrete and finite sequence of observable states $\Omega=(o_1,...,o_T)$. The sequence of the corresponding real states are represented by the sequence of latent states $Z=(z_1,...,z_{T})$. For this design we represent the observable state $o_{t}$  containing all the relevant information available for the agent at time $t$. We choose policies $\pi$ as a function of the approximated distribution of $Z$. A sequence of actions $A=(a_1,...,a_{T-1})$ given by the policy $\pi$ is sent to the environment, which in turn, yields a sequence of rewards $R=(r_1,...,r_T)$.

We aim to develop a model with the ability to represent a wide variety of functions responsible for the interaction between actions $A$, latent states $Z$, observable states $\Omega$ and rewards $R$. To achieve this goal we rely on artificial neural networks to behave as universal approximators \citep{Cybenko1989,Balazs2001, NIPS2017_7203}.
Before delving into the specifics of the model we develop a general intuition about how the RDMM is conceptualised. The complete specifications are given in the next section.

\begin{figure}[H] 
	\centering
		\includegraphics[width=0.75\textwidth]{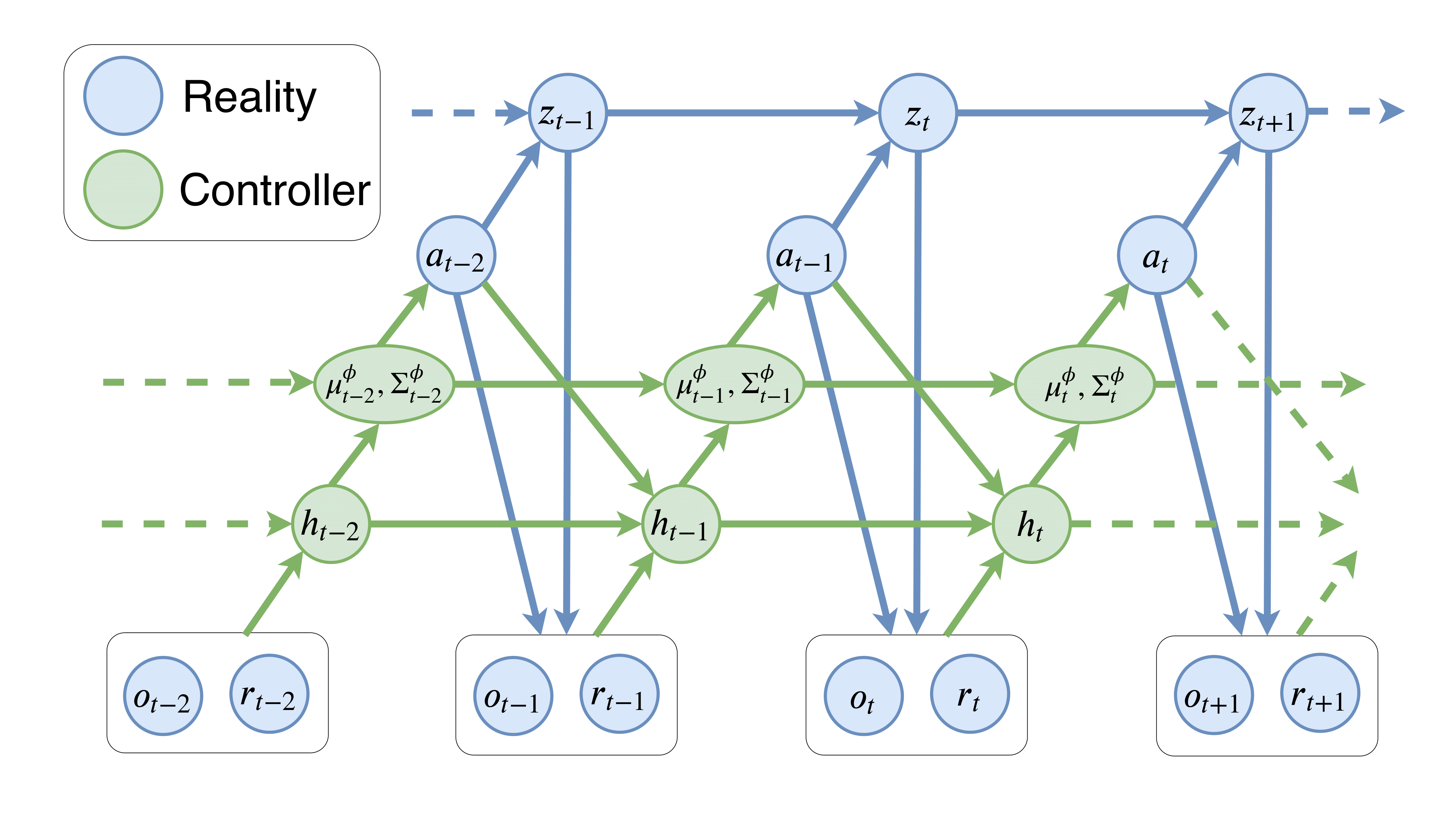}
	\caption{\small Graphical representation of the RDMM. The reality sub-diagram represents the interaction of the actions $a_{t-1}$, the environment $z_t$, and the emission of observable states $o_t$ and rewards $r_t$. After the emissions and actions are integrated into a summary of past information $h_t$ by an LSTM, the algorithm uses an approximation of the posterior to construct a new action $a_t$ as a deterministic function of the mean $\mu^{\phi}_{t}$ and covariance $\Sigma^{\phi}_{t}$ of the approximated posterior distribution.}
	\label{fig:arch}
\end{figure}

Figure \ref{fig:arch}, shows a  graphical representation of the RDMM approach where the sequence of events can be divided into two interconnected sections: \textbf{reality} and \textbf{controller}.
To explain the rationale of the architecture represented by Figure \ref{fig:arch} we start narrating the sequence of the events in the reality section at the instant $t$. At this point, we consider the action $a_{t-1}$  as it is observed --- the description of how the actions are created appears in the controller section. This action is sent to the environment where all the underlying dynamics of the system occur and are represented by the latent state $z_{t}$.  The environment is perturbed by that action, which in turn, emits an observable state $o_{t}$ and yields a reward $r_{t}$.  The visible state and reward are also considered to be directly influenced by the action $a_{t-1}$. All these processes are modelled by artificial neural networks and are represented in Figure \ref{fig:arch} by the blue arrows. This reality section is also considered the emission or decoding part of a variational auto-encoder (VAE).

Once the action $a_{t-1}$, the state $o_{t}$, and the reward $r_{t}$ are observed, the RDMM transitions to the controller section where  $a_{t-1}$,  $o_{t}$ and $r_{t}$  update an LSTM with a new hidden state $h_t$, which serves as a summary of all past information up to the instant $t$.  With the possession of $h_t$ the RDMM approximates the posterior distribution of the latent states, concluding the recognition portion of a VAE.  The parameters of this approximated distribution $\mu^{\phi}_{t}$ and  $\Sigma^{\phi}_{t}$ are the inputs of a deterministic function that outputs the next action  $a_t$. The parameters in the controller section are also modelled by artificial neural networks and are represented in Figure \ref{fig:arch} by the green arrows. Once the action $a_t$ is formed the system transitions from the state $z_t$  to the next state $z_{t+1}$ and the process repeats  itself.

The presence of edges from the actions $\vec{a}$ to the latent states $\vec{z}$ might be crucial in certain contexts. In our application we consider the possibility of the agent's actions affecting the environment; for instance, a trader buying or selling a significant amount of an asset compared to the other agent's positions in the market may alter other agents' perception of the asset's price.

It should be noted that the `reality' part of the RDMM differs from the deep Kalman filters (DKF) presented in \citet{Krishnan2015}  or the deep Markov models (DMM) in \citet{Krishnan2016}, where the authors consider only edges from actions to latent states to make counterfactual inference for medical data. In the RDMM we also have the incorporation of the reward nodes $\vec{r}$, the inclusion of edges from actions to observable states, and edges from actions to rewards.  Since the actions are assumed to be market orders, these edges are relevant to cause an immediate impact on the rewards, latent and observable states. These adaptations result in a more appropriate non-linear state-space model (SSM) to add a gradient-based policy learning in the `controller' part of the RDMM. In this part, the actions stem from the parameters of the approximate distribution of the belief states (approximation of the posterior).

In the next section, we use deep neural networks to further specify the architecture depicted in Figure \ref{fig:arch}.
\section{Model specification}
\label{modelespecs}

The definitions in this section rely heavily on the conditional independence properties of directed graphical models. Henceforth, observed states $o_t$ and latent states $z_t$ are considered as elements of $\mathds{R}^n$ and $\mathds{R}^m$, respectively, while rewards $r_t$ and actions $a_t$ scalars.
\nocite{Murphy2012}
\nocite{Bishop2006}

The latent state transitions are modelled as conditionally normal. Specifically, 
\begin{align} \label{eq:transitions}
\textit{Transitions:} \quad p_{\lambda}(z_{t} | z_{t-1},a_{t-1})=N \left( \mu^{\lambda}_t,\Sigma_t^{\lambda}\right).
\end{align}
As in DKF \citep{Krishnan2015}  the mean $ \mu^{\lambda}_t$  is an interpolation between linear transitions $\text{L}_{t}^{\lambda}$ and non-linear transitions $\text{NL}_{t}$ controlled by a gated unit $g_t$ (for further justification will be provided ahead), and is explicitly given by 
\begin{align*} 
\mu^{\lambda}_t&=g_t \odot \text{NL}_{t}+ (1-g_t) \odot \text{L}_t,
\end{align*}
where $A \odot B$  represents the Hadamard\footnote{The Hadamard product is the entry-wise product between matrices of same dimensions.} product between $A$ and $B$  and,
\begin{align}
\text{NL}_{t} &=W_{\text{NL}_{2}}   \left[  \texttt{ReLU} (W_{\text{NL}_{1}} (z_{t},a_{t-1})^{T} + b_{\text{NL}_{1}} )\right]   +  b_{\text{NL}_{2}} \quad &\textsf{non-linear,}\\
\text{L}_t&=W_{\text{L}} (z_{t},a_{t-1})^{T} + b_{\text{L}} \quad &\textsf{linear,}\\
g_t &=\texttt{sigmoid}  \left( W_{g_{2}}  \left[   \texttt{ReLU} (W_{g_{1}} (z_{t},a_{t-1})^{T} + b_{g_{1}} ) \right]  +  b_{g_{2}} \right) \quad &\textsf{gated unit,}  \label{gated}
\end{align}

The covariance function is modelled as
\begin{align*} 
\Sigma_t^{\lambda}&=\texttt{softplus}( W_{\Sigma^{\lambda}} \texttt{ReLU}(\text{NL}_{t} )+b_{\Sigma^{\lambda}}).
\end{align*}
The motivation behind using an interpolation for the mean  $ \mu_{\lambda t}$ between linear and non-linear components, resides on the fact that some datasets tend to achieve better approximations in terms of held-out likelihood with linear functions, while other datasets have better results with non-linear functions as noted in \citet{Krishnan2016}. The interpolation is inspired by the \textit{update gate} on gated recurrent units (GRUs) \citep{Cho2014},  and provides the model freedom to seek the best combination of the linear and non-linear components.

Observed states $o_t$ and rewards $r_t$ conditioned on $z_t$ and $a_t$ are assumed normally distributed and parametrised by a multi-layer perceptron (MLP). Explicitly,
\begin{align}  \label{price_emission}
\textit{Emissions:} \quad p_{\theta} (o_{t} | z_{t}, a_{t-1} )= N \left(\mu^{\theta}_{t}, \Sigma^{\theta}_{t}\right)
\quad \textit{and} \quad
 p_{\eta} (r_{t} | z_{t}, a_{t-1} )=N \left(\mu^{\eta}_{t}, \Sigma^{\eta}_{t} \right),
\end{align}
where,
\begin{align} 
\mu^{\theta}_{t}&=\texttt{ReLU}(W_{\mu^{\theta}}h^{\theta}_t+b_{\mu^{\theta}}), & 
\mu^{\eta}_{t}&=W_{\mu^{\eta}}h^{\eta}_t+b_{\mu^{\eta}},  \label{eq:mu_theta}\\
\Sigma^{\theta}_{t}&=\texttt{softplus}( W_{\Sigma^{\theta}}h^{\theta}_t+b_{\Sigma^{\theta}}), &
\Sigma^{\eta}_{t}&=\texttt{softplus}( W_{\Sigma^{\eta}}h^{\eta}_t+b_{\Sigma^{\eta}}),\\
h^{\theta}_t&=\texttt{tanh}( W_{\theta} (z_{t},a_{t-1})^{T}+b_{\theta}), &
h^{\eta}_t&=\texttt{tanh}( W_{\eta} (z_{t},a_{t-1})^{T}+b_{\eta}).
\end{align}
For $\Sigma^{\eta}_{t}$ and $\Sigma^{\theta}_{t}$ we use a softplus as an activation function, whereas, $\mu^{\theta}_{t}$ and $\mu^{\eta}_{t}$ are parametrised by an MLP with rectifier and linear activation. We do not use the rectifier for the reward as we must allow negative rewards. Since the posterior $p_{\lambda}(z|o,r,a)$ is intractable \citep{Krishnan2015}, we use  an approximation instead.

\subsection{Variational approximation to the posterior}
As in VAEs (see Appendix \ref{s:VAE}), the posterior distribution over latent states given observations is intractable and instead we introduce a variational approximation to the posterior. Our model uses the variational approximation
\begin{align}
q_{\phi}(z_{t+1}|z_{t},h_{t})=N\left(\mu^{\phi}_{t+1},\Sigma^{\phi}_{t+1}, \right), \label{vaep}
\end{align}
where,
\begin{align}
\mu^{\phi}_{t+1}&=W_{\mu^{\phi}}h+b_{\mu^{\phi}}, \label{eq:approx_dist1}\\
\Sigma^{\phi}_{t+1}&=\texttt{softplus} ( W_{\Sigma^{\phi}} h+b_{\Sigma^{\phi}}), \label{eq:approx_dist2}\\
h&=\tfrac{1}{2}(\texttt{tanh}(W_{\phi } (z_{t})^{T} + b_{\phi}) + h_{t}),
\label{eq:approx_dist}
\end{align}
and $ h_{t}$ is the hidden state of a long short-term memory RNN (LSTM) \citep{Hochreiter1997} \index{LSTM} summarising past information up to the instant $t$, i.e., $(o_{1:t}, r_{1:t}, a_{1:t-1})$ (see Appendix \ref{LSTM_ex}).  To ensure the hidden layer $h$ is confined to the interval $\left[ -1,1 \right]$, we divide the right hand side of (\ref{eq:approx_dist}) by 2, since $tanh$ and $h_t$ ranges are $\left[ -1,1 \right]$.
The equations defining the LSTM cell are
\begin{align}
i_{t}&=\texttt{sigmoid} (W_{i}  X_{t}+U_{i}h_{t-1}+b_{i}) \quad &\textsf{input gate,}\\
f_{t}&=\texttt{sigmoid} (W_{f} X_{t}+U_{f}h_{t-1}+b_{f}) \quad &\textsf{forget gate,}\\
o_{t}&=\texttt{sigmoid} (W_{o}X_{t}+U_{o}h_{t-1}+b_{o}) \quad &\textsf{output gate,}\\
\tilde{c}_{t}&=\texttt{tanh}(W_{c}X_{t}+U_{c}h_{t-1}+b_{c}) \quad &\textsf{candidate cell,} \\
c_{t}&=f_{t}\circ c_{t-1}+i_{t}\circ \tilde{c}_{t} \quad &\textsf{cell state,}\\
h_{t}&=o_{t}\circ \texttt{tanh}(c_{t}) \quad &\textsf{output/hidden state,} \label{lstm_hidden}
\end{align}
where $X_{t}=(o_t,r_t,a_{t-1})$ represents the input array containing  prices, inventory, rewards and actions.

\subsection{Policy}

\begin{figure}[H]
	\centering
\begin{tikzpicture}[scale=.5]
\draw[->] (-3,0) -- (7,0) node (xaxis) [right] {$x$};
\draw[->] (0,-1) -- (0,6) node (yaxis) [above] {$f(x)$};
\draw	(-0.2,0) node[anchor=north] {0}
(1,0) node[anchor=north] {1}
(2,0) node[anchor=north] {2}
(-0.2,1) node[anchor=north] {1}
(-0.2,2) node[anchor=north] {2};
\draw [thick,color=blue] (-3,0) coordinate (a_01) -- (0,0) coordinate (a_02);
\draw [thick,color=blue] (0,0) coordinate (a_11) -- (4,4) coordinate (a_12);
\draw [thick,color=blue] (4,4) coordinate (a_21) -- (7,4) coordinate (a_22);
\coordinate (c) at (4,4);
\draw[dashed] (yaxis |- c) node[left] {$q_t$}
-| (xaxis -| c) node[below] {$q_t$};
\end{tikzpicture}
	\caption{Rectified linear unit (ReLU) with a cap at $q_t$} \label{fig:caprelu}
\end{figure}
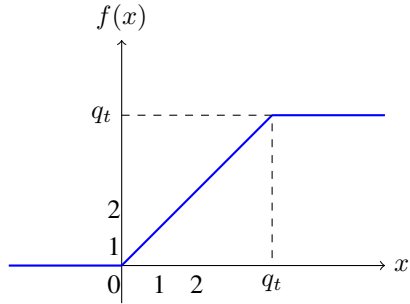

The policy $\pi$ is defined deterministically as an MLP using rectifier activation whose inputs are the mean and variance of the approximate posterior distribution over latent states  $q_{\phi}(z_{t}|z_{t-1},h_{t-1})$, specifically, the agent's actions are given by

\begin{align}
a_t=\pi(\mu^{\phi}_{t},\Sigma^{\phi}_{t},q_t|\psi),
\label{eq:pol}
\end{align}
where $q_t$ is the inventory available at the instant $t$ and
\begin{align}
\pi(\mu^{\phi}_{t},\Sigma^{\phi}_{t}|\psi) &=\min(\texttt{ReLU}(W_{\psi} h_{\psi}+b_{\psi}),q_t), \label{caprelu}\\
h_{\psi}&=\texttt{tanh}( W_{\psi I} (\mu^{\phi}_{t},\Sigma^{\phi}_{t})^T, b_{\psi I}),
\end{align}
where $\mu^{\phi}_{t},\Sigma^{\phi}_{t}$ are defined in Equation \eqref{eq:approx_dist1} and \eqref{eq:approx_dist2}. It is worth mentioning that, similarly to \cite{Krizhevsky10convolutionaldeep}, the function in Equation \ref{caprelu} is Rectified linear unit (ReLU) capped at $q_t$ (See Figure \ref{fig:caprelu}). 

\section{Model Optimisation}
\label{modeloptimisation}

Applying the conditional independence of $d$-separation to the graphical representation, we can factor the model in a convenient way, which helps with the optimisation problem. The graphical model representation of RDMM depicted in Figure \ref{fig:arch} has two objectives to maximise: (i) the conditional likelihood given a sequence of observed actions $a=(a_1,...,a_{t-1})$, and (ii) the unconditional expected reward for policy search. The first target fits the dynamical model parameters and the variational approximation of the posterior given frozen action parameters. The second target optimises actions given fixed dynamical model parameters.
\subsection{Maximization of the conditional likelihood}
We aim to maximise  $p_{\lambda}(\vec{o},\vec{r}|\vec{a})$ which is equivalent to maximising $\log p_{\lambda}(\vec{o},\vec{r}|\vec{a})$. First notice that,
\begin{align}
p_{\lambda}(\vec{z}|\vec{o},\vec{r},\vec{a}) &= \frac{p_{\lambda}(\vec{z},\vec{o},\vec{r},\vec{a})}{p_{\lambda}(\vec{o},\vec{r},\vec{a})} =\frac{p_{\lambda}(\vec{o},\vec{r}|\vec{z},\vec{a})p_{\lambda}(\vec{z}|\vec{a}) }{p_{\lambda}(\vec{o},\vec{r}|\vec{a})}
\end{align}
Therefore,
\begin{align} \label{eq:simple}
p_{\lambda}(\vec{o},\vec{r}|\vec{a}) &= \frac{p_{\lambda}(\vec{o},\vec{r}|\vec{z},\vec{a})p_{\lambda}(\vec{z}|\vec{a})}{p_{\lambda}(\vec{z}|\vec{o},\vec{r},\vec{a})}.
\end{align}

From variational inference theory and the identity in Equation \eqref{eq:simple}, we decompose $\log p_{\lambda}(\vec{o},\vec{r}|\vec{a})$ as,

\begin{align} 
\log p_{\lambda}(\vec{o},\vec{r}|\vec{a}) &= \log \frac{p_{\lambda}(\vec{o},\vec{r}|\vec{z},\vec{a})p_{\lambda}(\vec{z}|\vec{a})}{p_{\lambda}(\vec{z}|\vec{o},\vec{r},\vec{a})}\\
&=  \log{\frac{p_{\lambda}(\vec{o},\vec{r}|\vec{z},\vec{a})p_{\lambda}(\vec{z}|\vec{a})}{q_{\phi}(\vec{z}|\vec{o},\vec{r},\vec{a})}} - \log {\frac{p_{\lambda}(\vec{z}|\vec{o},\vec{r},\vec{a})}{q_{\phi}(\vec{z}|\vec{o},\vec{r},\vec{a})}} \\
&=\int q_{\phi}(\vec{z}|\vec{o},\vec{r},\vec{a}) \log{\frac{p_{\lambda}(\vec{o},\vec{r}|\vec{z},\vec{a})p_{\lambda}(\vec{z}|\vec{a})}{q_{\phi}(\vec{z}|\vec{o},\vec{r},\vec{a})}} d\vec{z} \\
&- \int q_{\phi}(\vec{z}|\vec{o},\vec{r},\vec{a}) \log {\frac{p_{\lambda}(\vec{z}|\vec{o},\vec{r},\vec{a})}{q_{\phi}(\vec{z}|\vec{o},\vec{r},\vec{a})}} d\vec{z} \nonumber \\
&=\mathcal{L}(q)+KL\left[q_{\phi}(\vec{z}|\vec{o},\vec{r},\vec{a})||p_{\lambda}(\vec{z}|\vec{o},\vec{r},\vec{a})\right] \label{vaepdeco}
\end{align}
where,
\begin{align}
\mathcal{L}(q):=\int q_{\phi}(\vec{z}|\vec{o},\vec{r},\vec{a}) \log{\frac{p_{\lambda}(\vec{o},\vec{r}|\vec{z},\vec{a})p_{\lambda}(\vec{z}|\vec{a})}{q_{\phi}(\vec{z}|\vec{o},\vec{r},\vec{a})}} d\vec{z}. 
\end{align}
Since $KL\left[q_{\phi}(\vec{z}|\vec{o},\vec{r},\vec{a})||p_{\lambda}(\vec{z}|\vec{o},\vec{r},\vec{a})\right] \geq 0$, we have
\begin{align}
\log p_{\lambda}(\vec{o},\vec{r}|\vec{a}) \geq  \mathcal{L}(q).
\end{align}

For this reason the term $\mathcal{L}$ is called the evidence lower bound (ELBO)\index{ELBO} on $\log p_{\lambda}(\vec{o},\vec{r}|\vec{a})$, and it is the objective function that we want to maximise during the learning process in order to maximise $\log p_{\lambda}(\vec{o},\vec{r}|\vec{a})$ . Another consequence of the ELBO maximisation is to provide a good approximation between $q_{\phi}(\vec{z}|\vec{o},\vec{r},\vec{a})$  and the intractable posterior $p_{\lambda}(\vec{z}|\vec{o},\vec{r},\vec{a})$.

More explicitly, we know that $KL\left[q_{\phi}(\vec{z}|\vec{o},\vec{r},\vec{a})||p_{\lambda}(\vec{z}|\vec{o},\vec{r},\vec{a})\right] = 0$ if and only if $q_{\phi}(\vec{z}|\vec{o},\vec{r},\vec{a})$ is equal to the posterior $p_{\lambda}(\vec{z}|\vec{o},\vec{r},\vec{a})$. Consequently, by Equation \eqref{vaepdeco} we see that maximising $\mathcal{L}$ implies minimising $KL\left[q_{\phi}(\vec{z}|\vec{o},\vec{r},\vec{a})||p_{\lambda}(\vec{z}|\vec{o},\vec{r},\vec{a})\right]$ since the model evidence $p_{\lambda}(\vec{o},\vec{r}|\vec{a})$ does not depend on $q_{\phi}$.

In the theorem below we further characterise the ELBO for the RDMM architecture summarised in Figure \ref{fig:arch}.

\textbf{Theorem.} The evidence lower bound $\mathcal{L}$ of the conditional log likelihood is given by

\begin{align} 
\mathcal{L}=&\underset{z_1 \sim q_{\phi}(z_{1}|\vec{o},\vec{r},\vec{a})}{\mathbb{E}}\left[\log p_{\theta}(o_{1}|z_{1},a_{0})\right] \nonumber \\
&+ \underset{z_1 \sim q_{\phi}(z_{1}|\vec{o},\vec{r},\vec{a})}{\mathbb{E}}\left[\log p_{\eta}(r_{1}|z_{1},a_{0})\right] \nonumber\\
&+ \sum_{t=2}^{T} \underset{z_{t} \sim q_{\phi}(z_{t}|z_{t-1},\vec{o},\vec{r},\vec{a})}{ \mathbb{E}}\left[\log p_{\theta}(o_{t}|z_{t},a_{t-1})\right] \nonumber\\
&+ \sum_{t=2}^{T} \underset{z_{t} \sim q_{\phi}(z_{t}|z_{t-1},\vec{o},\vec{r},\vec{a})}{ \mathbb{E}}\left[\log p_{\eta}(r_{t}|z_{t},a_{t-1})\right] \nonumber\\
&- KL(q_{\phi}(z_1|\vec{o},\vec{r},\vec{a})||p_{\lambda}(z_1|\vec{a})) \nonumber\\
&-\sum_{t=2}^{T} \underset{z_{t-1} \sim q_{\phi}(z_{t-1}|z_{t-2},\vec{o},\vec{r},\vec{a})}{\mathbb{E}}\left[ KL(q_{\phi}(z_{t}|z_{t-1},\vec{o},\vec{r},\vec{a})||p_{\lambda}(z_{t}|z_{t-1},\vec{a})) \right]. \label{eq:lb1}
\end{align}

\textbf{Proof:}
First we have by definition
\begin{align*}
\mathcal{L}(q)&=\int q_{\phi}(\vec{z}|\vec{o},\vec{r},\vec{a}) \log{p_{\lambda}(\vec{o},\vec{r}|\vec{z},\vec{a})}d\vec{z} +  \int q_{\phi}(\vec{z}|\vec{o},\vec{r},\vec{a}) \log \frac{p_{\lambda}(\vec{z}|\vec{a})}{q_{\phi}(\vec{z}|\vec{o},\vec{r},\vec{a})} d\vec{z}\\
&=\int q_{\phi}(\vec{z}|\vec{o},\vec{r},\vec{a})  \log{p_{\lambda}(\vec{o},\vec{r}|\vec{z},\vec{a})}d\vec{z}  - KL\left[q_{\phi}(\vec{z}|\vec{o},\vec{r},\vec{a}) ||p_{\lambda}(\vec{z}|\vec{a})\right]\\
&= \underset{\vec{z} \sim q_{\phi}(\vec{z}|\vec{o},\vec{r},\vec{a})}{\mathbb{E}}\left[\log p_{\lambda}(\vec{o},\vec{r}|\vec{z},\vec{a}) \right] - KL\left[q_{\phi}(\vec{z}|\vec{o},\vec{r},\vec{a}) ||p_{\lambda}(\vec{z}|\vec{a})\right].
\end{align*}

Applying $d$-separation and the conditional independence of $o_t$ and $r_t$ given $z_t$ and $a_t$, (see \eqref{price_emission}) we have,
\begin{align*}
p_{\lambda}(\vec{o},\vec{r}|\vec{z},\vec{a}) &= \prod_{t=1}^{T} p_{\theta}(o_{t}|z_{t},a_{t-1}) p_{\eta}(r_{t}|z_{t},a_{t-1}).
\end{align*}
Similarly $q_{\phi}(\vec{z}|\vec{o},\vec{r},\vec{a})$ can be factorised as
\begin{align*}
q_{\phi}(\vec{z}|\vec{o},\vec{r},\vec{a})&=q_{\phi}(z_{1}|\vec{o},\vec{r},\vec{a}) \prod_{t=2}^{T} q_{\phi}(z_{t}|z_{t-1},\vec{o},\vec{r},\vec{a}).
\end{align*}
and combining these two equalities we have
\begin{gather}
\begin{aligned}
\mathcal{L} =&\int...\int q_{\phi}(z_{1}|\vec{o},\vec{r},\vec{a}) \times \\
& \prod_{t=2}^{T} q_{\phi}(z_{t}|z_{t-1},\vec{o},\vec{r},\vec{a}) \left[ \sum_{t=1}^{T}  \log p_{\theta}(o_{t}|z_{t},a_{t-1}) + \log p_{\eta}(r_{t}|z_{t},a_{t-1}) \right] dz_1...dz_T \\
&- KL\left[q_{\phi}(\vec{z}|\vec{o},\vec{r},\vec{a})||p_{\lambda}(\vec{z}|\vec{a})\right]\\
&=\int q_{\phi}(z_{1}|\vec{o},\vec{r},\vec{a})\left[ \log p_{\theta}(o_{1}|z_{1},a_{0}) + \log p_{\eta}(r_{1}|z_{1},a_{0})\right]dz_1\\ 
&+ \sum_{t=2} ^{T}\int  q_{\phi}(z_{t}|z_{t-1},\vec{o},\vec{r},\vec{a}) \left[ \log p_{\theta}(o_{t}|z_{t},a_{t-1}) + \log p_{\eta}(r_{t}|z_{t},a_{t-1})\right] dz_t\\
& - KL\left[q_{\phi}(\vec{z}|\vec{o},\vec{r},\vec{a})||p_{\lambda}(\vec{z}|\vec{a})\right]\\
&=\underset{z_1 \sim q_{\phi}(z_{1}|\vec{o},\vec{r},\vec{a})}{\mathbb{E}}\left[\log p_{\theta}(o_{1}|z_{1},a_{0})\right] + \underset{z_1 \sim q_{\phi}(z_{1}|\vec{o},\vec{r},\vec{a})}{\mathbb{E}}\left[\log p_{\eta}(r_{1}|z_{1},a_{0})\right]\\
&+ \sum_{t=2}^{T} \underset{z_{t} \sim q_{\phi}(z_{t}|z_{t-1},\vec{o},\vec{r},\vec{a})}{ \mathbb{E}}\left[\log p_{\theta}(o_{t}|z_{t},a_{t-1})\right] \\
&+ \sum_{t=2}^{T} \underset{z_{t} \sim q_{\phi}(z_{t}|z_{t-1},\vec{o},\vec{r},\vec{a})}{ \mathbb{E}}\left[\log p_{\eta}(r_{t}|z_{t},a_{t-1})\right]\\
&- KL\left[q_{\phi}(\vec{z}|\vec{o},\vec{r},\vec{a})||p_{\lambda}(\vec{z}|\vec{a})\right].
\end{aligned}
\label{eq:lb2}
\end{gather}
\normalsize

Finally, a similar factorization can be applied to $ KL\left[q_{\phi}(\vec{z}|\vec{o},\vec{r},\vec{a})||p_{\lambda}(\vec{z}|\vec{a})\right]$ to obtain
\begin{gather}
\begin{aligned}
&KL\left[q_{\phi}(\vec{z}|\vec{o},\vec{r},\vec{a})||p_{\lambda}(\vec{z}|\vec{a})\right] =-\int ...\int q_{\phi}(z_{1}|\vec{o},\vec{r},\vec{a})\times\\
&\prod_{t=2}^{T} q_{\phi}(z_{t}|z_{t-1},\vec{o},\vec{r},\vec{a}) \log \left[\frac{ p_{\lambda}(z_1|\vec{a}) \prod_{t=2}^{T} p_{\lambda}(z_{t}|z_{t-1},\vec{a}) }{q_{\phi}(z_{1}|,\vec{o},\vec{r},\vec{a}) \prod_{t=2}^{T} q_{\phi}(z_{t}|z_{t-1},\vec{o},\vec{r},\vec{a})}  \right] dz_1...dz_T\\
=& -\int q_{\phi}(z_{1}|\vec{o},\vec{r},\vec{a}) \log \frac{ p_{\lambda}(z_1|a)}{q_{\phi}(z_{1}|\vec{o},\vec{r},\vec{a})} dz_1\\
&- \sum_{t=2}^{T} \int \int q_{\phi}(z_{t-1}|z_{t-2},\vec{o},\vec{r},\vec{a}) q_{\phi}(z_{t}|z_{t-1},\vec{o},\vec{r},\vec{a}) \log \frac{p_{\lambda}(z_{t}|z_{t-1},\vec{a})}{q_{\phi}(z_{t}|z_{t-1},\vec{o},\vec{r},\vec{a})} dz_{t-1}dz_{t-2}\\
=&KL(q_{\phi}(z_1|\vec{o},\vec{r},\vec{a})||p_{\lambda}(z_1|\vec{a}))\\
&+ \sum_{t=2}^{T} \underset{z_{t-1} \sim q_{\phi}(z_{t-1}|z_{t-2},\vec{o},\vec{r},\vec{a})}{\mathbb{E}}\left[ KL(q_{\phi}(z_{t}|z_{t-1},\vec{o},\vec{r},\vec{a})||p_{\lambda}(z_{t}|z_{t-1},\vec{a})) \right].
\end{aligned}
\label{eq:lb3}
\end{gather}
\normalsize
Combining the results in Equations \eqref{eq:lb2} and \eqref{eq:lb3} we obtain Equation \eqref{eq:lb1}. \qed

In \eqref{eq:lb1} the expectations are with respect to the latent states $z_t$ and are approximated using the reparametrization trick (see Appendix \ref{s:VAE}). On the other hand, the KL divergence terms has a closed form expression,  as the prior $p_\lambda$ and the posterior $q_\phi$ are normally distributed. 

As a consequence of Equation \eqref{eq:lb1} the gradients of $\mathcal{L}$ with respect to the parameters will propagate to the entire architecture via reverse-mode automatic differentiation (See Appendix \ref{autodiff}). Indeed, from Equation \eqref{eq:lb1} the emission functions \eqref{price_emission} in the ELBO the parameters $\theta$ and $\eta$, represent the weights and biases of their respective MLPs, i.e.,  $\theta$ through $\left\lbrace W_{\mu^\theta}, W_{\Sigma^\theta}, W_{\theta} \right\rbrace$ and $\eta$ through $\left\lbrace W_{\mu^\eta}, W_{\Sigma^\eta}, W_{\eta} \right\rbrace$.
Likewise, for the $KL$ divergence in Equation \eqref{eq:lb1} the transition/prior distribution $p_{\lambda}$ \eqref{eq:transitions}, contain the parameters $\lambda$ that summarises all weights and biases of the neural networks: $\lambda = \left\lbrace W_{NL_2}, W_{NL_1}, W_{L}, W_{g_2} , W_{g_1}, W_{\Sigma^\lambda}, b_{NL_2}, b_{NL_1}, b_{L}, b_{g_2}, b_{g_1} ,b_{\Sigma^\lambda}  \right\rbrace $. Similarly, $\phi$ represents  all parameters of the neural networks in the posterior/approximated distribution, including the LSTM, i.e., $\phi$ contains $\left\lbrace W_{\mu^\phi}, W_{\Sigma^\phi}, W_{\phi}, W_i, W_f, W_o, W_c, U_i, U_f, U_o, U_c, b_{\mu^\phi}, b_{\Sigma^\phi}, b_{\phi}, b_i, b_f, b_o, b_c \right\rbrace $. For the objective functions defined by Equation \eqref{eq:lb1} the reverse-mode AD computational time for computing the gradient of $\mathcal{L}$ is similar to the computational time of the ELBO.

Recall the actions $\vec{a}$ in $\mathcal{L}$ are considered to be observed and the corresponding network parameters considered fixed. The role of the actions, with regards to $\mathcal{L}$ optimisation, is to allow the model to ``learn'' the interplay between the actions and the emissions of prices and rewards through Equation \eqref{eq:lb1}.

\subsection{Policy Search}

In this section, the agent's actions are the outputs of an MLP whose inputs are the parameters of the posterior approximated distribution, as defined in Equation \eqref{eq:pol}, and the parameters are chosen to maximise the unconditional expected reward\index{uncoditional expected reward}
\begin{align} \label{policyobjective}
\mathcal{J}= \mathbb{E}_{p(r)} \left[ R \right],
\end{align}
where $R=\sum_{t=1}^{T} r_{t}$ and $p(r)=p(r_1,..,r_T)$ is the joint distribution of rewards. 

Through iterated expectations we have
\begin{align}
\mathcal{J}&= \mathbb{E}_{p(r)} \left[ \sum_{t=1}^{T} r_{t} \right] \nonumber\\
 &=  \sum_{t=1}^{T}  \mathbb{E}_{p(z_t,a_{t-1})} \left[\, \mathbb{E}_{p(r_{t}|z_t,a_{t-1})} \left[r_{t} \right]\, \right] \nonumber\\
&=  \sum_{t=1}^{T}  \mathbb{E}_{p(z_t,a_{t-1})} \left[\,\mu^{\eta}_{t} \,\right] \nonumber\\
 &=\sum_{t=1}^{T}  \mathbb{E}_{p(z_t,a_{t-1})}  \left[\, MLP_{\mu^{\eta}_{t}}(z_{t},a_{t-1})\,\right]  \nonumber\\
&= \sum_{t=1}^{T}  \mathbb{E}_{p(z_t,a_{t-1})}  \left[\, MLP_{\mu^{\eta}_{t}}(z_{t},\pi(\mu^{\phi}_{t-1},\Sigma^{\phi}_{t-1}|\psi)) \,\right]
\label{eq:rl_app} 
\end{align}
where $MLP_{\mu^{\eta}_{t}}(z_{t},a_{t-1})$ represents the multi-layer perceptron parametrising $\mu^{\eta}_{t}$ as defined in Equation \eqref{price_emission}.

The unconditional expected reward $\mathcal{J}$ in \eqref{eq:rl_app}  is approximated by

\begin{align}
\mathcal{J} & \approx  \sum_{t=1}^{T} MLP_{\mu^{\eta}_{t}}(z_{t},\pi(\mu^{\phi}_{t-1},\Sigma^{\phi}_{t-1}|\psi)).
\label{eq:rl_app11} 
\end{align}

Therefore, our goal is to find $\psi^{*}$ such that,

\begin{align}
\psi^{*} & =  \underset{\psi}{\mathrm{argmax}}~  \sum_{t=1}^{T} MLP_{\mu^{\eta}_{t}}(z_{t},\pi(\mu^{\phi}_{t-1},\Sigma^{\phi}_{t-1}|\psi)).
\label{eq:rl_app2} 
\end{align}

As in the ELBO maximisation, we use reverse-mode AD to compute the gradient of $\mathcal{J}$ with respect to $\psi := \left\lbrace W_{\psi}, W_{\psi I}, b_{\psi}, b_{\psi I} \right\rbrace$, while the set of parameters  $\theta$, $\eta$, $\lambda$ and $\phi$ are kept fixed.

\section{Learning}
\label{RDMMlearning}

The learning algorithm has two  goals. The first goal is to obtain a good model representation of the observed data given the actions taken. In this phase we maximise the lower bound in Equation \eqref{eq:lb1} in order to maximise the conditional log-likelihood  $\log p_{\lambda}(\vec{o},\vec{r}|\vec{a})$.  The second goal is to perform a \emph{policy search}\index{policy search} by maximising unconditional expected reward in Equation \eqref{eq:rl_app}, and extract the best policy using a gradient-based method.

Algorithm \ref{alg:RDMM} summarises the steps in our approach, which can be broken into three main parts . The first part defines  the functions of our RDMM model. It is important to follow the correct order of construction of the graphical architecture in Figure \ref{fig:arch}. In our implementation, Theano will compile this graph in native machine instructions into a callable object. The second part is the ELBO optimisation through the  application of Equation \eqref{eq:lb1}, where gradients and the parameters of the compiled graph are updated using the Adam optimisation algorithm \citep{DBLP:journals/corr/KingmaB14}. The third part of the algorithm is implemented after the ELBO convergence is achieved. In this part we perform policy search to maximise the unconditional expected reward $\mathcal{J}$ approximation in Equation \eqref{eq:rl_app11}. 

\def\HiLi{\leavevmode\rlap{\hbox to \hsize{\color{gray!30}\leaders\hrule height .8\baselineskip depth .5ex\hfill}}}

\begin{algorithm}[H] 
\HiLi	\tcc{PART 1: RDMM graph architecture}
	Define  the LSTM for hidden state $h_{t}$ summarising past information \eqref{lstm_hidden}\;
	Define the MLPs in  $\mu^{\phi}_{t}$ and $\Sigma^{\phi}_{t}$ for the inference layer $q_{\phi}(z_{t}|z_{t-1},h_{t-1})$  \eqref{vaep}\;
	Define MLPs $\mu^{\theta}_t$, $\Sigma^{\theta}_t$, $\mu^{\eta}_t$ and $\Sigma^{\eta}_t$ of the emission functions \eqref{price_emission}\;
	Define the MLPs $\mu^{\lambda}_t$ and $\Sigma^{\lambda}_t$ for transitions  \eqref{eq:transitions}\;
	Build MLP of the actions  $a_t$ \eqref{eq:pol}\;
	Initialise all parameters\;
\HiLi    \tcc{PART 2: ELBO  Optimisation} 
  \For{$i \in 0:epochs1$}{
    \For{all minibatches}{
    Compute $h_{t}$, $\mu^{\phi}_{t}$ and $\Sigma^{\phi}_{t}$ of PART 1 for all $t$ in the mini-batch\;
    Sample $z_{t}= \mu^{\phi}_{t} + \epsilon \Sigma^{\phi}_{t}$ where $\epsilon \sim N(0,1)$ for all $t$ in the mini-batch\;
    Compute $\mu^{\theta}_t$, $\Sigma^{\theta}_t$, $\mu^{\eta}_t$, $\Sigma^{\eta}_t$, $\mu^{\lambda}_t$ and $\Sigma^{\lambda}_t$ of PART 1 for all $t$ in the mini-batch\;
    Compute $\mathcal{L}$ using \eqref{eq:lb1} and check convergence\;
    Compute gradients of $\mathcal{L}$ using reverse-mode AD (Theano)\; 
	Update parameters $\lambda$, $\eta$, $\theta$ and $\phi$  related to $\mathcal{L}$ optimisation using ADAM\;
}
  }
 \HiLi   \tcc{PART 3: Policy Search Optimisation}
    \For{$i \in 0:epochs2$}{
    \For{all minibatches}{  
    Compute actions $a_t$ using  \eqref{eq:pol}\;
    Compute $\mathcal{J}$ using \eqref{eq:rl_app11}\; 
    Compute gradients of $\mathcal{J}$  using reverse-mode AD (Theano)\; 
    Update parameters  $\psi$ for policy search using ADAM\;
  }
  }    
 \caption{Overview of the RDMM learning}
 \label{alg:RDMM}
\end{algorithm}

\section{Optimal Liquidation Problem}
\label{liquidation}

To test the algorithms presented in this article, we evaluate their ability to perform a liquidation task. By liquidation, we mean to sell all stocks of an inventory where the agent's goal is to optimally sell a certain number of stocks with prices subject to a model dynamics. This situation can be easily converted to an acquisition problem. For a more formal treatment of these financial problems, we recommend \citet{Casgrain2017}.

\nocite{CarteaJaimungal2016}
\nocite{Cartea2017}
\nocite{Rudin1976}

Let us represent the inventory and the trader's actions at time $t$  by $q_{t}$ and $a_{t}$, respectively. In the liquidation problem, we assume that the trader starts with a positive inventory of $q_{0} \geq 0$ shares of an asset. The mid-price at instant $t$ is represent by $x_t$. In our reinforcement setting the pair $s_t=(x_t,q_t)$ comprised of price and inventory represent the the agent's state. Most of the derivations on this article could be readily expanded such that $x_t$ could represent a matrix containing all the information available in the limit order book at instant $t$. For our purposes $x_t$, $a_t$ and $q_t$ $\in \mathbb{R}^{+}$.

When a certain number of shares $a_{t}$ is liquidated at instant $t$, it is reasonable to assume a \emph{permanent price impact} due to the volume traded, or in the financial jargon, where the order executed ``walk the book'' causing a decrement of the mid-prices. As shown in \citet{CarteaJaimungal2016}, the permanent price impact can be satisfactorily approximated by a linear model. For simplicity purposes let us represent the permanent impact by $c_{1} x_{t}$, with $c_{1}$ being a scalar.


We may also consider a \emph{temporary price impact }due to cost transactions, which can also be approximated by a linear model as shown in \citet{Frei2015}. Thus, we represent the temporary price impact by a scalar multiplied by the number of stocks sold, i.e., $c_{2} a_{t}$. Consequently, the return obtained by an action of $a_{t}$ shares liquidated at step $t$ can be represented as
\begin{equation}
r_t=(x_{t}- c_{2} a_{t})a_{t}
\label{simplereward}
\end{equation}

\section{Synthetic Data Experiments}
\label{simulations_and_results}

We conduct simulations to solve the liquidation problem where the stock mid-price $x_{t}$ are mean-reverting, \citep{Cartea2017}, \citep{Casgrain2017}, which assumes that prices revert back towards the average price. The  mid-price dynamics of mean reversion can be formulated as follows,
\begin{align} 
x_{t+1}=-c_1 a_t + \theta + e^{-\kappa \Delta t}(x_{t}-\theta)+  vol \times \epsilon_{t}  \quad \text{with} \quad \epsilon_{t} \sim N(0,1) \quad i.i.d. 
\label{eq:mean_rev2}
\end{align} 
where $x_t$ represents the mid-price of a stock at instant $t$, the constant $\theta$ is the mean price, the term $c_1 a_t$ represents the permanent price impact, $\kappa$ is the mean-reversion rate, and 
\begin{equation}
vol= \sigma \sqrt{((1-\exp(-2\kappa \Delta t))/(2 \kappa))}  \approxeq \sigma \sqrt{\Delta t} 
\end{equation}
is the asset's volatility term that makes the price deviate from the mean.  The reward is defined by
\begin{equation}
r_t=(x_{t}- c_{2} a_{t})a_{t} - c_3 q_t^2,
\label{simplereward22}
\end{equation}
which corresponds to the proceeds from trading, a penalty for speed of trading as a proxy for liquidity costs, and a penalty for holding inventory as a proxy for urgency or quadratic variation.

In our simulations we used $\theta=10$, $\kappa=0.005$, $\sigma=0.02$, $c_1=c_2=c_3=0.001$, and $\Delta t=1$.

The training set for the RDMM model has 400 trajectories of size $T=500 s$ (200,000 data points in total) with prices following the mean reversion dynamics. Each trajectory starts with the asset's price equal to \$10 and inventory 0 units. After t=40s a sequence of a fictional inventory, actions and the respective rewards is added up to t=400s. The details of this construction are presented in Algorithm  \ref{alg:inv_act}.

\begin{algorithm}[H]
  $x_0=10$\;
  $q_0=0$\; 
  $i=20$\;
  \For{$t \in 0:200,000$}{
  \If{t mod i=0}{
   	 $i=i+n$  where $n \sim \mathcal{U}_{d}\{7,13\}$\footnote{$\mathcal{U}_{d}\{a,b\}$ stands for a uniform discrete distribution over $a, a+1, .... ,b$}\;
 	 \uIf{$40<i<400$}{
 	 $q_t=u$ where $u \sim \mathcal{U}_{d}\{0,10\}$\;
  }
  \uElse{
 	 $q_t=0.0$ \;
 	 }
  }
  \If{t mod 500=0}{
  $x_t=10.0$\;
  $q_t=0.0$\;  
  }
  $a_t=a$ where $a \sim \mathcal{U}_{d}\{0,q_t\}$\;
  $r_t=x_t a_t-c_2 a_t^2-c_3 q_t^2$\;
  $x_{t+1}=f(x_t,c_1,\kappa,\theta,\sigma,a_t, \epsilon_{t})$ where $f$ is the mean reversion in Equation \eqref{eq:mean_rev2}\;
  $q_{t+1}=q_t-a_t$\;
  }
 \caption{Simulation training data set creation}
 \label{alg:inv_act}
\end{algorithm}

\vspace{1cm}

Each trajectory of length 500 has zero activity in the first 40 timestamps and the last 100. In this manner we constrain the actions to not take place near the extremes of the time series. The reason for this choice is that we do not want the terminal inventory to become a major concern for the agents and outbalance the other strategies involved in the liquidation process.

Similarly, the choice of the intervals of the discrete uniform distributions is not completely arbitrary. 
Adding inventory in intervals spaced by $\mathcal{U}_{d}\{7,13\}$ is more complex than equally spaced intervals. If these intervals are increased, hypothetically, to $\mathcal{U}_{d}\{40,60\}$, more than 400 trajectories would be needed to train the model satisfactorily.

During the first part of the training phase, the RDMM model learns not only the price dynamics but also the interplay among prices, the inventory, the actions and the rewards. A good approximation of the training set is vital to a proper policy search driven by Equation \eqref{eq:rl_app} to provide good results.

 We compare the RDMM model against three benchmarks; Q-learning, DynaQ with an autoregressive integrated moving average model ARIMA (DynaQ-ARIMA) and DynaQ with long short-term memory network (DynaQ-LSTM). The details of the benchmark implementation may be found in Appendix \ref{qlearning} on Algorithms \ref{alg:qlearning} (Q-learning) and \ref{alg:dynaq} (DynaQ and variations).

\begin{figure}[H]
\centering
\begin{minipage}{.5\textwidth}
  \centering
  \includegraphics[width=0.7\textwidth]{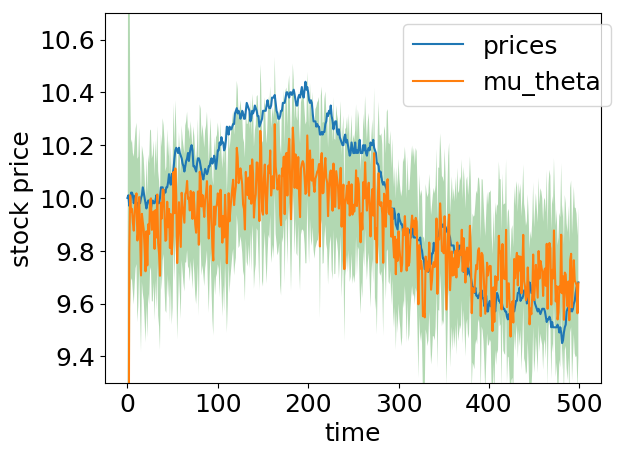}
\end{minipage}%
\begin{minipage}{.5\textwidth}
  \centering
  \includegraphics[width=0.7\textwidth]{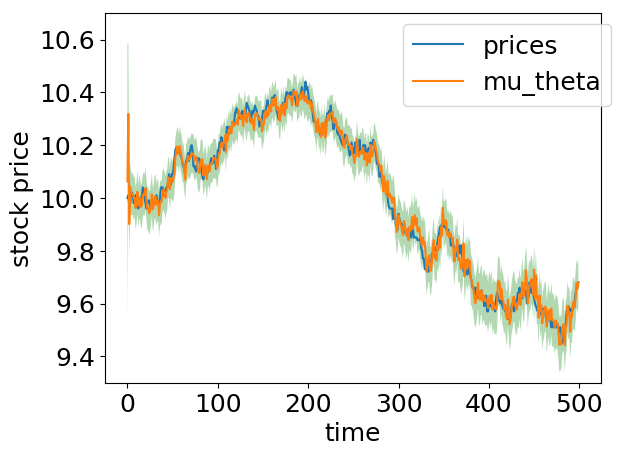}
\end{minipage}
\caption{Price approximations of a sampled single training batch with horizon T=500. Model approximation at 25 (left) and 225 (right) epochs}
\label{fig:prices_approx}
\end{figure}

After training, we test the  policies in 10,000 simulated time series with a trading horizon $T=500$ generated by Equation \eqref{eq:mean_rev2}. We use Algorithm \ref{alg:inv_act} from the training set, to create 10,000 batches of size 500 each for the test set. Figure \ref{fig:prices_approx} shows the price approximations(see Equation \eqref{eq:mu_theta}) for a single training batch with horizon $T=500$ with 25 epochs on the left panel and 225 epochs on the right panel. The green shaded area corresponds to  $\mu_{\theta} \pm 2 \sigma_{\theta}$ (see Equation \eqref{price_emission}).

\begin{figure}[H]
\centering
\begin{minipage}{.5\textwidth}
  \centering
  \includegraphics[width=0.7\textwidth]{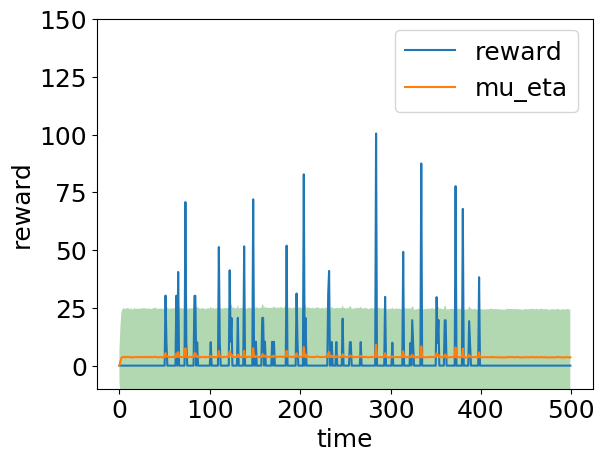}
\end{minipage}%
\begin{minipage}{.5\textwidth}
  \centering
  \includegraphics[width=0.7\textwidth]{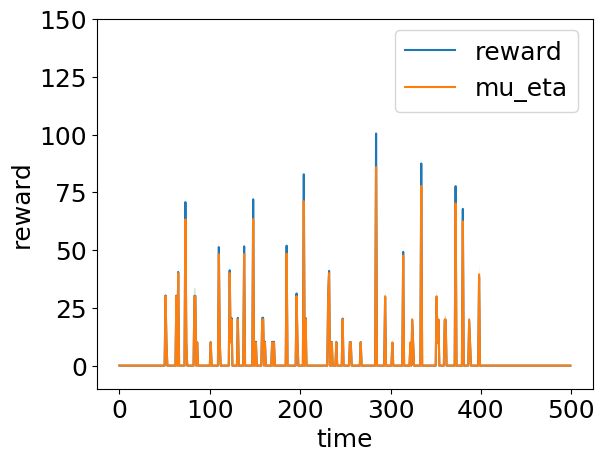}
\end{minipage}
\caption{Reward approximation of a sampled single training batch with horizon T=500. Model approximation at 25 (left) and 225 (right) epochs}
\label{fig:reward_approx}
\end{figure}

In Figure \ref{fig:reward_approx} we plot the reward approximations (see Equation \eqref{eq:mu_theta}) for the same training batch as in Figure \ref{fig:prices_approx}.

\begin{figure}[H] 
	\centering
		\includegraphics[width=0.40\textwidth]{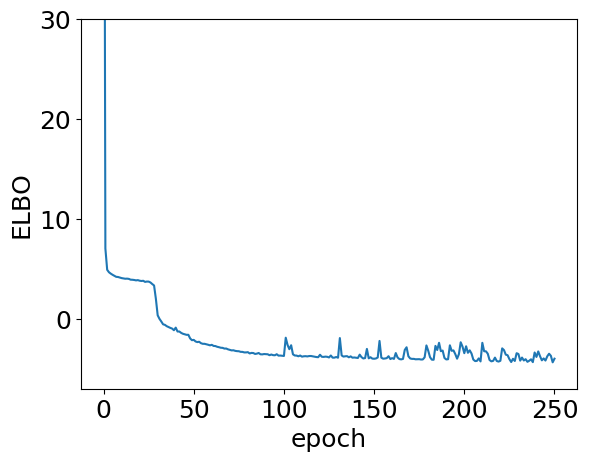}
	\caption{Convergence of the negative ELBO with $1 \times 10^{-4}$ learning rate}
	\label{fig:ELBO}
\end{figure}

The convergence of the negative evidence lower bound (ELBO) of the conditional log-likelihood $\mathcal{L}$ (see Equation \eqref{eq:lb1}) during the training phase is shown by Figure \ref{fig:ELBO}. We used ADAM with a learning rate of $1 \times 10 ^{-4}$ for stochastic optimisation.

\begin{figure}[H] 
	\centering
		\includegraphics[width=0.40\textwidth]{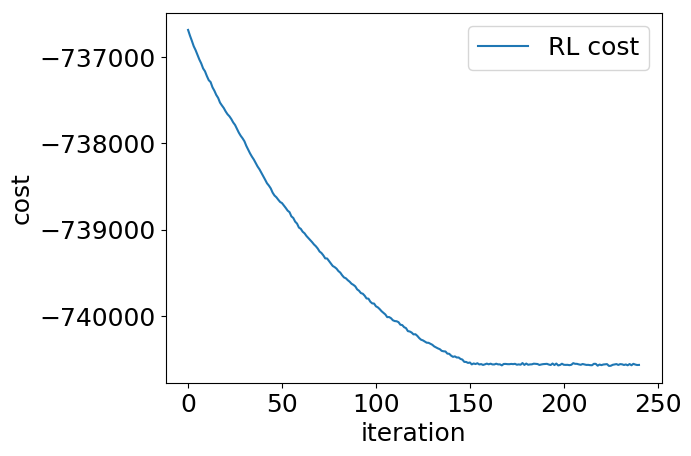}
	\caption{Convergence of unconditional expected reward approximation in Equation \eqref{eq:rl_app}  with ADAM optimisation using a learning rate of $2 \times 10^{-7}$.}
	\label{fig:RL_Cost}
\end{figure}

The convergence of the negative approximation of the  unconditional expected reward (see Equation \eqref{eq:rl_app}) is shown by Figure \ref{fig:RL_Cost}. Similarly to the ELBO,  we use ADAM with a learning rate  of $2 \times 10 ^{-7}$ for stochastic optimisation.

\begin{figure}[H] 
	\centering
		\includegraphics[width=0.5\textwidth]{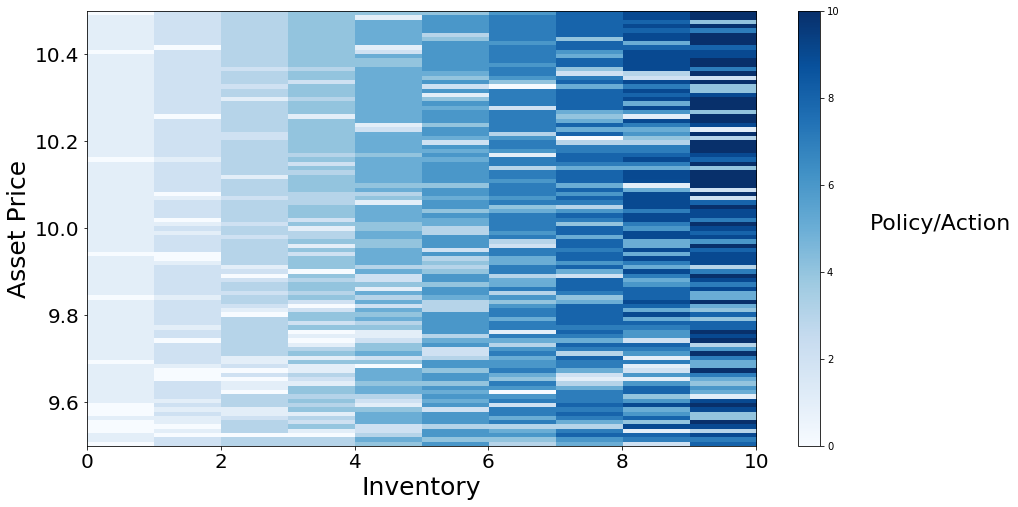}
	\caption{Q-learning policy heatmap across price and inventory. Darker shades represent more stocks sold.}
	\label{fig:qlearning_policy}
\end{figure}

\begin{figure}[H] 
	\centering
		\includegraphics[width=0.5\textwidth]{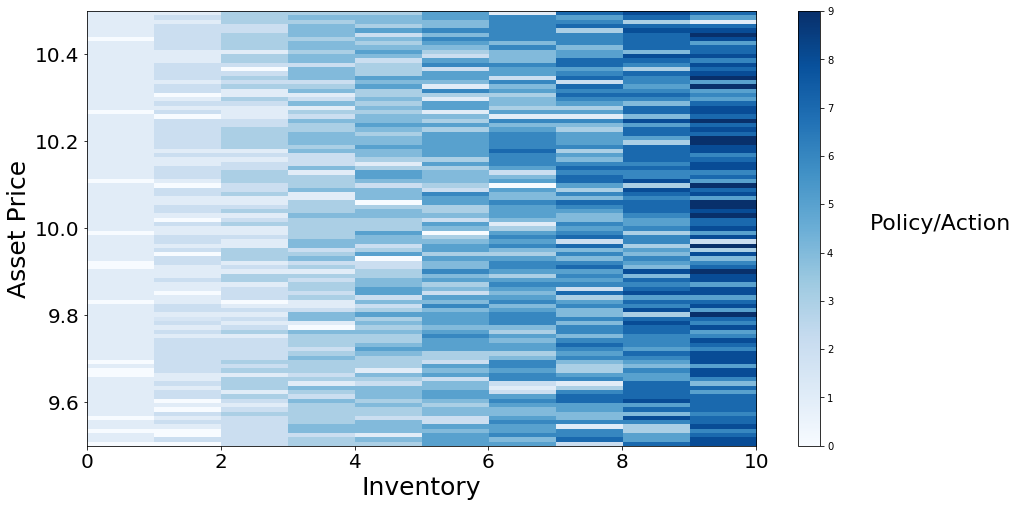}
	\caption{DynaQ-ARIMA policy heatmap across price and inventory. Darker shades represent more stocks sold.}
	\label{fig:dynaq-arima_policy}
\end{figure}

\begin{figure}[H] 
	\centering
		\includegraphics[width=0.50\textwidth]{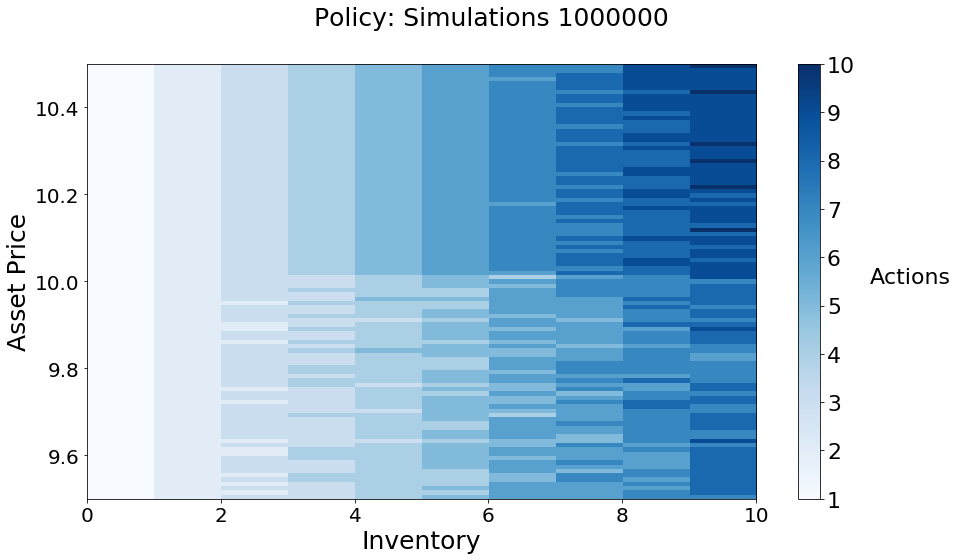}
	\caption{DynaQ-LSTM policy heatmap across price and inventory. Darker shades represent more stocks sold.}
	\label{fig:dynaq-lstm_policy}
\end{figure}

To compare the RDMM method against the benchmarks, we record and keep track of all actions taken by the algorithms, asset prices and the rewards.
To better visualise and understand the policies behind each algorithm we create heatmaps of actions across states, prices, and inventory. The heatmaps contain different shades of blue representing the average action taken (i.e. the number of stocks executed) by the algorithm in question across the 10,000 time-series batches created for the test set. Figures \ref{fig:qlearning_policy}, \ref{fig:dynaq-arima_policy} and \ref{fig:dynaq-lstm_policy} contain the policy heatmaps for Q-learning, DynaQ-ARIMA and DynaQ-LSTM, respectively.
The actions taken by these benchmarks (Q-learning and DynaQ algorithms) use an $\epsilon$-greedy strategy exploration, where a random action is taken with a frequency $(100  \epsilon) \%$  of the time and they vary only with price and inventory. In other words, the benchmarks are Markov in $s_t=(x_t,q_t)$. The RDMM model, however, is not Markov in $(x_t,q_t)$ alone, as each action depends on the approximated distribution which in turn depends on the entire path. As shown before, the model architecture summarises past information by an LSTM hidden state shown by the graphical representation in Figure \ref{fig:arch} and in more detail in Equations \eqref{eq:pol} and \eqref{eq:approx_dist}. 

\begin{figure}[H] 
	\centering
		\includegraphics[width=0.50\textwidth]{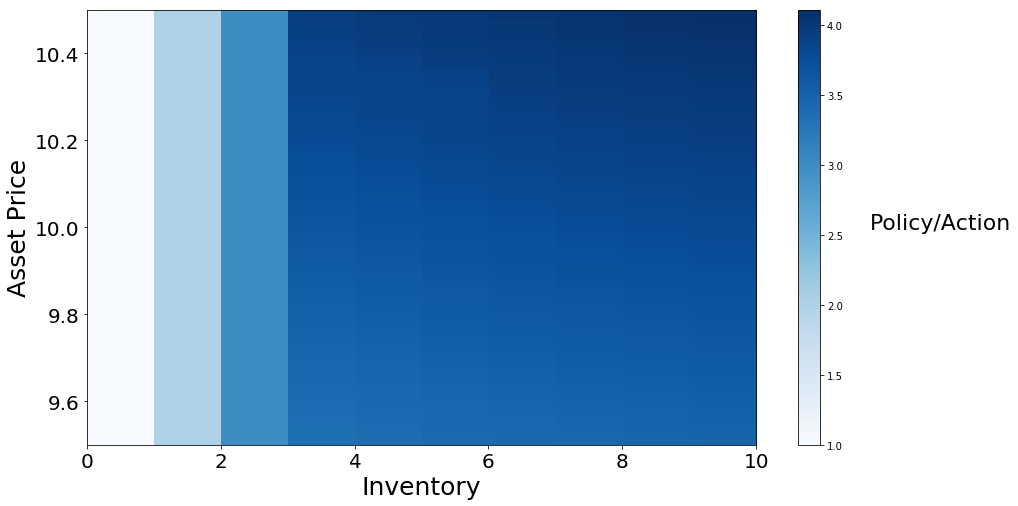}
	\caption{\small Policy heatmap showing the average of actions taken by the RDMM approach on the training set. Darker shades represent more stocks sold.}
	\label{fig:RDMM_policy}
\end{figure}

The heatmaps provide an accurate representation of policies based on value functions like Q-learning. In the RDMM, however, we have an LSTM summarising past information. In other words, the policy derived from the RDMM architecture does not depend only on the current price and inventory; it also depends on all historical data such as prices, inventory, rewards, and actions. For that reason, the in Figure \ref{fig:RDMM_policy} we have a heatmap showing the average of actions taken by the RDMM approach for every pair (price, inventory) available in the training set.

\begin{figure}[H] 
	\centering
		\includegraphics[width=0.40\textwidth]{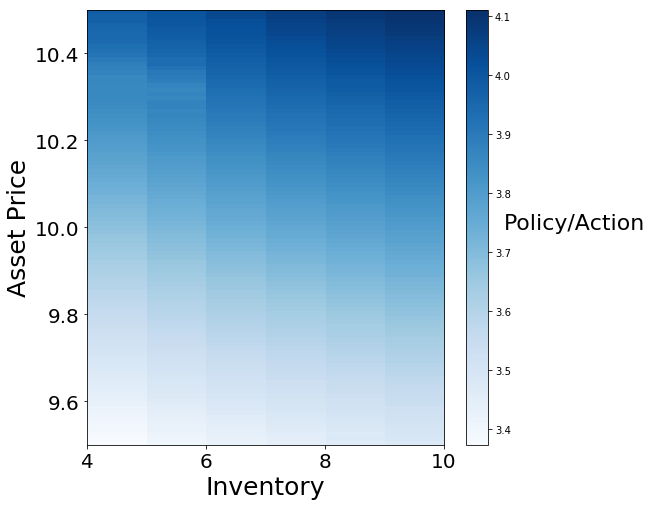}
	\caption{\small RDMM policy snippet extracted from Figure \ref{fig:RDMM_policy}. States ranging from \$9.5 to \$10.5 and from $4$ to $10$ units of inventory.Darker shades represent more stocks sold.}
	\label{fig:RDMM_sec_policy} 
\end{figure}

Figure \ref{fig:RDMM_policy} shows that the RDMM generates a policy with far smoother transitions between states than any of the Q-learning approaches. In order to better visualise the policy we plot a cropped version of Figure \ref{fig:RDMM_policy} in Figure \ref{fig:RDMM_sec_policy} where the inventories range from 4 to 10 units to showcase the policies sensitivity to price and inventory. 

The heatmap pattern where the policy is to, with all else the same, execute larger shares for higher prices and inventories.

An important measure to compare to execution strategies is the relative savings in basis point \index{relative savings in basis point} defined as
\begin{equation}
RS= \frac{R_{RDMM}-R_{benchmark}}{R_{benchmark}} \times 10^4,
\end{equation}
where $R_{RDMM}:=\sum_t^T r_{RDMM,t}$ and similarly for $R_{benchmark}$.

\begin{figure}[h] 
	\centering
		\includegraphics[width=0.7\textwidth]{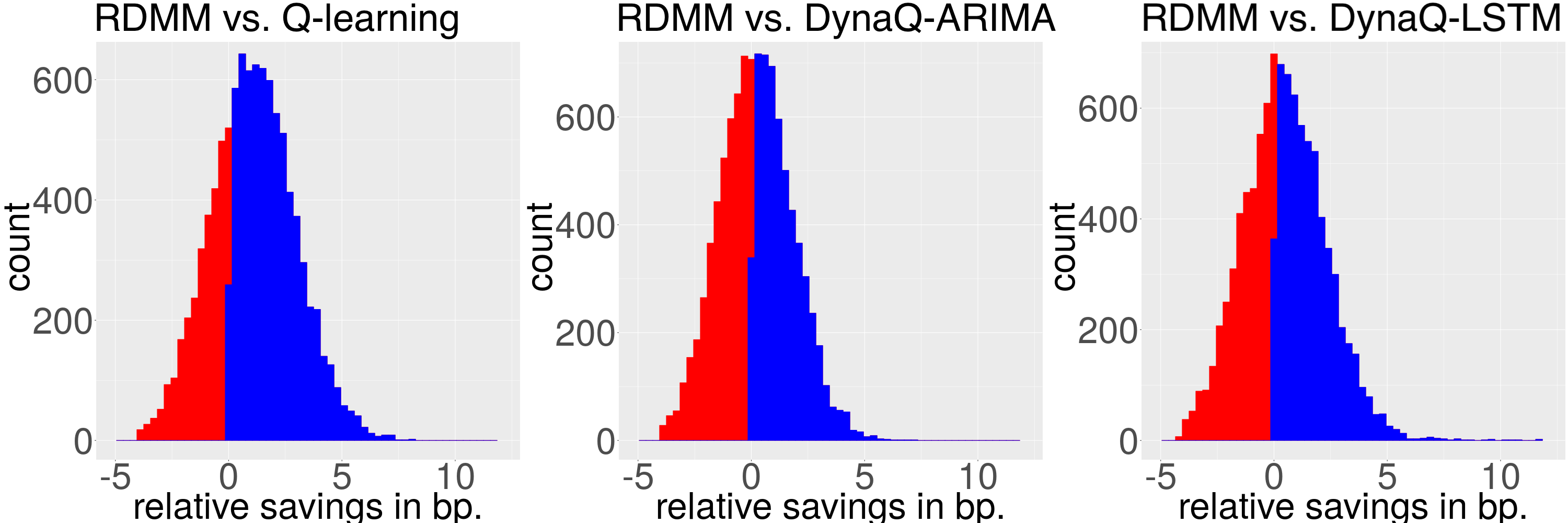}
	\caption{Histogram of relative savings in basis points of total reward. The red area highlights the part of the histogram where the RDMM underperformed the benchmark  while the blue area  highlights the part of the histogram where the RDMM outperformed}
	\label{fig:hist_reward}
\end{figure}

In Figures \ref{fig:hist_reward}  we plot the histograms for the  $RS$ of 10,000 simulated time series. The red area highlights the part of the histogram where the RDMM underperformed the benchmark (negative savings)  while the blue area  highlights the part of the histogram where the RDMM outperformed the benchmarks (positive savings). As we can see, the model bias \index{model bias} introduced by the LSTM has a negative impact in the results compared to the other benchmarks. Model bias, as \cite{deisen2010} pointed out, is a known problem in model-based RL where the model, in our example, the LSTM, fails to provide an accurate representation of the environment. As a consequence, the policies obtained from model-based RL tend to exploit the shortcomings of the model and its misrepresented environment leading to poor results.

Table \ref{results_table} provides a summary of the performance of each strategy and Table \ref{pairedtest} provides a pairwise comparison between the methods.

\begin{table}[H]
  \centering
  \begin{tabular}{llll}
\toprule
    \thead{Algorithm}     & \thead{Time-Series (Training)}    & \thead{$\mathbb{E}(\sum r_t)$} & \thead{$sd (\sum r_t)$}  \\
  \midrule  
    Q-learning & 1 million  & 2104.90 & 217.76\\ 
    DynaQ-ARIMA     & 1 million & 2105.08  & 217.79 \\
    DynaQ-LSTM     & 1 million      & 2104.89& 218.11\\
      RDMM     & 400      & 2105.30 & 217.76\\ 
      \bottomrule
  \end{tabular}
    \caption{Average reward on test set}
  \label{results_table}
\end{table}

\begin{table}[H]
  \centering
  \begin{tabular}{lcll}
\toprule
    \thead{Comparison}     & \thead{Mean of the differences}    & \thead{t-Statistic} & \thead{p-value} \\
    \midrule
    RDMM vs Q-learning &  \$0.38  &  6.71 & 0.000\\
    RDMM vs DynaQ-ARIMA     & \$0.21 & 4.47 & 0.000 \\
    RDMM vs DynaQ-LSTM     & \$0.40     & 4.52  & 0.000 \\
    \bottomrule
  \end{tabular}
    \caption{Paired t-test results for mean reward difference}
  \label{pairedtest}
\end{table}
The results demonstrate that, our method slightly outperforms the benchmarks in terms average reward.

\section{Real Data Experiment}
\label{realex}

\begin{figure}[H]
	\centering
	\begin{subfigure}[b]{0.49\textwidth}
		\centering
		\includegraphics[width=\textwidth]{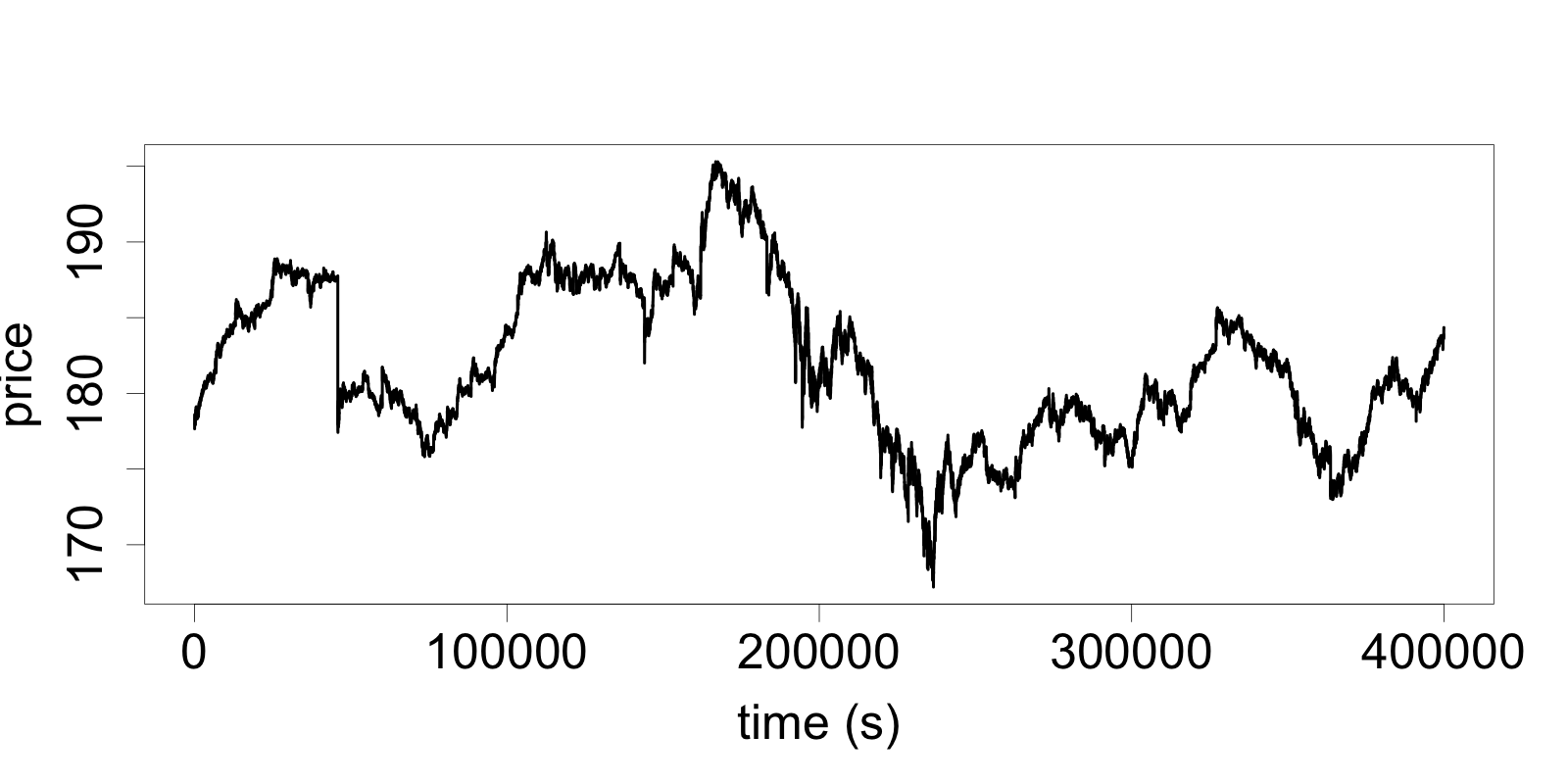}
		\caption[Network2]%
		{{\scriptsize Facebook}}    
		\label{fig:FBts}
	\end{subfigure}
	\begin{subfigure}[b]{0.49\textwidth}  
		\centering 
		\includegraphics[width=\textwidth]{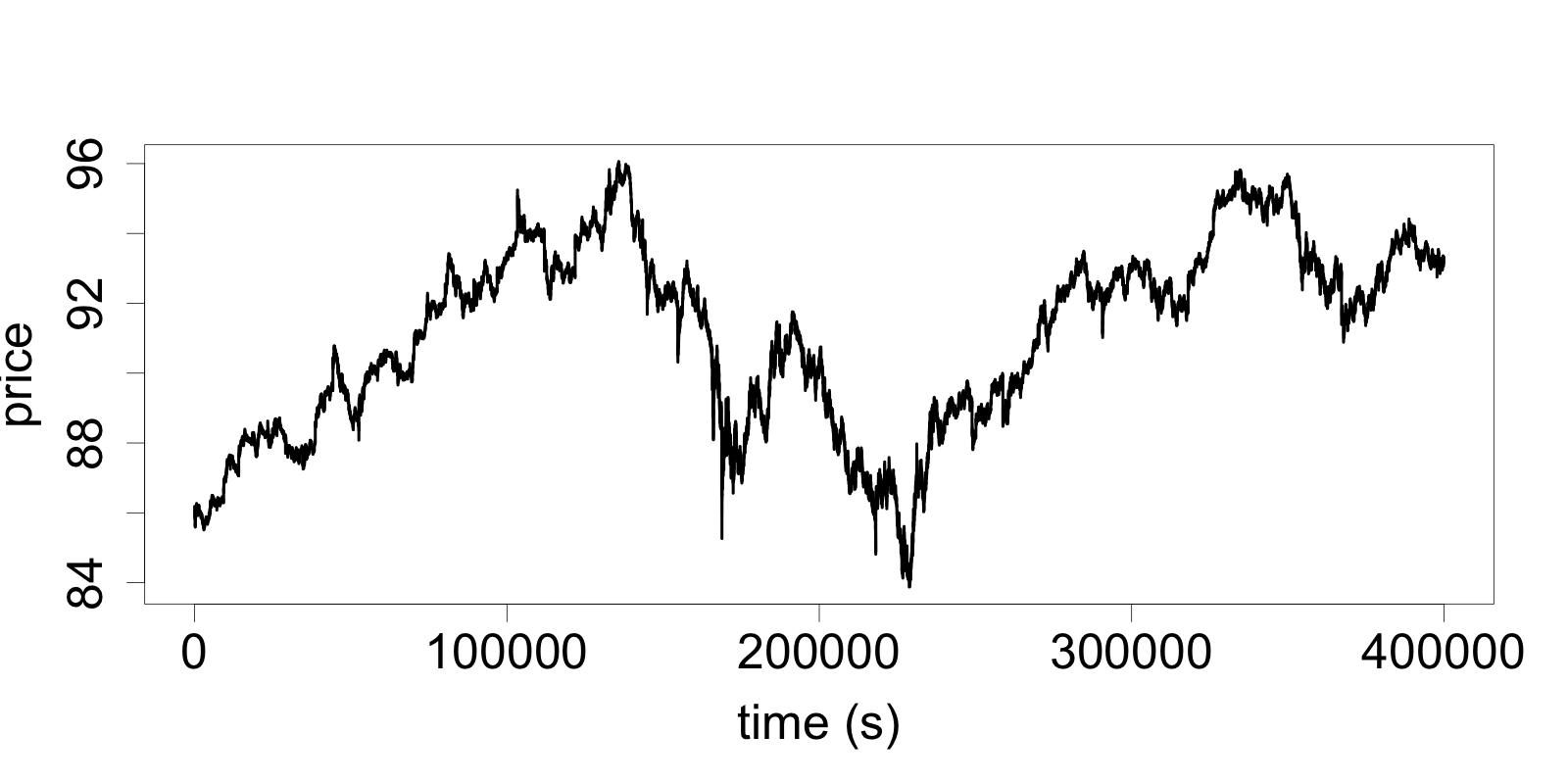}
		\caption[]%
		{{\scriptsize Microsoft}}    
		\label{fig:MSFTts}
	\end{subfigure}
	\vskip\baselineskip
	\begin{subfigure}[b]{0.49\textwidth}   
		\centering 
		\includegraphics[width=\textwidth]{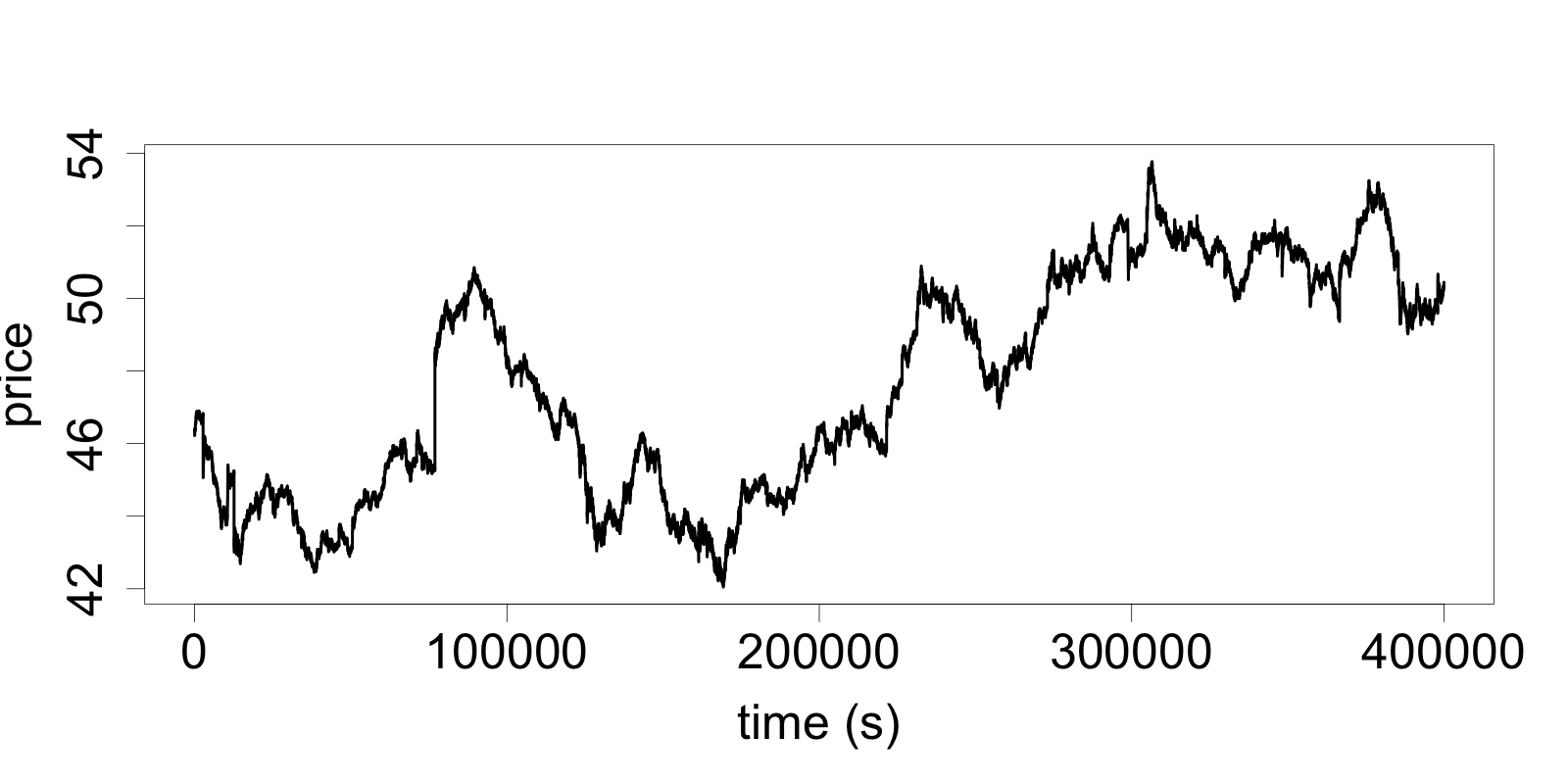}
		\caption[]%
		{{\scriptsize Intel}}    
		\label{fig:Intelts}
	\end{subfigure}
	\begin{subfigure}[b]{0.49\textwidth}   
		\centering 
		\includegraphics[width=\textwidth]{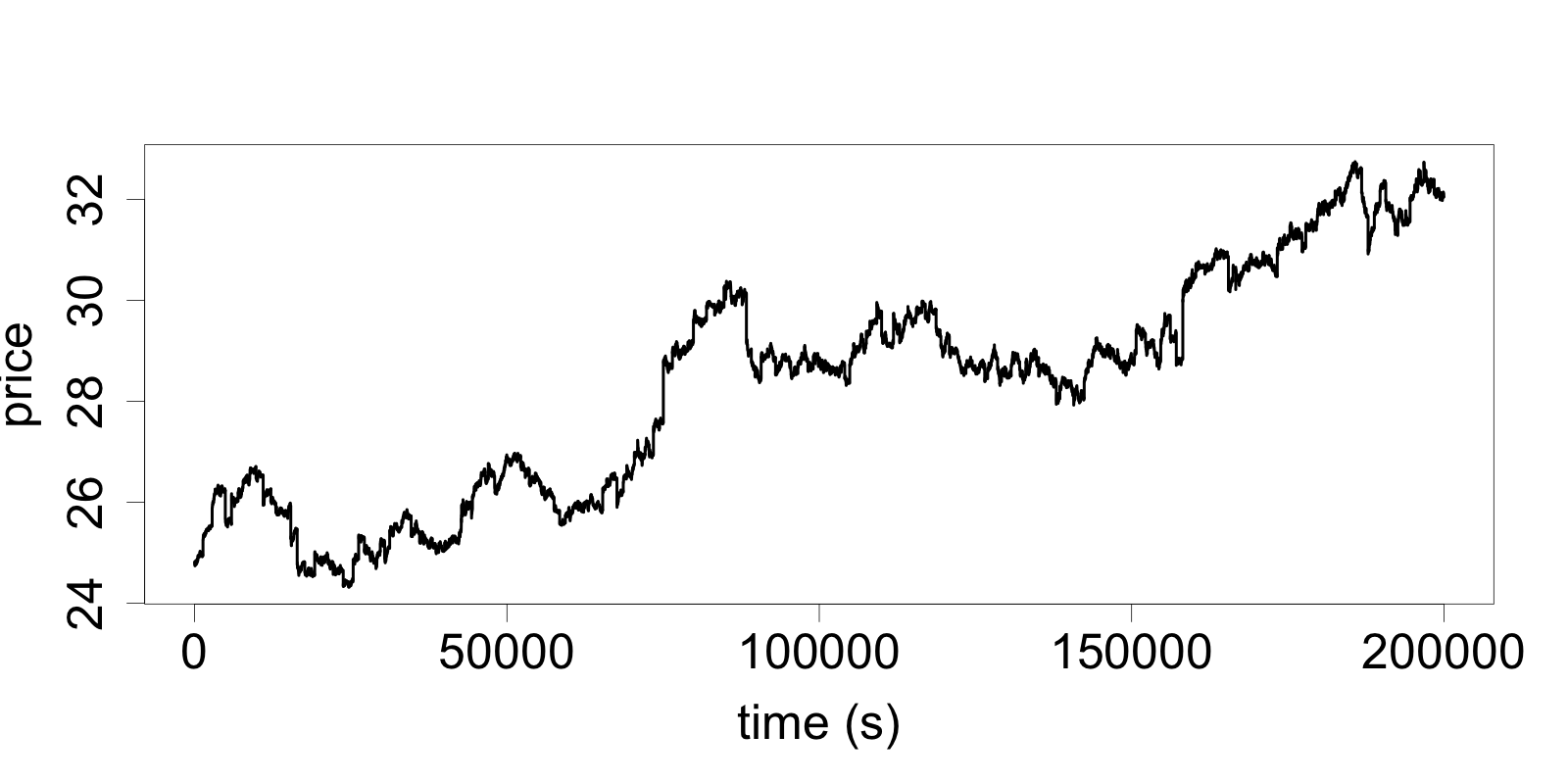}
		\caption[]%
		{{\scriptsize Vodafone}}    
		\label{fig:VODts}
	\end{subfigure}
	\caption[]
	{\small Price time series plot in US Dollars, calculated using order executions from the limit order book tick-by-tick data and sampling mid  prices at every second} 
	\label{fig:timeseries}
\end{figure}

In this section,  we repeat the experiments from Section \ref{simulations_and_results} with real stock prices data from Intel (INTC), Microsoft (MSFT), Facebook (FB), traded between January 2018 and March 2018, and Vodafone (VOD) traded in 2017.  To do so, we extract all order executions from the limit order book tick-by-tick data, and sample the mid price every second. For training and testing we select the first 400,000 data points for INTC, MSFT and FB, and the first 200,000 data points for VOD. The remainder of each time-series is used as validation set for parameter tuning.  Figure \ref{fig:timeseries} shows the time series plot for the first 400,000 data points (200,000 for VOD).

We use the same reward defined  in Equation \eqref{simplereward22}:
\begin{equation}
r_t=(x_{t}- c_{2} a_{t})a_{t} - c_3 q_t^2.
\label{simplereward22_real}
\end{equation}

Here, however, the value of the constants $c_2$ and $c_3$ are set to be proportional to the magnitude of the average price of the stock considered. In real-time execution, this may be replaced by realised execution costs. We adopted $c_2=c_3=0.017$ for FB, $c_2=c_3=0.008$ for MSFT, $c_2=c_3=0.004$ for INTC, and $c_2=c_3=0.002$ for VOD.

\begin{algorithm}[H] \small
	Prices $(x_0,x_1,x_2,....)$ are taken sequentially from the real data set.\;
	$q_0=0$\; 
	$i=20$\;
	\For{$t \in 0:200,000$}{
		\If{t mod i=0}{
			$i=i+n$  where $n \sim \mathcal{U}_{d}\{7,13\}$\footnote{$\mathcal{U}_{d}\{a,b\}$ stands for a uniform discrete distribution over $a, a+1, .... ,b$}\;
			\uIf{$40<i<400$}{
				$q_t=u$ where $u \sim \mathcal{U}_{d}\{0,10\}$\;
			}
			\uElse{
				$q_t=0.0$ \;
			}
		}
		$a_t=a$ where $a \sim \mathcal{U}_{d}\{0,q_t\}$\;
		$r_t=x_t a_t-c_2 a_t^2-c_3 q_t^2$\;
		$q_{t+1}=q_t-a_t$\;
	}
	\caption{Real training data set; random inventory, actions and rewards creation}
	\label{alg:inv_act2}
\end{algorithm}
\vspace{1cm}

The first 200,000 data points from FB, MSFT, and INTC stock prices are split into batches of 500 observations and used as the training set for the RDMM model. For VOD we use only the first 150,000 observations for the training set since we have less data available. Similarly to the experiments for mean reversion dynamics, we use Algorithm \ref{alg:inv_act2} to create a sequence of a fictional inventory, actions, and their respective rewards.

For the real data experiment, we compare the RDMM model against seven benchmarks: Q-learning, DynaQ with an autoregressive integrated moving average model (DynaQ-ARIMA), DynaQ with long short-term memory network (DynaQ-LSTM), time-weighted average price with three seconds for execution (TWAP3), RDMM without state uncertainty $\Sigma^{\phi}_{t}$ as input in the deterministic policy as in Equation \ref{eq:pol}  (RDMM-NoU), a DynaQ using a deep Markov model (DynaQ-DMM) as the model M of the simulated experience in the DynaQ Algorithm \ref{alg:dynaq2},{
	\color{black}   and an RDMM with the policy search inputs augmented with extra simulated price 
The architecture of the DMM used in the DynaQ-DMM approach is the architecture in the RDMM without the feedback of actions.

We include TWAP approach because it is a well-known strategy in finance for mitigating the adverse effects of executing a large number of stocks at once. The choice of 3 seconds of execution for TWAP was motivated by the average execution time observed on other strategies on the same task.

The RDMM can be seen as a combination of deep learning structures arranged to provide desirable properties for handling sequential data as a trading system. Gauging which structure or guideline provides a significant increment in financial performance is a challenge. In that sense, we follow the rationale found in \cite{Mnih2015},  where the authors attempt to assess the importance of constituent parts of an RL approach by disabling some individual core components of the deep Q-Network agent to show the detrimental effects on the agent's performance. This procedure is known as ablation studies\index{ablation studies}.

For this reason, we add two models to further investigate the strengths of the RDMM model. The RDMM-NoU model (A RDMM without state uncertainty $\Sigma^{\phi}_{t}$ as input in the deterministic policy)  allows us to determine the contribution of the state uncertainty in the RDMM model. In the DynaQ-DMM approach, we borrow the predictions made by the DMM part in the RDMM and use it to make predictions on a DynaQ algorithm. This benchmark informs us whether DMM alone can provide a significant improvement compared to other model-based RL methods such as DynaQ-ARIMA and DynaQ-LSTM. 

Figure \ref{fig:timeseries} shows that the price range of the real time-series varies considerably more than the price range in our mean-reversion simulation. The wider the price range interval, the larger the number of possible states that Q-learning and DynaQ algorithm have to visit to generate their policies. To cope with this issue, we modify Algorithm \ref{alg:dynaq} in the Appendix \ref{qlearning}, making the number of visits proportional to the time-series price range. Algorithm \ref{alg:dynaq2} contains the general form for the Q-learning and DynaQ models used in this section. Notice that in the first 2 lines we first compute the number of possible states by multiplying the price range over 400,000 seconds (200,000 for Vodafone), by the number of bins of size one cent, and by the maximum number of stocks provided to the agent (10 units in our experiments). Next, we multiply the estimated number of states by 200. 
trajectories drawn from the generative part of the RDMM (RDMMx) .
}

\begin{algorithm} [H]\small 
	$nstates= \left[ max(TS)-min(TS)\right]  \times 100 \times 10 $ (price range x 100 cents x 10 inventory)\;
	$simN=nstates \times 200 $ (number of states x number of visits)\;
	Initialize Q(x,q,a)\;
	$\epsilon=0.9,\alpha=0.9$\;
	\For{$i\leftarrow 1$ \KwTo $simN$}{
		\If{i mod (simN/30)==0}{
			$\epsilon=\epsilon \times 0.9$\;
			$\alpha=\alpha \times 0.9$\;
		}
		$x \sim  \text{real mid-price time-series}, q \sim U(0,10)$\;
		$x=round(s,2),q=round(q,0)$\;
		\While{$q>0$}{
			\tcc{choose action using $\epsilon$-greedy} 
			\uIf{$U(0,1)<\epsilon$}{
				$a=\underset{a'}{\mathrm{argmax}} Q(x,q,a')$\;
			}
			\uElse{$a \sim U(0,q)$\;
				$a=round(a,0)$\;}
			$r=x*a- c_2 a^2 - c_3 q^2$\;
			$x'=f(x,a)$\;
			$q'=q-a$\;
			$Q(x,q,a) \leftarrow Q(x,q,a)+\alpha \left( r + \max_{a'} Q(x',q',a')-Q(x,q,a) \right)$\;
			$q=q', x=x'$\;
			\HiLi	\tcc{\Large Simulated experience (DynaQ only)} 
			\For{$i\leftarrow 1$ \KwTo $5$}{
				$x,q \leftarrow$ random previously observed state\;
				$a \leftarrow$ random action taken on state $x,q$\;
				$r,x',q'=M(x,q,a)$ (model) \;
				$Q(x,q,a) \leftarrow Q(x,q,a)+\alpha \left( r + \max_{a'} Q(x',q',a')-Q(x,q,a) \right)$\;
			}
		}
	}
	\caption{Q-learning/DynaQ - version 2 - adapted from Sutton, 1998}
	\label{alg:dynaq2}
\end{algorithm}
\vspace{1cm}

During policy learning, we give the advantage to the benchmark RL agents \textbf{to access the entire data set} represented in Figure \ref{fig:timeseries} during the training phase, whereas the RDMM has to complete the same task using only the first half of the data set (i.e., 200,000 points for INTC, FB, and MSFT and 150,000 points for VOD). 

Table \ref{dataset} summarises some facts about our dataset. The column ``$\mathbf{simN}$'' refers to the number of trajectories sampled for Algorithm \ref{alg:dynaq2}. The column ``$\mathbf{c_2},\mathbf{c_3}$'' contains constants adopted for the reward function \eqref{simplereward22_real}. Finally, the last two columns indicate the size of the training and test sets for the RDMM model.

\begin{table}[h]
  \centering \small
  \begin{tabular}{lrrrrrrr}
  	\toprule
   \thead{Stock}    & \thead{Range} &\thead{Min}  & \thead{Max}&$\mathbf{simN}$  &  $\mathbf{c_2},\mathbf{c_3}$ & \thead{Train Set}& \thead{Test Set}\\  
 \midrule
   FB  &28.09  & 167.21 & 195.30 &5,618,000  & 0.017 & 200,000 & 200,000  \\
   INTC  & 11.72 & 42.05 & 53.77& 2,344,000  & 0.004 &  200,000 & 200,000 \\
   MSFT & 12.18  &83.88& 96.06 &2,436,000 & 0.008  &  200,000 & 200,000  \\
   VOD & 8.43  &24.32&32.75 &1,686,000 &  0.002  & 150,000 & 50,000    \\
   \bottomrule
  \end{tabular}
    \caption{Data set summary.}
      \label{dataset}
\end{table}

The 200,000  (50,000 for VOD) data points  in the test set are split into batches of 500 observations. We also adopt the same inventory and actions generation as described in Algorithm \ref{alg:inv_act2}.

\begin{figure}[H] 
	\centering
	\includegraphics[width=0.95\textwidth]{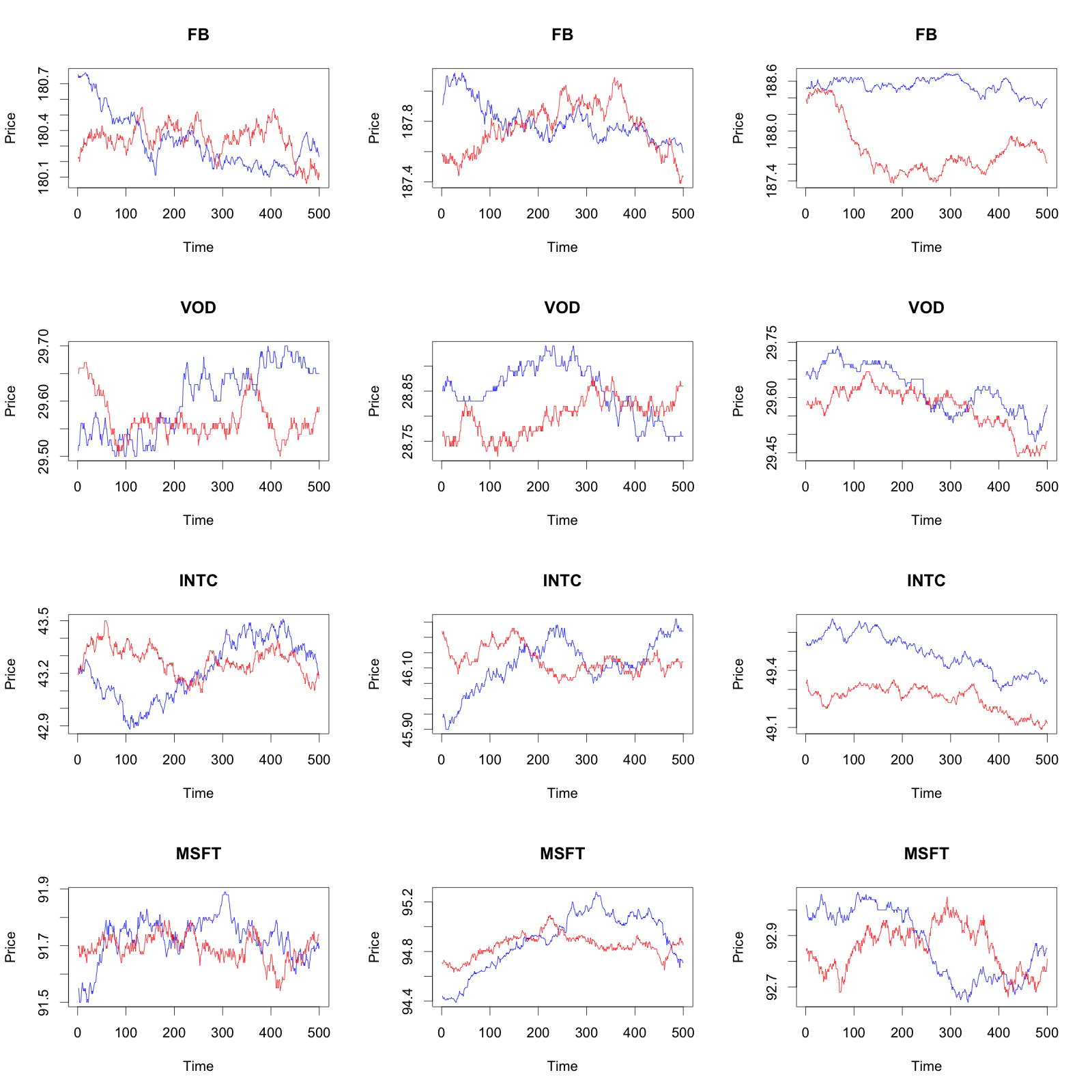}
	\caption{\small Samples of price trajectories. Real prices in blue and generated trajectory in red.}
	\label{fig:trajsim}
\end{figure}

{\color{black}  In the RDMMx, we exploit the generative nature of the variational autoencoder in the RDMM, creating a price sample trajectory drawn from the combination of distribution of the transitions in Equation \ref{eq:transitions} and the distribution of the price emissions in Equation \ref{price_emission}. This price trajectory is generated after the ELBO optimisation of Algorithm \ref{alg:RDMM}, and, analogous to how the real prices of the training set are treated, is combined with a sequence of a fictional inventory, actions, and their respective rewards using Algorithm \ref{alg:inv_act2}. The resulting simulated path is used to augment the training set for the policy search optimisation part of  Algorithm \ref{alg:RDMM}. 
More specifically, for every real data batch $b$ of size 500, $(x_{b1},x_{b2},...,x_{b500})$, an extra batch of the same size is simulated, taking the last latent state $z_{b500}$ of the inference part of the RDMM to be the initial state of the simulated trajectory (see Figure \ref{fig:trajsim}). The next states are computed using the transitions in Equation \ref{eq:transitions} and emissions in Equation \ref{price_emission}. The size of the training set for the policy search in the RDMMx is double that of the size of the training set for the policy search in the regular RDMM.
The goal of RDMMx is to investigate if artificially increasing the number of trajectories enhances the policy search.   
}

\begin{figure}[H]
	\centering
	\begin{subfigure}[b]{0.4\textwidth}
		\centering
		\includegraphics[width=\textwidth]{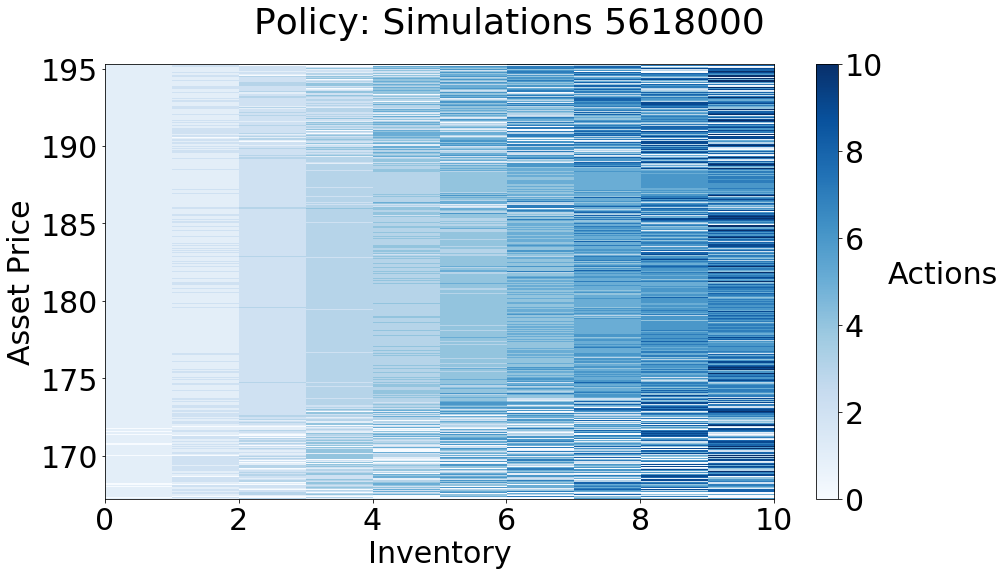}
		\caption[]%
		{{\scriptsize Q-learning}}    
	\end{subfigure}
	\begin{subfigure}[b]{0.4\textwidth}  
		\centering 
		\includegraphics[width=\textwidth]{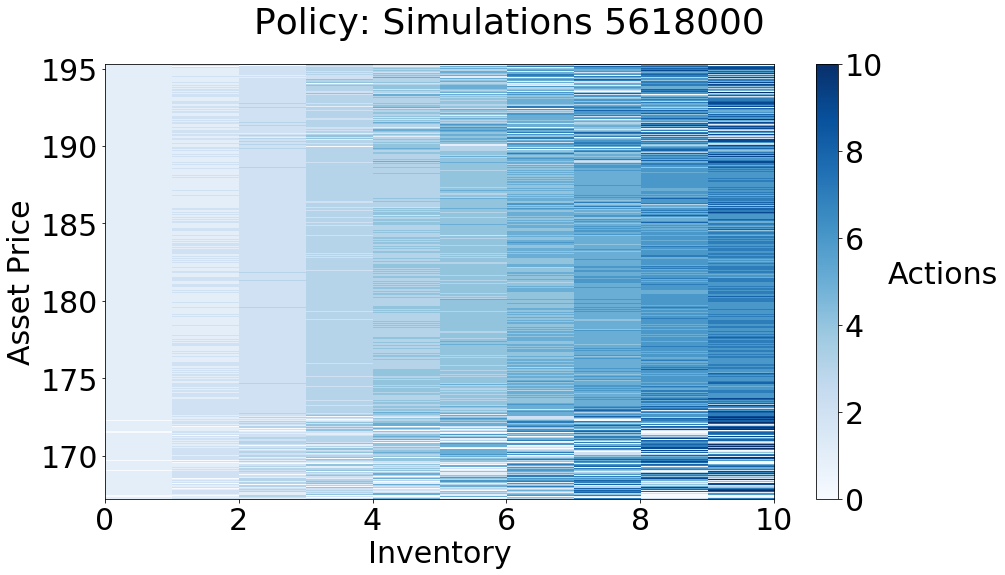}
		\caption[]%
		{{\scriptsize DynaQ-ARIMA}}    
	\end{subfigure}
	\vskip\baselineskip
	\begin{subfigure}[b]{0.4\textwidth}   
		\centering 
		\includegraphics[width=\textwidth]{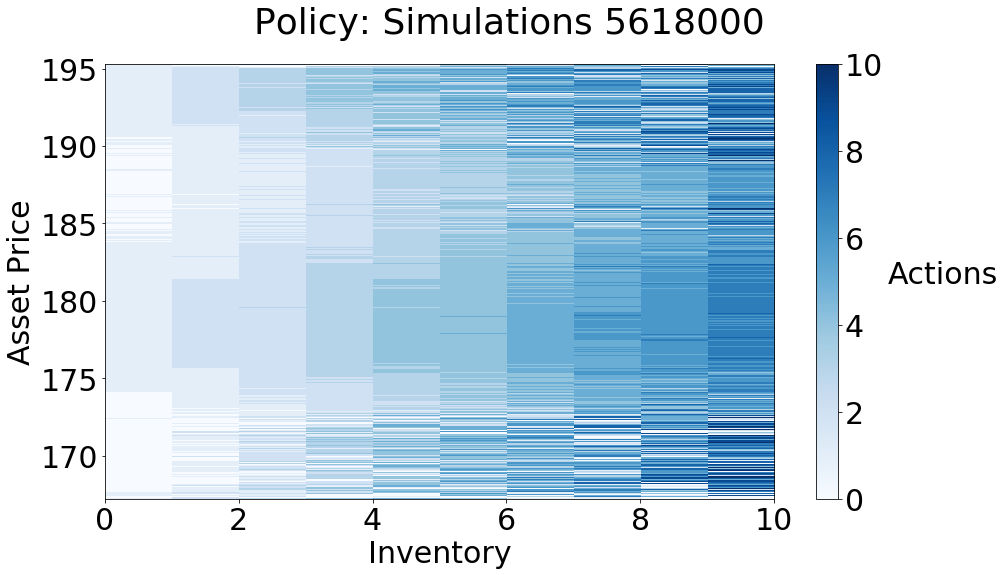}
		\caption[]%
		{{\scriptsize DynaQ-LSTM}}    
	\end{subfigure}
	\begin{subfigure}[b]{0.4\textwidth}   
		\centering 
		\includegraphics[width=\textwidth]{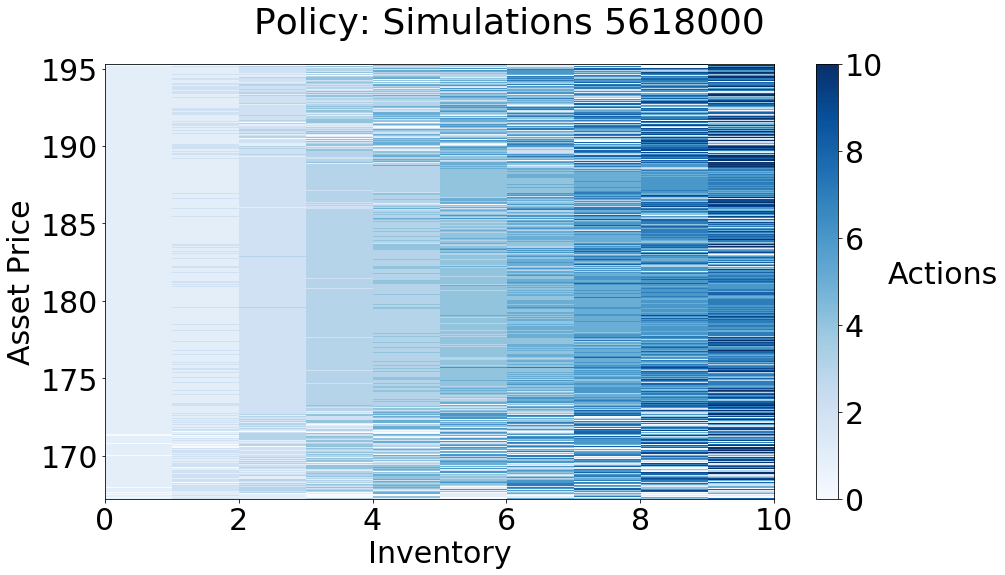}
		\caption[]%
		{{\scriptsize DynaQ-DMM}}    
	\end{subfigure}
	\caption[]
	{\small Facebook policy heatmaps for states where the stock prices go from \$167.21 to \$195.30 and inventory ranges from 0 to 10 units. In this heatmap, the actions - or the number of stocks to be sold by the agent - are correlated to the tonality indicated by the colourbar on the right side of each heatmap.} 
	\label{fig:FB_policies}
\end{figure}

\begin{figure}[h]
	\centering
	\begin{subfigure}[b]{0.4\textwidth}
		\centering
		\includegraphics[width=\textwidth]{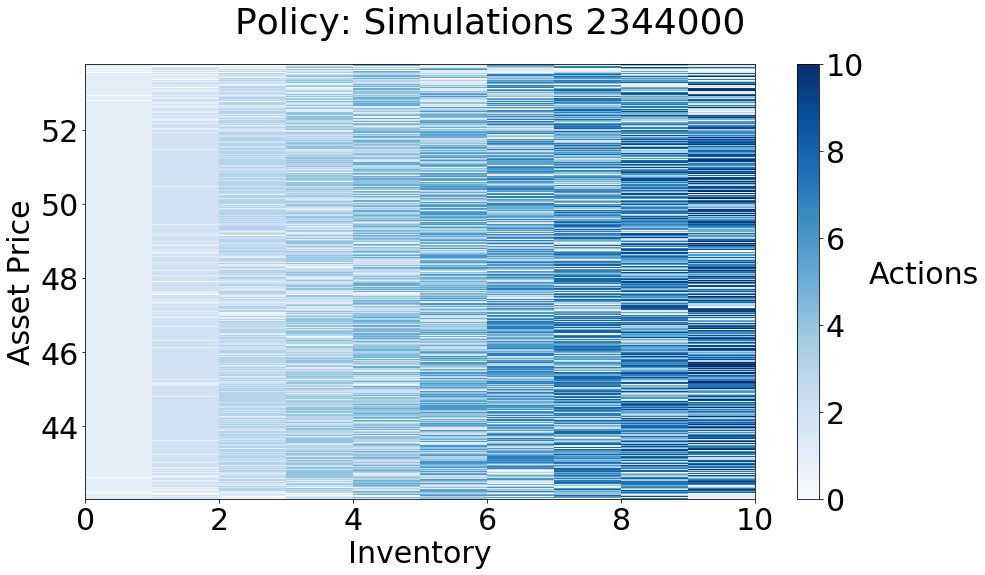}
		\caption[]%
		{{\scriptsize Q-learning}}    
	\end{subfigure}
	\begin{subfigure}[b]{0.4\textwidth}  
		\centering 
		\includegraphics[width=\textwidth]{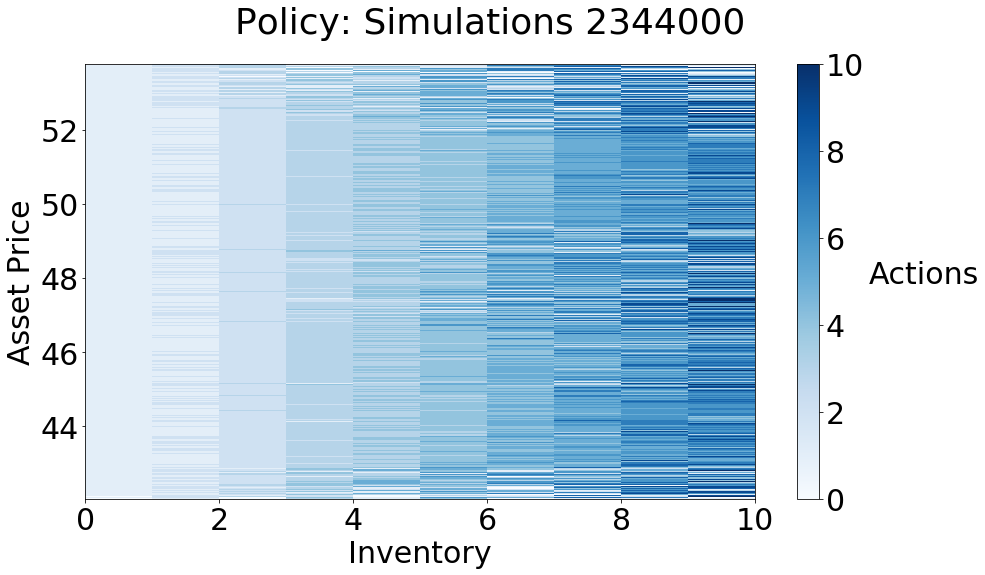}
		\caption[]%
		{{\scriptsize DynaQ-ARIMA}}    
	\end{subfigure}
	\vskip\baselineskip
	\begin{subfigure}[b]{0.4\textwidth}   
		\centering 
		\includegraphics[width=\textwidth]{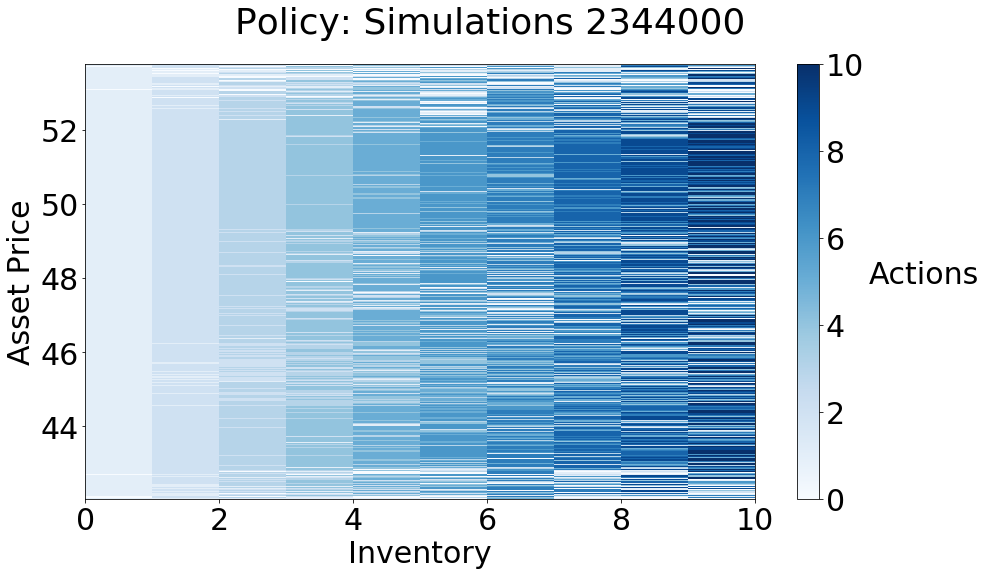}
		\caption[]%
		{{\scriptsize DynaQ-LSTM}}    
	\end{subfigure}
	\begin{subfigure}[b]{0.4\textwidth}   
		\centering 
		\includegraphics[width=\textwidth]{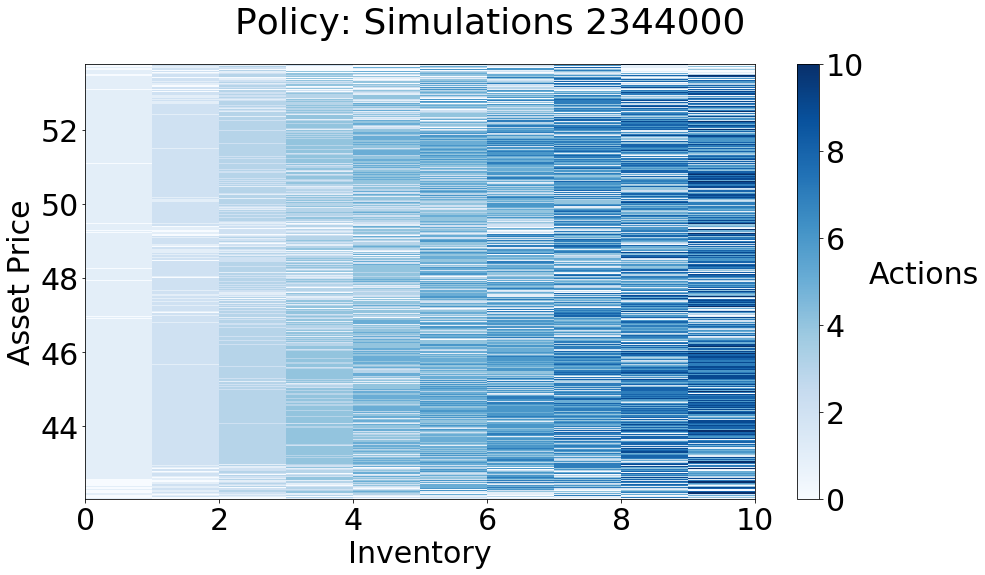}
		\caption[]%
		{{\scriptsize DynaQ-DMM}}    
	\end{subfigure}
	\caption[]
	{\small Intel policy heatmaps for states where the stock prices go from \$42.05 to \$53.77 and inventory ranges from 0 to 10 units. In this heatmap the actions, or the number of stocks to be sold by the agent, are correlated to the tonality indicated by the colour bar on the right side of each heatmap.} 
\label{fig:Intel_policies}
\end{figure}

\begin{figure}[H]
	\centering
	\begin{subfigure}[b]{0.4\textwidth}
		\centering
		\includegraphics[width=\textwidth]{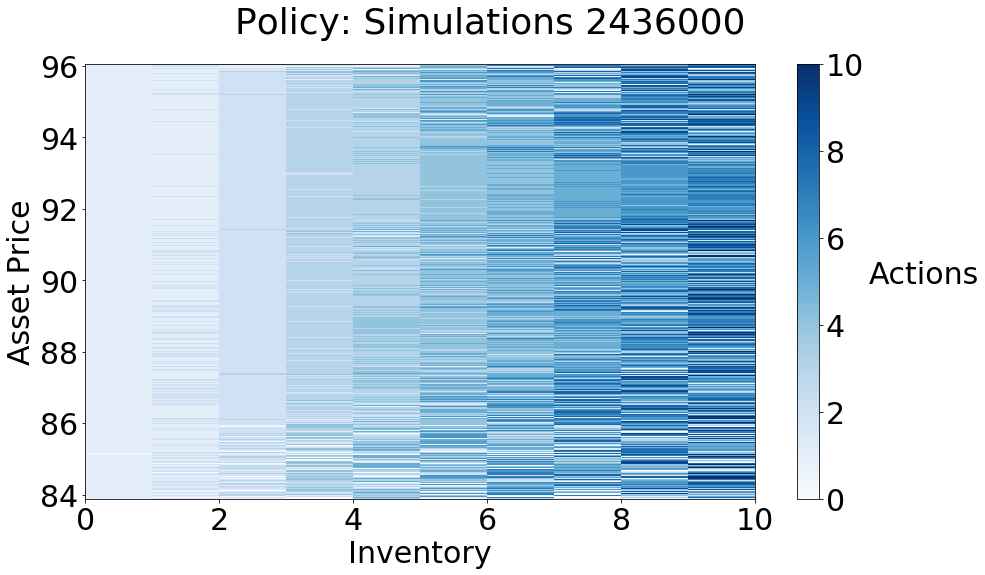}
		\caption[]%
		{{\scriptsize Q-learning}}
	\end{subfigure}
	\begin{subfigure}[b]{0.4\textwidth}  
		\centering 
		\includegraphics[width=\textwidth]{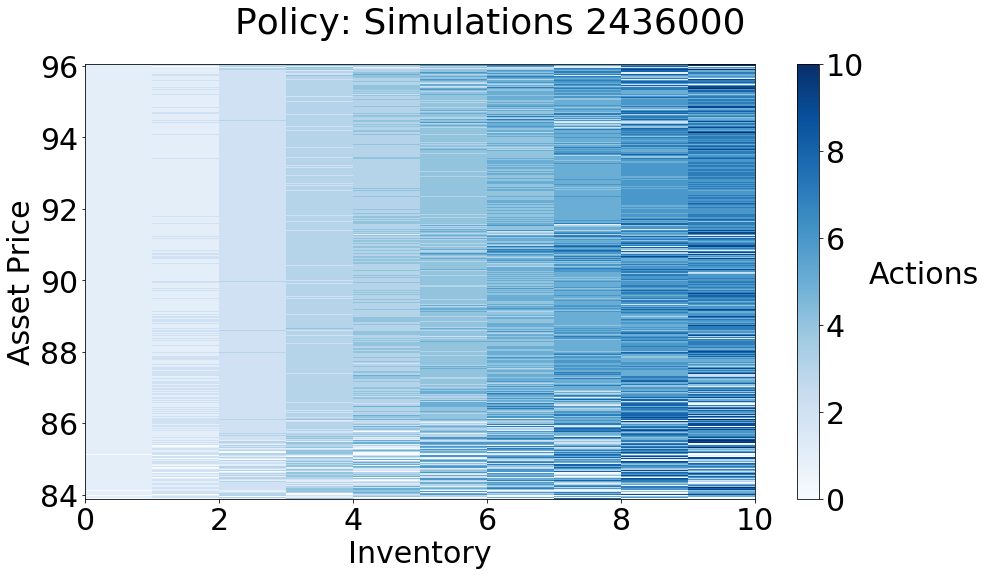}
		\caption[]%
		{{\scriptsize DynaQ-ARIMA}}    
	\end{subfigure}
	\vskip\baselineskip
	\begin{subfigure}[b]{0.4\textwidth}   
		\centering 
		\includegraphics[width=\textwidth]{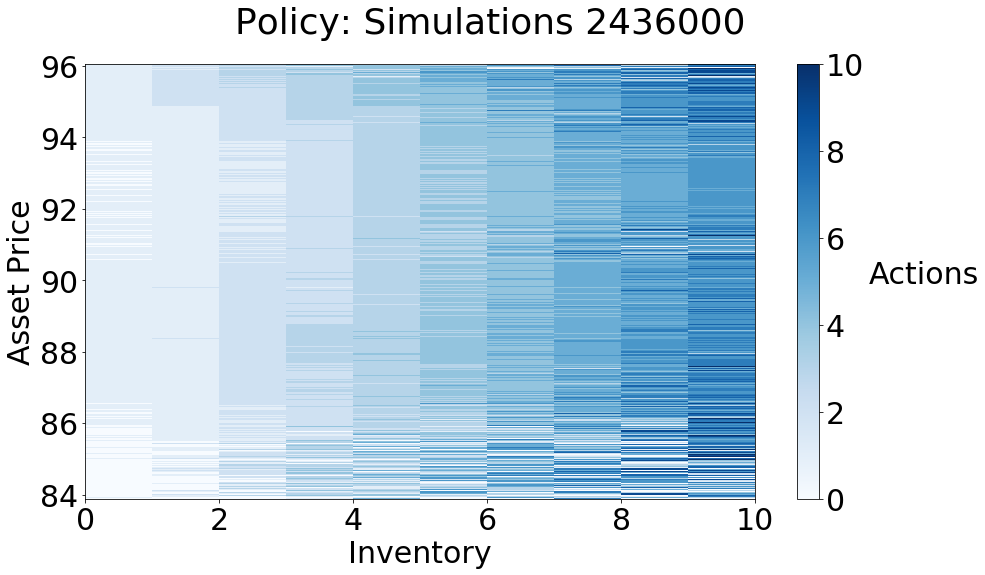}
		\caption[]%
		{{\scriptsize DynaQ-LSTM}}    
	     \label{fig:MSFT_policies_lstm}
	\end{subfigure}
	\begin{subfigure}[b]{0.4\textwidth}   
		\centering 
		\includegraphics[width=\textwidth]{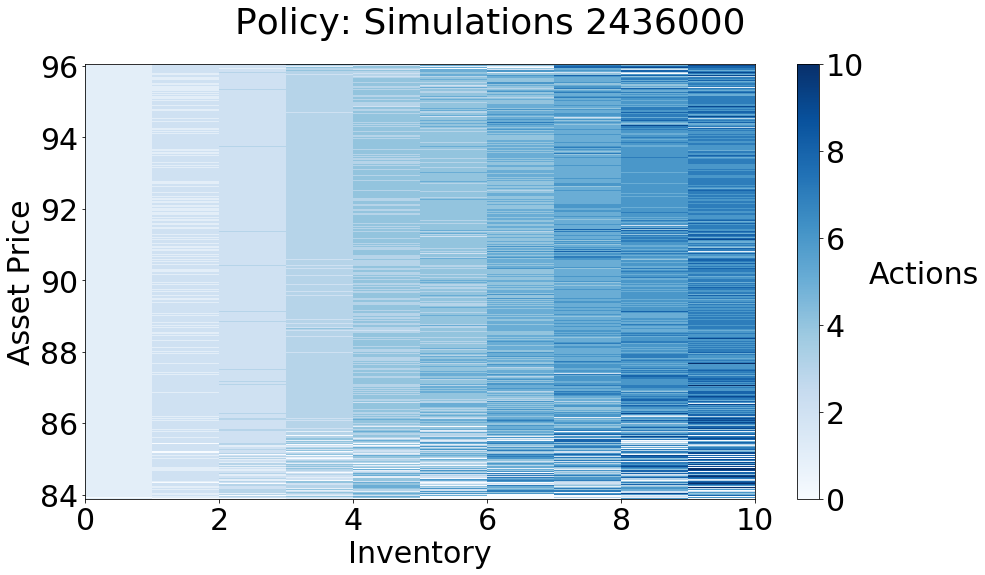}
		\caption[]%
		{{\scriptsize DynaQ-DMM}}    
	\end{subfigure}
	\caption[]
	{\small Microsoft policy heatmaps for states where the stock prices range from \$83.88 to \$96.06 and inventory ranges from 0 to 10 units. In this heatmap, the actions, or the number of stocks to be sold by the agent, are correlated to the tonality indicated by the colour bar on the right side of each heatmap.} 
   \label{fig:MSFT_policies}
\end{figure}

\begin{figure}[H]
	\centering
	\begin{subfigure}[b]{0.4\textwidth}
		\centering
		\includegraphics[width=\textwidth]{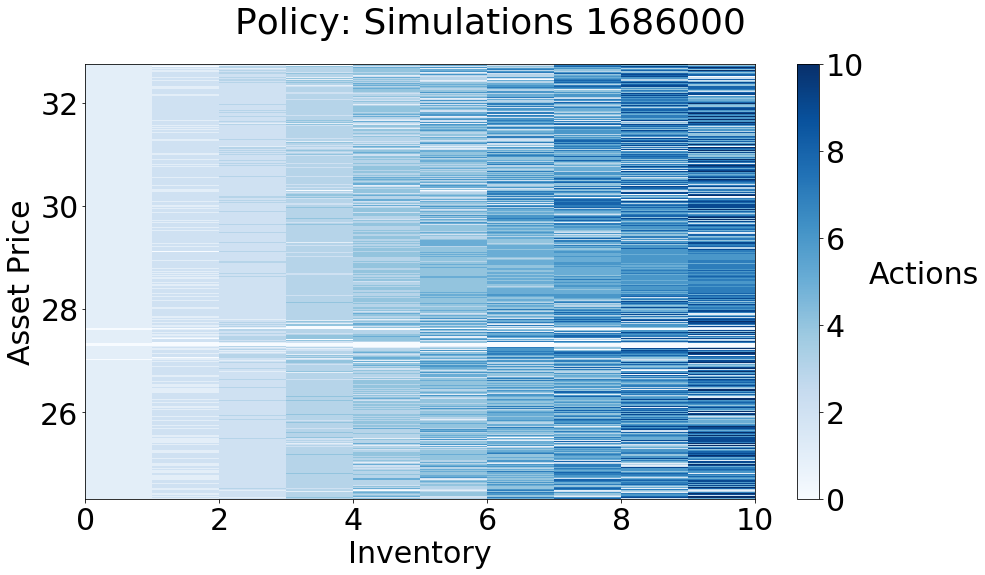}
		\caption[]%
		{{\scriptsize Q-learning}}    
	\end{subfigure}
	\begin{subfigure}[b]{0.4\textwidth}  
		\centering 
		\includegraphics[width=\textwidth]{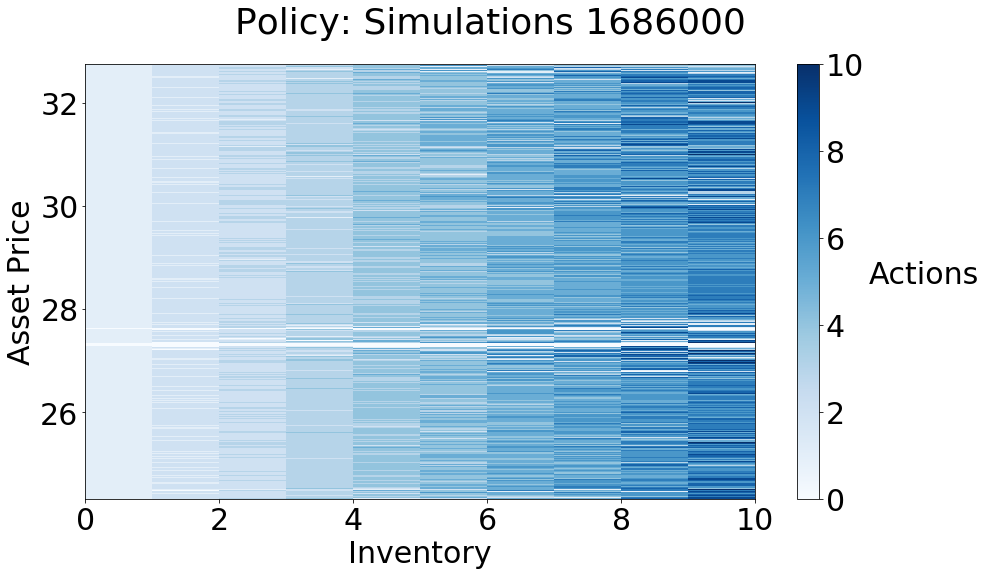}
		\caption[]%
		{{\scriptsize DynaQ-ARIMA}}    
	\end{subfigure}
	\vskip\baselineskip
	\begin{subfigure}[b]{0.4\textwidth}   
		\centering 
		\includegraphics[width=\textwidth]{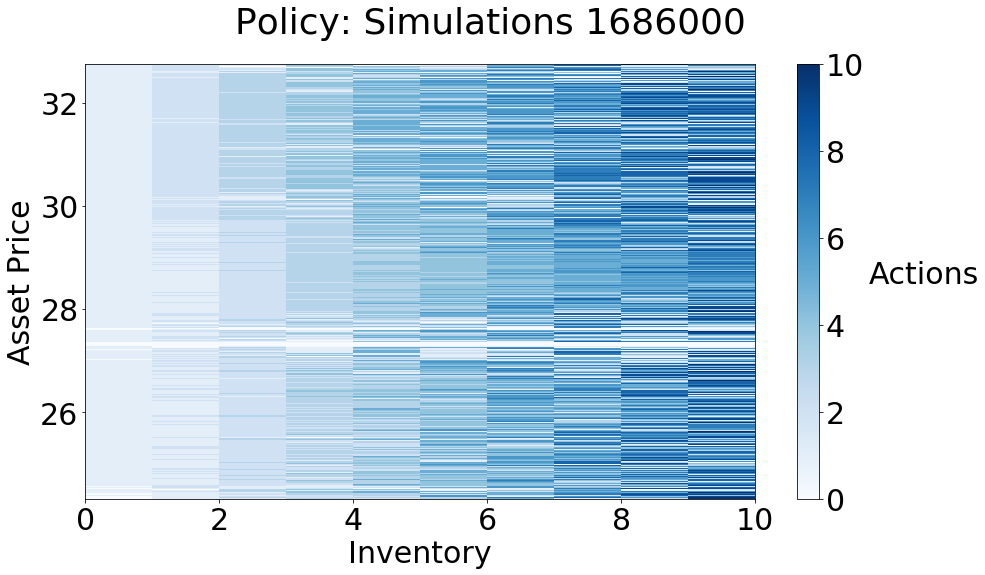}
		\caption[]%
		{{\scriptsize DynaQ-LSTM}}    
	\end{subfigure}
	\begin{subfigure}[b]{0.4\textwidth}   
		\centering 
		\includegraphics[width=\textwidth]{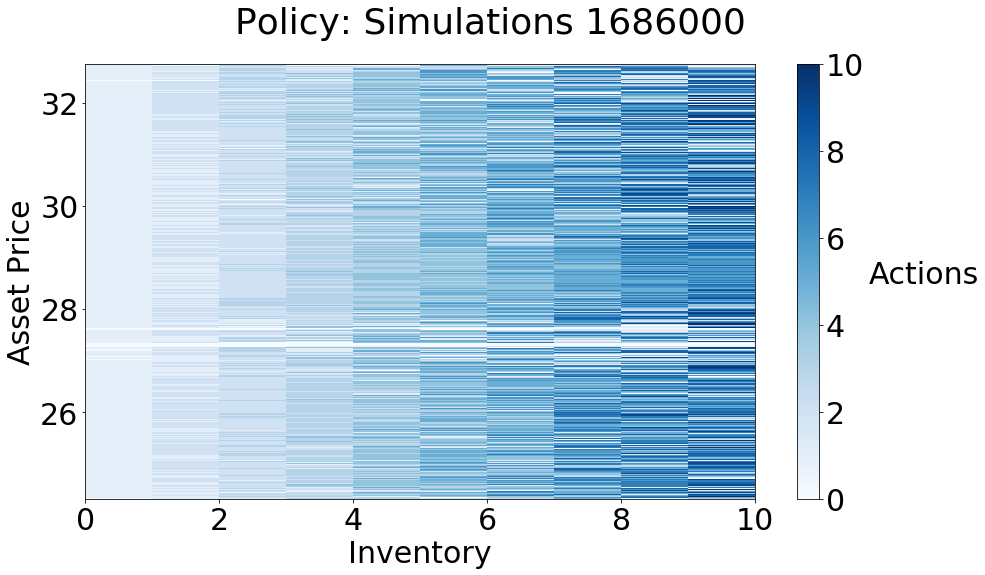}
		\caption[]%
		{{\scriptsize DynaQ-DMM}}    
	\end{subfigure}
	\caption[]
	{\small Vodafone policy heatmaps for states with prices ranging from \$24.32 to \$32.75 and inventory ranging from 0 to 10 units. In this heatmap the actions, or the number of stocks to be sold by the agent, are correlated to the tonality indicated by the colourbar on the right side of each heatmap.} 
	\label{fig:VOD_policies}
\end{figure}

The heatmap representations of the policies generated by Q-learning, DynaQ-ARIMA, DynaQ-LSTM and DynaQ-DMM are presented in Figures \ref{fig:FB_policies}, \ref{fig:Intel_policies}, \ref{fig:MSFT_policies} and \ref{fig:VOD_policies}. The colourbar to the right side of each plot represents the number of stocks sold, with the number increasing with darker shades.

In the simulated price dynamics, we observed that higher prices trigger higher-order executions. For the real price dynamics, we observe the same characteristic in the heatmaps to a certain extent, but not as pronounced as in the simulations. There is, however, an observed inventory dependence, as before.
In the Facebook policy heatmaps of  Figure \ref{fig:FB_policies}, we notice that  DynaQ-ARIMA has the smoothest shade transitions between contiguous states, followed by DynaQ-DMM and Q-learning. In general, we observe, for all cases, that prices that appear less frequently in the time-series, such as those in the neighbourhood of the minimum and the maximum value, tend to result in more abrupt transitions between adjacent states. In the DynaQ-LSTM heatmap we observe clusters of states with diminished actions proportional to its inventory in the asset price ranges of \$165 to \$175 and \$185 to \$190. From the heatmap, we suspect that the policy search is trying the exploit shortcomings of a possible model bias introduced by the LSTM,
where the DynaQ-LSTM outputted an over-optimistic policy on those cluster states, meaning that the agent decided to hold the stocks in expectation of a price reversion in the near future as a consequence of model bias discussed previously.

In the Intel policy heatmaps of Figure \ref{fig:Intel_policies}, we notice that DynaQ-ARIMA the smoothest shade transitions between adjacent states, followed by Q-learning and DynaQ-DMM.  Compared with Facebook policies, the abrupt transitions between contiguous states is less noticeable in the neighbourhood of the minimum and the maximum prices. This may be explained by the time-series plotted in Figure \ref{fig:timeseries}, where we notice that the price trajectory of the Intel stocks approaches the maximum and the minimum values more frequently than the Facebook time-series. In the DynaQ-LSTM heatmap we see a more aggressive policy for the darker cluster on states with prices around \$50.  Similarly to the Facebook case, we suspect that it is a negative result of model bias introduced by the LSTM.
The Microsoft policy heatmaps displayed in Figure \ref{fig:MSFT_policies} seem to be similar to each other compared to the Facebook and Intel policies. We observe the smoothest shade transitions between contiguous states in DynaQ-DMM followed closely by DynaQ-ARIMA.  Similar to Intel stocks,  we notice, in the time-series plotted in Figure \ref{fig:timeseries}, that the Microsoft time-series approaches the surroundings of the maximum value more than once - we refer to the two peaks around t=130,000 and t=325,000. This seems to result in less abrupt transitions between adjacent states in the top part of the heatmaps, i.e., near the maximum price.  The high contrast between contiguous states in the bottom part of the heatmaps is consistent with the fact that the prices trajectory has a sharp dive to the minimum price around t=220,000. The presence of different clusters is less noticeable in the Microsoft DynaQ-LSTM heatmap compared with the previous cases.

In the Vodafone policy heatmaps of Figure \ref{fig:VOD_policies} , we observe a white band (no stocks executed) for all cases located between \$27 and \$28. The reason for this particular pattern is that, in the VOD dateset, there is a 12 cents gap between the prices \$27.27 and \$27.39; therefore, the initial state ($Q=0$) of the Q-table does not change because states whose price lies between \$27.27 and \$27.39 cannot be sampled. For the very same reason, we can be assured that this white band does not affect the performance of the benchmarks since the agents are not required to perform on those states in our experiments.
We also notice that Q-learning, DynaQ-LSTM and DynaQ-DMM policy heatmaps displayed in Figure \ref{fig:VOD_policies} appear similar. DynaQ-ARIMA seem  to be the one with the smoothest shade transitions between contiguous states.
As in the Microsoft case, the presence of uncommon clusters is less noticeable in the DynaQ-LSTM heatmap compared to Facebook and Intel DynaQ-LSTM heatmaps.

\subsection{RDMM learning - real data }
\label{RDMM_learning_real_data}

To achieve reasonable results during the learning phase, the RDMM requires an increment of the size of the neural network in the real data set compared to the  mean reversion case. To facilitate our investigation an additional strategy is added during model optimisation; instead of completing the learning in a single session, where the model approximation of the VAE  (ELBO optimisation) is followed by the RL optimisation (policy search), we introduced the possibility to have an intermission between the ELBO optimisation and the policy search . The goal of this modification is three-fold i) to allow the execution of multiple learning sessions during ELBO optimisation until we achieve the desired approximation using different learning rates for each session; ii) to facilitate parameters fine tuning and avoid overfitting; and  iii) to start the RL optimisation (or continue training the model approximation) directly from a set of previously saved parameters.

In addition to the three goals mentioned above, we should consider that this work has, at the present moment, experimental and academic inclinations. These ideas were influenced by the guidelines described in \cite{DBLP:journals/corr/Smith17}. The strategy of breaking down the training phase into segmented sessions would be replaced by an additional module implemented to control the learning process for cases where the RDMM is used for commercial purposes. From our experience of training and testing the RDMM on the simulated mean reversion and real data sets, this module should coordinate the interplay among  i) the complexity of the neural networks used and the volatility of the dataset; and ii) the number of epochs adopted for ELBO and policy search optimisations along with an intelligent management of the learning rates. We revisit this topic at the end of this subsection. 

With the above remark in mind, we provide a summary of the RDMM learning for the four stocks mentioned previously (INTC, FB, VOD and MSFT). The summary contains a few plots of the learning phase for each stock to give the reader a brief notion of some aspects of the model training relative to the intensity of learning rates, the number of epochs used for the stochastic gradient descent and their relation with the convergence of the objective functions. We start in Table \ref{realtraining}, with detailed information regarding the number of sessions used as well as the learning rates and epochs.
\begin{table}[h]
	\centering \footnotesize
	\begin{tabular}{l||c|c|c||c|c|c}
		 & \multicolumn{3}{c||}{\textbf{ELBO} $\mathcal{L}$}&   \multicolumn{3}{c} {\textbf{Uncond. Exp. Reward $\mathcal{J}$}} \\ 
		 		\hline  
		\textbf{Stock}    & \textbf{sessions} &\textbf{learning rates}  & \textbf{epochs} &  \textbf{sessions}  &  \textbf{rates} & \textbf{epochs}\\  
		\hline  
		FB  & 3  & $1 \times 10^{-4}$, $1 \times 10^{-6}$, $1 \times 10^{-8}$ & 500, 500, 50 & 1  & $5 \times 10^{-5}$ & 150  \\
		INTC  & 1 & $1 \times 10^{-4}$ & 500  & 1  & $5 \times 10^{-4}$&  100 \\
		MSFT & 3  &$1 \times 10^{-4}$, $1 \times 10^{-6}$, $1 \times 10^{-7}$ & 500, 250, 10&1 & $1 \times 10^{-7}$ &  130  \\
		VOD  & 2 & $1 \times 10^{-4}$,$1 \times 10^{-6}$ & 150, 20  & 1  & $3 \times 10^{-5}$&  20\\
	\end{tabular}
	\caption{RDMM model training summary - real data}
	\label{realtraining}
\end{table}

In the last three columns of Table \ref{realtraining} we see that a single session was enough to achieve good convergence of the approximated unconditional expected reward (see Equation \eqref{eq:rl_app11}). During the ELBO optimisation, we divided the learning phase into three sessions for the Facebook and Microsoft data, where the intensity of the learning rates decreases as we progress through the learning sessions. Intel, on the other hand, achieved satisfactory results in only one session. We use two short sessions for Vodafone to achieve ELBO convergence.
\begin{figure}[H]
	\centering
	\begin{subfigure}[b]{0.4\textwidth}
		\centering
		\includegraphics[width=\textwidth]{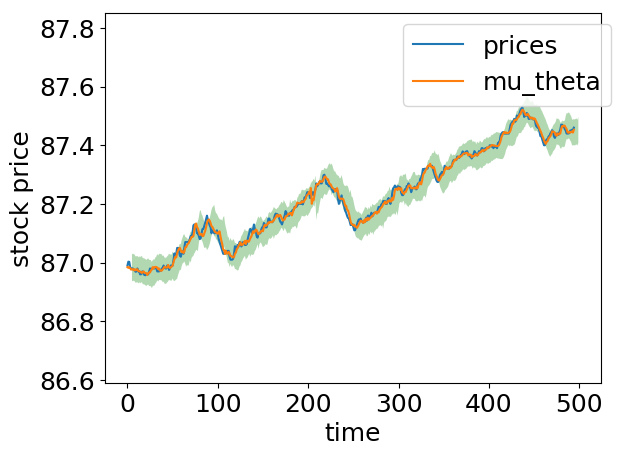}
		\caption[]%
		{{\footnotesize Prices approximation - training}} 
		\label{fig:MSFT_optimization1.1}   
	\end{subfigure}
	\begin{subfigure}[b]{0.4\textwidth}   
		\centering 
		\includegraphics[width=\textwidth]{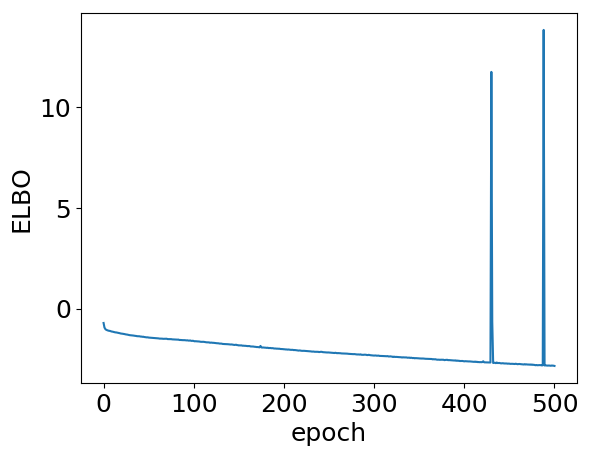}
		\caption[]%
		{{\footnotesize ELBO convergence}}  
		\label{fig:MSFT_optimization1.2}  
	\end{subfigure}
\caption[]
{\small RDMM optimisation for the MSFT dataset. (a) Price approximation of a sampled batch from the training set. (b) ELBO convergence on its second training session.} 
\label{fig:MSFT_optimization1}
\end{figure}
In Figure \ref{fig:MSFT_optimization1.1}  we have plotted the real prices of Microsoft stocks in blue and the model's approximation in orange for a sampled batch from the training set. For each price approximation, we add and subtract its estimated standard deviation (shaded area in green). The RDMM seems capable to provide a smoother approximation of the original time series. In Figure \ref{fig:MSFT_optimization1.2} we observe two spikes in the ELBO convergence (see Equation \eqref{eq:lb1}) during the second session (from a total of three sessions) where the learning rate is $1 \times 10^{-6}$ with 500 epochs (see Table \ref{realtraining}). Since learning is not destabilized too frequently we decided to not use any technique to address high loss in bad batches such as gradient clipping or adaptive learning rate clipping as in  \cite{DBLP:journals/corr/abs-1906-09060}.
\begin{figure}[H]
	\centering
	\begin{subfigure}[b]{0.4\textwidth}  
		\centering 
		\includegraphics[width=\textwidth]{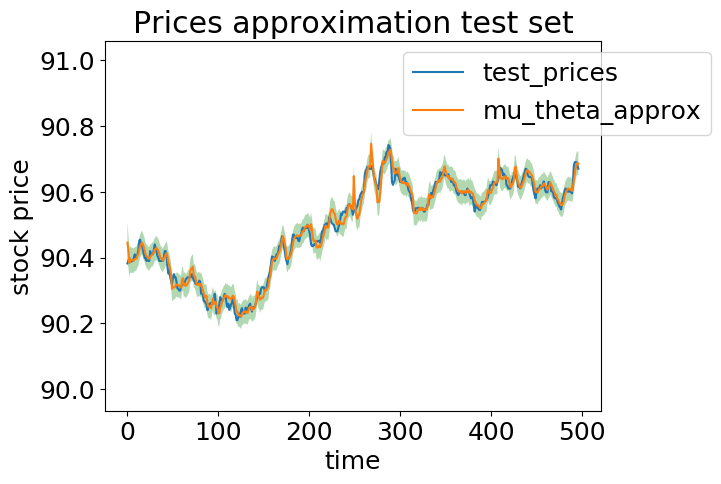}
		\caption[]%
		{{\footnotesize Predicted prices - test set}}    
		\label{fig:MSFT_optimization2.1}
	\end{subfigure}
	\begin{subfigure}[b]{0.4\textwidth}   
		\centering 
		\includegraphics[width=\textwidth]{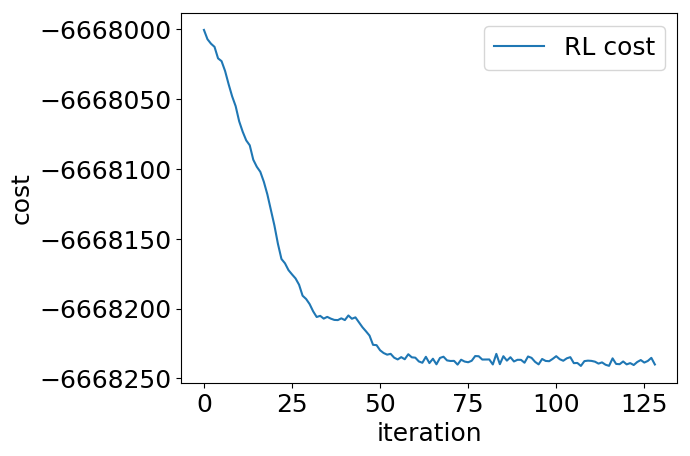}
		\caption[]%
		{{\footnotesize  Negative total expected reward}}
		\label{fig:MSFT_optimization2.2}
	\end{subfigure}
	\caption[]
	{\small RDMM optimisation for the MSFT dataset. (a) Price approximation of a sampled batch from the test set. (b) Negative total expected reward approximation convergence.}
	\label{fig:MSFT_optimization2}
\end{figure}
Figure \ref{fig:MSFT_optimization2.1}  presents the predicted prices for a sampled batch of the test set of Microsoft stocks. Figure \ref{fig:MSFT_optimization2.2} shows the approximation of the negative unconditional expected total reward (see Equation \eqref{eq:rl_app11}), performed in a single session with a learning rate of $1 \times 10^{-7}$ and 125 epochs.
\begin{figure}[H]
	\centering
	\begin{subfigure}[b]{0.35\textwidth}
		\centering
		\includegraphics[width=\textwidth]{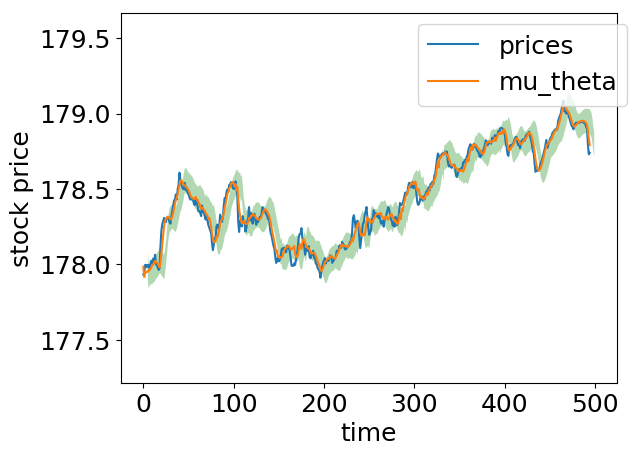}
		\caption[]%
		{{\footnotesize Prices approximation - training}} 
		\label{fig:FB_optimization1.1}   
	\end{subfigure}
	\begin{subfigure}[b]{0.35\textwidth}   
		\centering 
		\includegraphics[width=\textwidth]{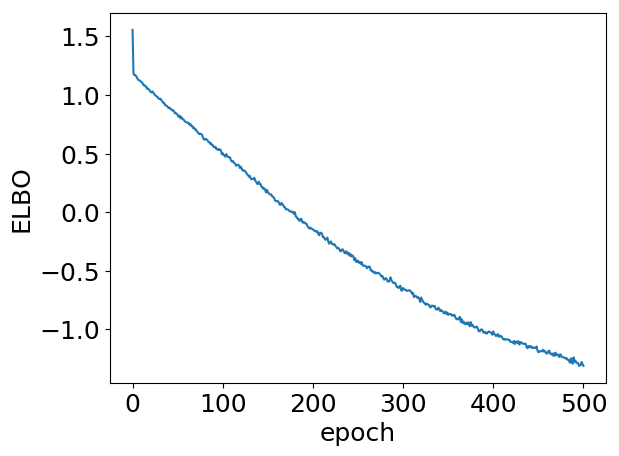}
		\caption[]%
		{{\footnotesize ELBO convergence}}  
		\label{fig:FB_optimization1.2}  
	\end{subfigure}
	\caption[]
	{\small RDMM optimisation for the FB dataset. (a) Price approximation of a sampled batch from the  training set. (b) ELBO convergence on its second training session.} 
	\label{fig:FB_optimization1}
\end{figure}
Similarly, in Figure \ref{fig:FB_optimization1.1}  we have plotted the real prices of Facebook stocks and the model's approximation for a sampled batch from the training set with the estimated variance in shaded green area. As before the RDMM seems to reduce the noise from the observed data. Figure \ref{fig:FB_optimization1.2} presents the ELBO convergence (see Equation \eqref{eq:lb1}) during the second session (from a total of three sessions) using a learning rate of $1 \times 10^{-6}$ and 500 epochs (see Table \ref{realtraining}) which seems to provide a smooth decrease in the objective function.
\begin{figure}[H]
	\centering
	\begin{subfigure}[b]{0.4\textwidth}  
		\centering 
		\includegraphics[width=\textwidth]{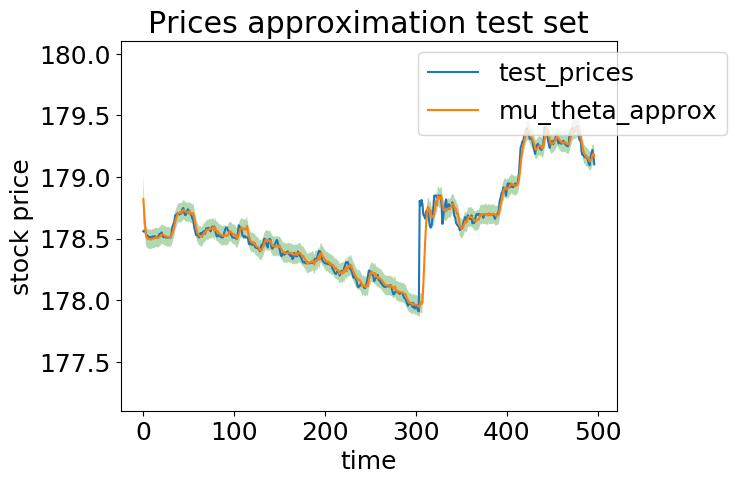}
		\caption[]%
		{{\footnotesize Predicted prices - test set}}    
		\label{fig:FB_optimization2.1}
	\end{subfigure}
	\begin{subfigure}[b]{0.4\textwidth}   
		\centering 
		\includegraphics[width=\textwidth]{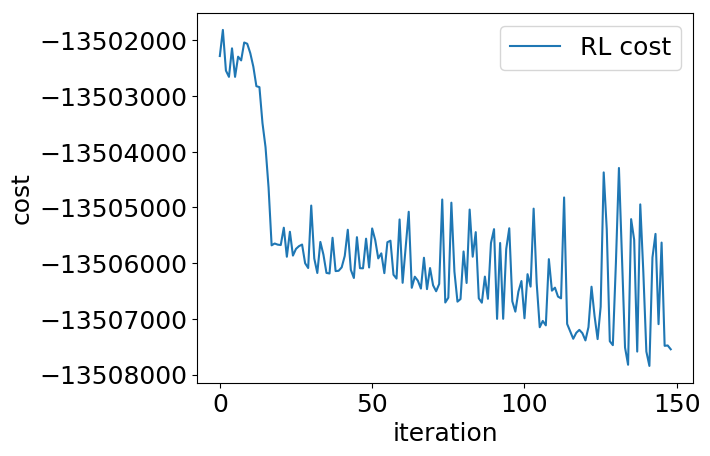}
		\caption[]%
		{{\footnotesize  Negative total expected reward}}
		\label{fig:FB_optimization2.2}
	\end{subfigure}
	\caption[]
	{\small RDMM optimisation for the FB dataset. (a) Price approximation of a sampled batch from the test set. (b) Negative total expected reward approximation convergence.}
	\label{fig:FB_optimization2}
\end{figure}
In Figure \ref{fig:FB_optimization2.1}  presents the predicted prices for a sampled batch of the test set of Facebook stock prices. Figure \ref{fig:FB_optimization2.2} shows the convergence of the approximated negative unconditional expected total reward (see Equation \eqref{eq:rl_app11}) performed in a single session using 150 epochs and a learning rate of $5 \times 10^{-5}$. We notice that for this learning rate the cost function of the policy search decreases as the number of epochs increases, but oscillates considerably more than the MSFT case, which uses a smaller learning rate.
\begin{figure}[H]
	\centering
	\begin{subfigure}[b]{0.4\textwidth}
		\centering
		\includegraphics[width=\textwidth]{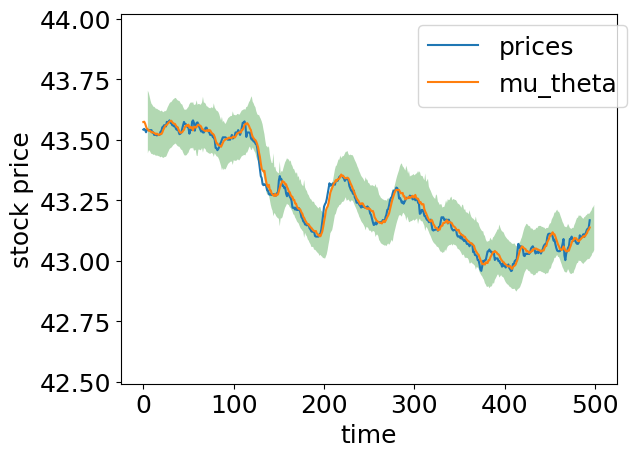}
		\caption[]%
		{{\footnotesize Prices approximation - training}} 
		\label{fig:Intel_optimization1.1}   
	\end{subfigure}
	\begin{subfigure}[b]{0.4\textwidth}   
		\centering 
		\includegraphics[width=\textwidth]{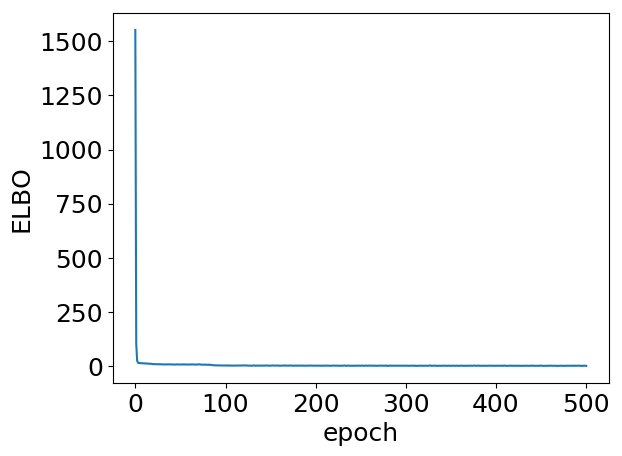}
		\caption[]%
		{{\footnotesize ELBO convergence}}  
		\label{fig:Intel_optimization1.2}  
	\end{subfigure}
	\caption[]
	{\small RDMM optimisation for the INTC dataset. (a) Price approximation of a sampled batch from the training set. (b) ELBO convergence on its first training session.} 
	\label{fig:Intel_optimization1}
\end{figure}
Intel stock prices and the RDMM approximation for a sampled batch of the training set is shown by Figure \ref{fig:Intel_optimization1.1}. The approximation seems to reduce noise fairly well.  In Figure \ref{fig:Intel_optimization1.2} we observe an abrupt drop in the ELBO convergence in the first training session (from a total of two sessions) where the learning rate adopted is $1 \times 10^{-4}$  (see Table \ref{realtraining}).
\begin{figure}[H]
	\centering{\tiny }
	\begin{subfigure}[b]{0.4\textwidth}  
		\centering 
		\includegraphics[width=\textwidth]{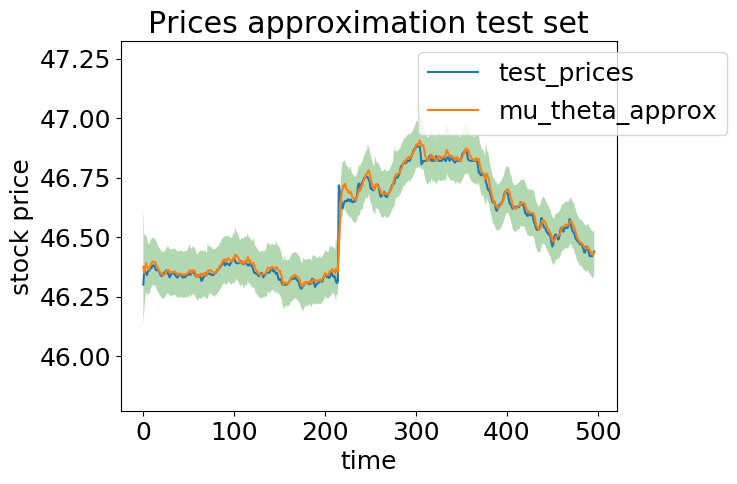}
		\caption[]%
		{{\footnotesize Predicted prices - test set}}    
		\label{fig:Intel_optimization2.1}
	\end{subfigure}
	\begin{subfigure}[b]{0.4\textwidth}   
		\centering 
		\includegraphics[width=\textwidth]{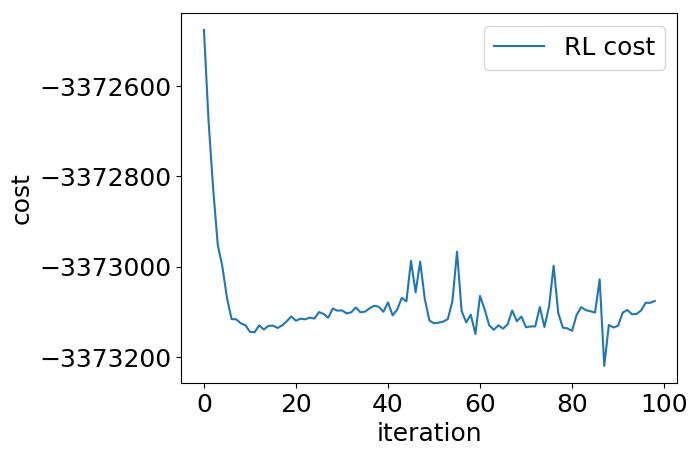}
		\caption[]%
		{{\footnotesize  Negative total expected reward}}
		\label{fig:Intel_optimization2.2}
	\end{subfigure}
	\caption[]
	{\small RDMM optimisation for the Intel dataset. (a) Price approximation of a sampled batch from the test set. (b) Negative total expected reward approximation convergence.}
	\label{fig:Intel_optimization2}
\end{figure}
Figure \ref{fig:Intel_optimization2.1}  presents the predicted prices for a sampled batch of the Intel test set. Figure \ref{fig:Intel_optimization2.2} shows the approximation of the negative unconditional expected total reward (see Equation \eqref{eq:rl_app11}) performed in a single session with a learning rate of $5 \times 10^{-4}$ and 100 epochs.
\begin{figure}[H]
	\centering
	\begin{subfigure}[b]{0.4\textwidth}
		\centering
		\includegraphics[width=\textwidth]{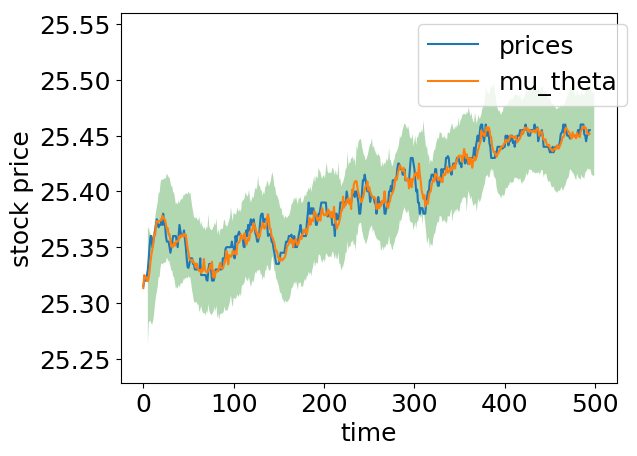}
		\caption[]%
		{{ \footnotesize Prices approximation - training}} 
		\label{fig:VOD_optimization1.1}   
	\end{subfigure}
	\begin{subfigure}[b]{0.4\textwidth}   
		\centering 
		\includegraphics[width=\textwidth]{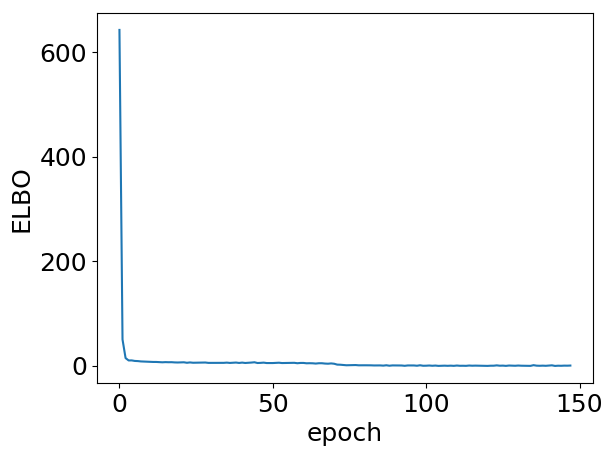}
		\caption[]%
		{{\footnotesize ELBO convergence}}  
		\label{fig:VOD_optimization1.2}  
	\end{subfigure}
	\caption[]
	{\small RDMM optimisation for the VOD dataset (a) Price approximation a of sampled batch from the training set. (b) ELBO convergence on its first training session.} 
	\label{fig:VOD_optimization1}
\end{figure}
In Figure \ref{fig:VOD_optimization1.1}  we have, for a sampled batch from the training set, plotted in blue the real prices of Vodafone stocks, whereas the model's approximation for those prices are plotted in orange. For each price approximation, we add and subtract its estimated standard deviation (shaded area in green).  Figure \ref{fig:VOD_optimization1.2} presents the ELBO convergence during the first training session (from a total of two sessions) using a learning rate of $1 \times 10^{-4}$ and 150 epochs (see Table \ref{realtraining}).
\begin{figure}[H]
	\centering
	\begin{subfigure}[b]{0.4\textwidth}  
		\centering 
		\includegraphics[width=\textwidth]{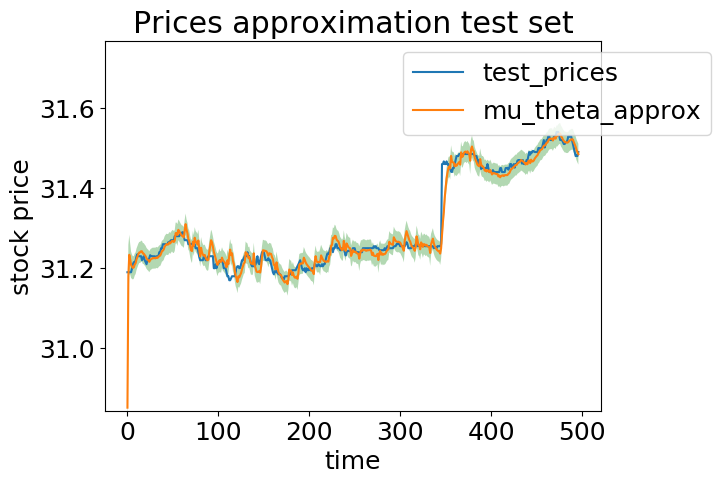}
		\caption[]%
		{{\footnotesize Predicted prices - test set}}    
		\label{fig:VOD_optimization2.1}
	\end{subfigure}
	\begin{subfigure}[b]{0.4\textwidth}   
		\centering 
		\includegraphics[width=\textwidth]{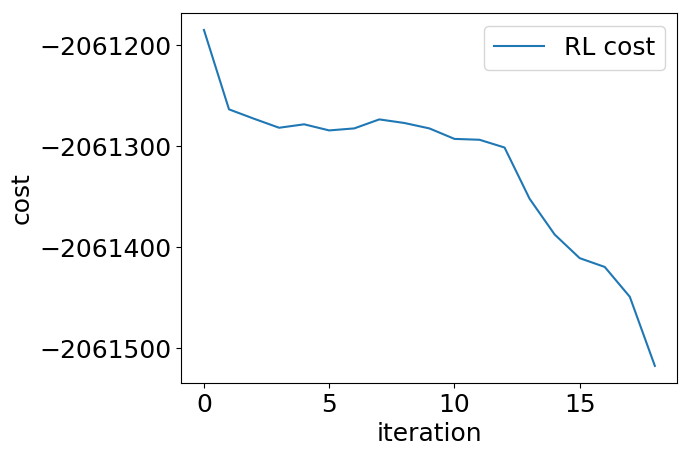}
		\caption[]%
		{{\footnotesize  Negative total expected reward}}
		\label{fig:VOD_optimization2.2}
	\end{subfigure}
	\caption[]
	{\small RDMM optimisation for the VOD dataset. (a) Price approximation a sampled batch from the test set. (b) Negative total expected reward approximation convergence.}
	\label{fig:VOD_optimization2}
\end{figure}
Figure \ref{fig:VOD_optimization2.1}  presents the predicted prices for a sampled batch of the test set of Vodafone stock prices. Figure \ref{fig:VOD_optimization2.2} shows the convergence of the approximated negative unconditional expected total reward (see Equation \eqref{eq:rl_app11}) where we chose to perform it in a single session using and a large learning rate of $3 \times 10^{-5}$ achieving good results with only 20 epochs.

The challenge of finding adequate values for hyperparameters is nearly ubiquitous to any machine learning algorithm as it is for the RDMM. In fact, as we suggested at the beginning of Subsection \ref{RDMM_learning_real_data}, it would be advantageous for commercial uses, the implementation of an extra module in the algorithm to manage and automate the selection of the size and complexity of the neural networks, number of epochs, and learning rates. This automation should take into account that the ideal size for the neural networks seems to be associated with the volatility of the dataset. The number of epochs adopted and the learning rates are closely associated, with the policy search part of the RDMM algorithm, suggesting that the optimisation could be facilitated by the approach proposed by \cite{Smith2017CyclicalLR}, where the author recommends allowing the learning rates to vary cyclically within a range of values, rather than using an exponential decay or adaptative learning rate such as ADAM used in our project.  As \cite{DBLP:journals/corr/DauphinVCB15} states; saddle points are the main obstacles in optimising large, deep neural networks with non-convex objective functions. \cite{Smith2017CyclicalLR} suggests that a cyclical approach, where the learning rate is periodically increased, allows the algorithm to transverse saddle point plateaus. Consequently, this could be a better technique for dealing with two conflicting situations, where a small learning rate makes the objective converge slowly while a large learning rate can destabilize the optimisation.

\subsection{RDMM financial performance on real data }
To evaluate the performance of the RDMM approach,  we compare our method against the six benchmarks mentioned before: Q-learning, DynaQ-ARIMA, DynaQ-LSTM, DynaQ-DMM, RDMM-NoU, and TWAP3, using the test data set, i.e., the last 200,000 data points (50,000 for VOD) of the time-series. The financial performance is evaluated by calculating the relative savings $RS_{R}$ in basis points as in the simulated price dynamics section, where for every batch $b$ of 500 data points we compute:
\begin{equation}
RS_{R^{(b)}}= \frac{R^{(b)}_{RDMM}-R^{(b)}_{benchmark}}{R^{(b)}_{benchmark}} \times 10^4,
\end{equation}
with
\begin{equation} \label{acc_rewards}
R^{(b)}=\sum_{t=1}^{500} r^{(b)}_{t}
\end{equation}
where $r_{t}$ is given by
\begin{equation}
r^{(b)}_t=(x^{(b)}_{t}- c_{2} a^{(b)}_{t})a^{(b)}_{t} - c_3 {q^2_t}^{(b)},
\label{simplereward22_real_batch}
\end{equation}
which is the reward for instant $t$, defined in Equation \eqref{simplereward22_real}, applied to batch $b$.

\begin{figure}[H] 
	\centering
	\includegraphics[width=0.7\textwidth]{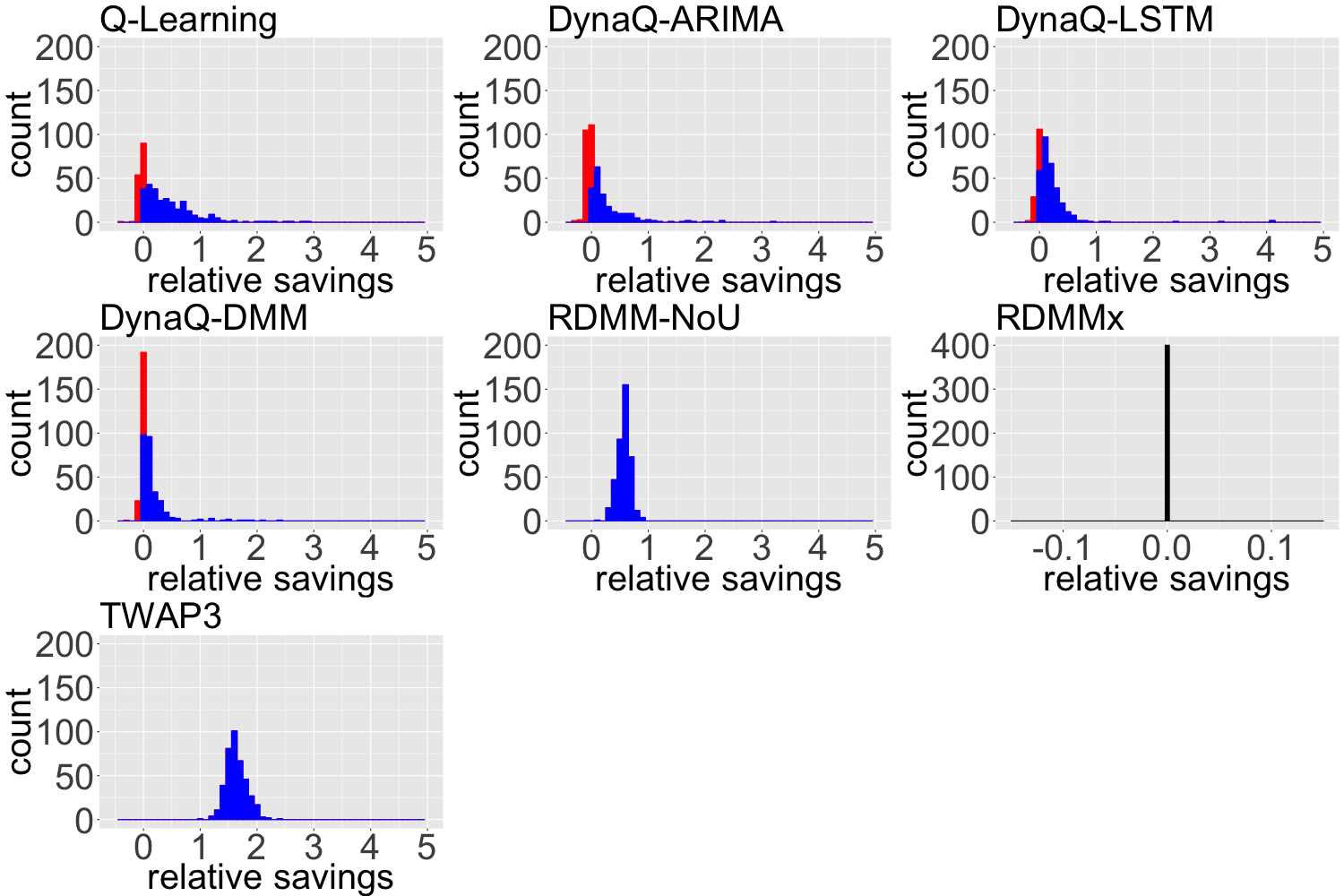}
	\caption{MSFT histogram of the relative savings in basis points of the total reward. Notice that the RDMMx is in a different scale.}
	\label{fig:MSFT_hist_reward}
\end{figure}

\begin{figure}[H] 
	\centering
	\includegraphics[width=0.7\textwidth]{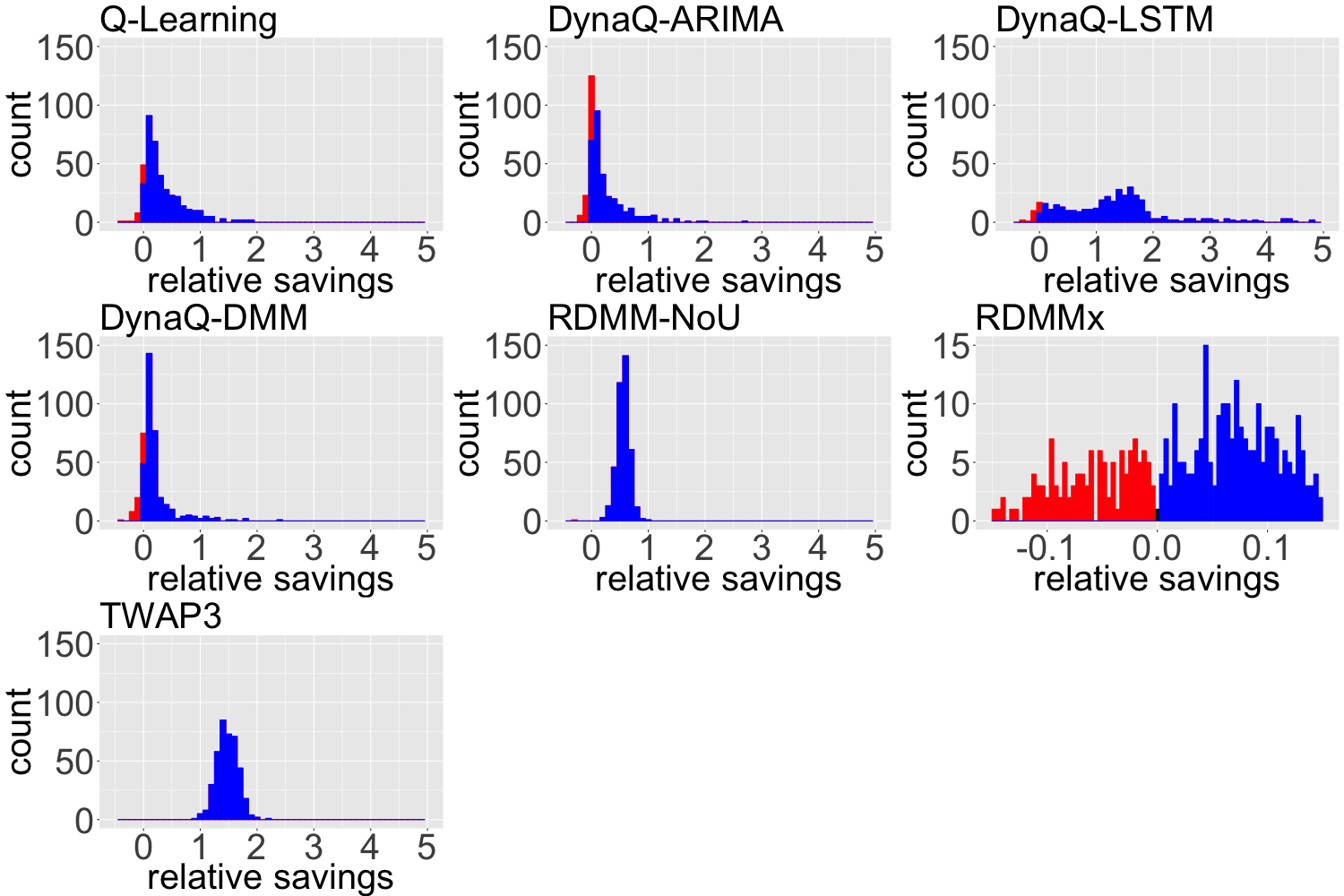}
	\caption{INTC histogram of the relative savings in basis points of the total reward. Notice that the RDMMx is in a different scale.}
	\label{fig:Intel_hist_reward}
\end{figure}

\begin{figure}[h] 
	\centering
	\includegraphics[width=0.7\textwidth]{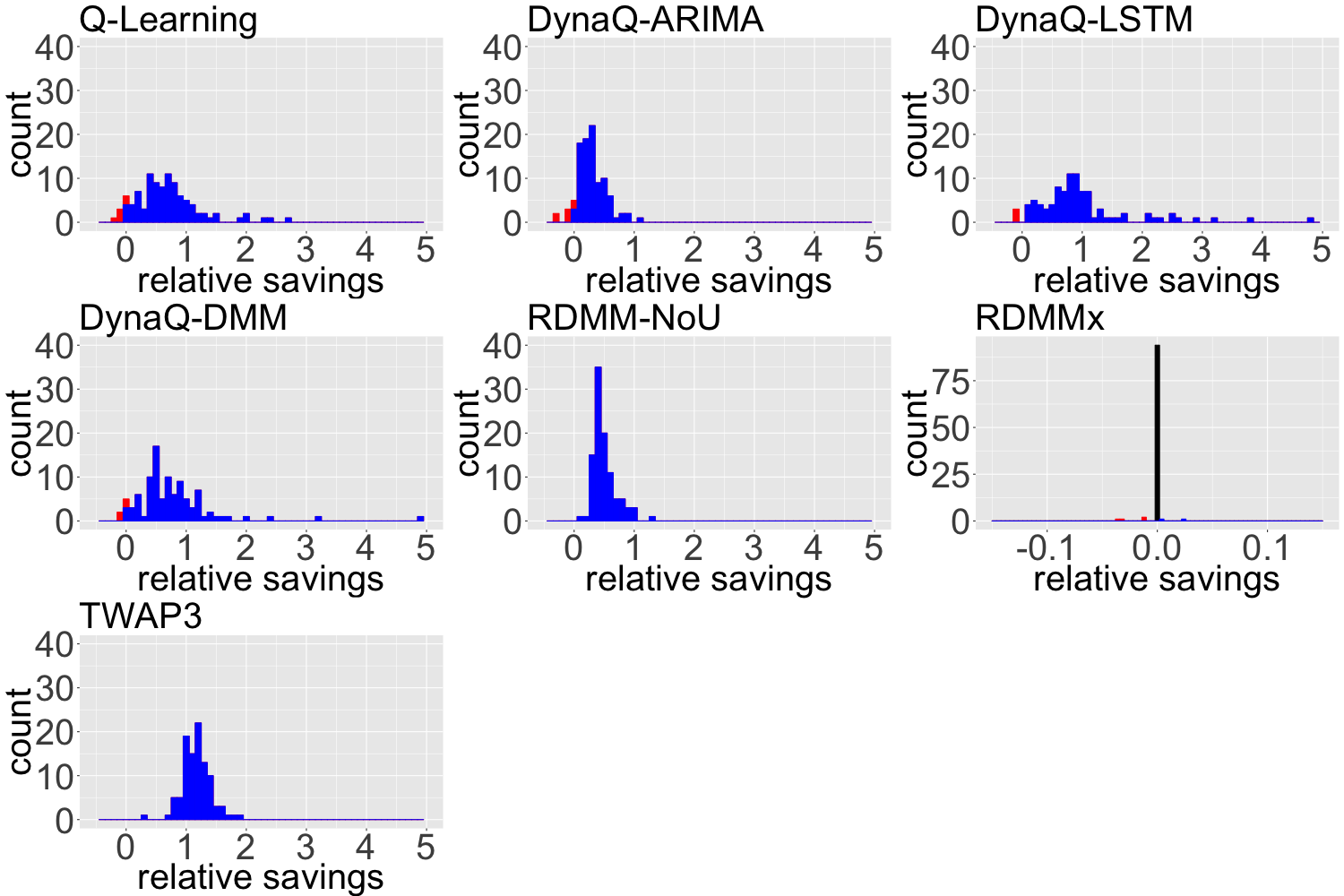}
	\caption{VOD histogram of the relative savings in basis points of the total reward. Notice that the RDMMx is in a different scale.}
	\label{fig:VOD_hist_reward}
\end{figure}

\begin{figure}[h] 
	\centering
	\includegraphics[width=0.7\textwidth]{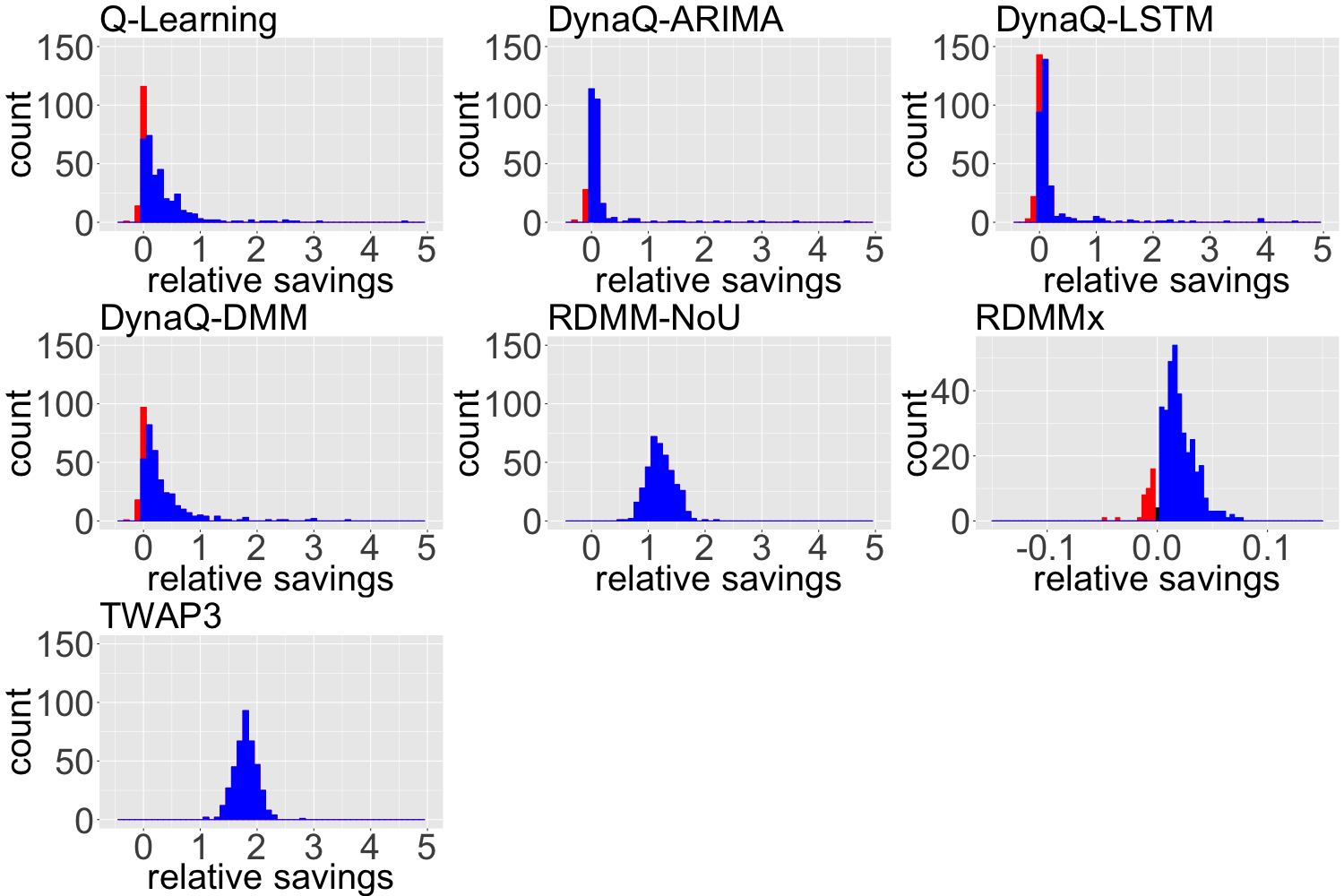}
	\caption{FB histogram of the relative savings in basis points of the total reward. Notice that the RDMMx is in a different scale.}
	\label{fig:FB_hist_reward}
\end{figure}

The histograms for the $RS_R$ of the 400 time series (100 for VOD) are plotted in Figures \ref{fig:MSFT_hist_reward}, \ref{fig:Intel_hist_reward}, \ref{fig:VOD_hist_reward}, and \ref{fig:FB_hist_reward}. The batches where the RDMM is outperforming the benchmark (i.e. $RS_{R}>0$) are represented in blue, and when the RDMM is outperformed by the benchmark (i.e. $RS_{R} \leq 0$) we represent it with red. {\color{black}  Due to the asymmetric nature of the distribution of the relative savings in basis points of the total reward, we add a table with the quantiles and the mean of the distribution above.}

\begin{table}[H]
	\centering
	\begin{tabular}{lrrrrrr}
		\toprule
		\thead{Approach} & \thead{10\%} & \thead{25\% }& \thead{Median} & \thead{75\%} & \thead{90\%} & \thead{Mean} \\ 
		\midrule
Q-learning & -0.07 & -0.01 & 0.18 & 0.54 & 0.90 & 0.34 \\ 
DynaQ-ARIMA & -0.09 & -0.06 & 0.02 & 0.21 & 0.57 & 0.16 \\ 
DynaQ-LSTM & -0.04 & 0.02 & 0.12 & 0.25 & 0.44 & 0.68 \\ 
DynaQ-DMM & -0.04 & -0.01 & 0.04 & 0.11 & 0.29 & 0.12 \\ 
RDMM-NoU & 0.42 & 0.51 & 0.58 & 0.64 & 0.69 & 0.57 \\ 
RDMMx & 0.00 & 0.00 & 0.00 & 0.00 & 0.00 & 0.00 \\ 
TWAP3 & 1.43 & 1.51 & 1.60 & 1.74 & 1.88 & 1.63 \\ 
		\bottomrule
	\end{tabular}
	\caption{MSFT quantiles and mean of the relative savings in basis points of the total reward}
	\label{tab:MSFT_hist_reward}   
\end{table}

In Figure \ref{fig:MSFT_hist_reward} we observe that the RDMM outperforms the benchmarks in almost all batches of the MSFT test set. DynaQ-LSTM produces inferior results in a few samples due to model bias, as we suspect during the discussion of the policies obtained, where the DynaQ-LSTM seems to generate a policy with no actions assigned to some states (see Figure \ref{fig:MSFT_policies_lstm}). One possible explanation is that the model bias introduces, for those states, some form of overly optimistic expectation that the stock prices will rise in the near future. This issue resulted in three out of 400 batches (0.75\%) where the inventory was not fully executed at the terminal state leaving a small remainder. For those batches, we see a large $RS_{R}$ in favour of the RDMM. {\color{black}  In the RDMMx histogram, we observe that all relative savings are equal to zero, i.e., there is no difference in performance compared to RDMM. After a closer inspection, we notice that RDMMx and RDMM performed the same actions in the MSFT dataset. Table \ref{tab:MSFT_hist_reward} confirms that the majority of the distributions are rightly skewed.}

\begin{table}[H]
	\centering
	\begin{tabular}{lrrrrrr}
		\toprule
		\thead{Approach} & \thead{10\%} & \thead{25\% }& \thead{Median} & \thead{75\%} & \thead{90\%} & \thead{Mean} \\ 
		\midrule
Q-learning & 0.02 & 0.09 & 0.22 & 0.49 & 0.85 & 0.34 \\ 
DynaQ-ARIMA & -0.03 & 0.01 & 0.10 & 0.29 & 0.66 & 0.22 \\ 
DynaQ-LSTM & 0.09 & 0.60 & 1.31 & 1.69 & 2.67 & 1.69 \\ 
DynaQ-DMM & -0.03 & 0.05 & 0.11 & 0.21 & 0.46 & 0.19 \\ 
RDMM-NoU & 0.42 & 0.50 & 0.56 & 0.62 & 0.69 & 0.55 \\ 
RDMMx & -0.09 & -0.02 & 0.05 & 0.10 & 0.14 & 0.04 \\ 
TWAP3 & 1.24 & 1.35 & 1.47 & 1.61 & 1.70 & 1.47 \\ 
		\bottomrule
	\end{tabular}
	\caption{INTC quantiles and mean of the relative savings in basis points of the total reward}
	\label{tab:Intel_hist_reward}   
\end{table}

In Figure \ref{fig:Intel_hist_reward} we, once again, observe that the RDMM outperforms the benchmarks in almost all batches of the Intel stocks test set. As with the MSFT stocks, the DynaQ-LSTM seems to suffer from the same model bias problem discussed previously and results in three  out of  400 batches (0.75\%) where the inventory is not fully executed, generating a large $RS_{R}$ in favour of the RDMM.  {\color{black}  For the INTC dataset, the RDMMx histogram shows a slightly better performance in favour of the RDMM approach which is corroborated by Table \ref{tab:Intel_hist_reward}}.

\begin{table}[H]
	\centering
	\begin{tabular}{lrrrrrr}
		\toprule
		\thead{Approach} & \thead{10\%} & \thead{25\% }& \thead{Median} & \thead{75\%} & \thead{90\%} & \thead{Mean} \\ 
		\midrule
Q-learning & 0.07 & 0.36 & 0.63 & 0.89 & 1.26 & 0.68 \\ 
DynaQ-ARIMA & 0.05 & 0.13 & 0.26 & 0.40 & 0.58 & 0.28 \\ 
DynaQ-LSTM & 0.22 & 0.58 & 0.86 & 1.18 & 2.63 & 2.07 \\ 
DynaQ-DMM & 0.16 & 0.44 & 0.67 & 0.96 & 1.31 & 0.81 \\ 
RDMM-NoU & 0.34 & 0.37 & 0.44 & 0.56 & 0.82 & 0.50 \\ 
RDMMx & 0.00 & 0.00 & 0.00 & 0.00 & 0.00 & -0.00 \\ 
TWAP3 & 0.92 & 1.03 & 1.19 & 1.30 & 1.43 & 1.18 \\ 
		\bottomrule
	\end{tabular}
	\caption{VOD quantiles and mean of the relative savings in basis points of the total reward}
	\label{tab:VOD_hist_reward}   
\end{table}

Figure \ref{fig:VOD_hist_reward} shows the dominance of the RDMM on the Vodafone stocks test set. 
As in the other test sets, DynaQ-LSTM yielded poor results in some batches, but the inventory is fully executed in 100\% of the cases leaving no remainder at the terminal states for all 50 batches of the test set.
On the top left and top centre panels we see that Q-learning and DynaQ-ARIMA produce better results than the RDMM in only a few batches.   {\color{black}  Similar to the MSFT case, the RDMMx histogram for the VOD dataset shows almost no difference in performance compared to the RDMM. In Table \ref{tab:VOD_hist_reward}, the small advantage of the RDMMx approach (RS mean equals to -0.00067) was suppressed due to rounding to 2 decimals places. After a closer inspection, we noticed that the RDMM and RDMMx actions differ only in a small fraction (16/8077) of the total number of actions executed.}

\begin{table}[H]
	\centering
	\begin{tabular}{lrrrrrr}
		\toprule
		\thead{Approach} & \thead{10\%} & \thead{25\% }& \thead{Median} & \thead{75\%} & \thead{90\%} & \thead{Mean} \\ 
\midrule
Q-learning & -0.02 & 0.03 & 0.14 & 0.39 & 0.72 & 0.30 \\ 
DynaQ-ARIMA & -0.04 & -0.01 & 0.02 & 0.08 & 0.15 & 0.11 \\ 
DynaQ-LSTM & -0.03 & 0.02 & 0.07 & 0.14 & 0.95 & 1.44 \\ 
DynaQ-DMM & -0.02 & 0.03 & 0.16 & 0.38 & 0.72 & 0.30 \\ 
RDMM-NoU & 0.92 & 1.06 & 1.21 & 1.36 & 1.54 & 1.22 \\ 
RDMMx & -0.00 & 0.01 & 0.02 & 0.03 & 0.04 & 0.02 \\ 
TWAP3 & 1.54 & 1.67 & 1.79 & 1.93 & 2.05 & 1.80 \\ 
 		\bottomrule
	\end{tabular}
	\caption{FB quantiles and mean of the relative savings in basis points of the total reward}
\label{tab:FB_hist_reward}   
\end{table}

In Figure \ref{fig:FB_hist_reward}, the superior performance of the RDMM is validated one more time as it significantly outperforms the benchmarks in almost all batches of the Facebook stocks test set. DynaQ-LSTM leaves a small remainder in the final state in eight out of 400 batches (2\%), where the inventory was not fully executed.
The shape of the histograms appears, in general, more positively skewed for the Facebook dataset compared to the other stocks, indicating higher financial gains for the RDMM compared to the other methods.  {\color{black}  We see a slightly better performance by the original RDMM according to the RDMMx histogram in the Facebook dataset.}

To better assess the financial gain produced by RDMM we compute the accumulated reward for every batch $b$ on the test set (size of 500) with Equation \ref{acc_rewards}.

Next we estimate the mean batch reward $\bar{R}$ of the test set with
\begin{equation}
\bar{R}=\frac{1}{B}\sum_{b=1}^{B} R^{b}
\label{avg_total_reward}
\end{equation}
where B is the total number of batches in the test set, i.e., $B=400$  ($B=100$ for VOD). For every algorithm, the resulting estimation of Equation \eqref{avg_total_reward} applied to Intel, Microsoft, Vodafone and Facebook stocks, can be found in Table \ref{rewardavg}.


\begin{table}[H]
	\centering  {\color{black} 
	\begin{tabular}{lcccc} 
		\toprule
		\thead{Approach}	& \thead{INTC} & \thead{MSFT} & \thead{VOD} & \thead{FB}  \\ 
		\midrule
		Q-learning & 10534.60 & 19241.95 & 6550.63 & 37588.15\\ 
		DynaQ-ARIMA& 10534.73 & 19242.30 & 6550.89 & 37588.90 \\ 
		DynaQ-LSTM & 10533.16 & 19241.40 & 6549.62 & 37584.17 \\ 
		DynaQ-DMM & 10534.75 & 19242.37 & 6550.54 & 37588.18\\ 
		RDMM-NoU & 10534.37 & 19241.50 & 6550.76 & 37584.70 \\ 
		TWAP3 &10533.40 & 19239.44 & 6550.31 & 37582.52 \\ 
		RDMMx  & 10534.91 & 19242.59 & 6551.09 & 37589.24\\ 
		RDMM & 10534.96 & 19242.59 & 6551.08 & 37589.30\\
		\bottomrule
	\end{tabular}}
	\caption{Mean batch reward $\bar{R}$ in US dollars on all test sets given by Equation \eqref{avg_total_reward} }
	\label{rewardavg}   
\end{table}

Table \ref{rewardavg} shows that the RDMM approach generated the highest reward in almost all test sets considered. {\color{black}  The exception is the RDMMx approach, which compared to the RDMM yielded the same mean batch reward for the MSFT dataset and a slightly superior mean batch reward  for the VOD dataset.}
 For the Intel and Microsoft test sets, DynaQ-DMM was the third place followed by DynaQ-ARIMA. For the Facebook and Vodafone test sets, DynaQ-DMM and DynaQ-ARIMA switch places in terms of their performances. Overall, TWAP3 and DynaQ-LSTM delivered inferior results compared with the other methods. 

One might also consider the total reward accumulated $R_{acc}$ in the test set which is computed by:
\begin{equation}
R_{acc}=\sum_{b=1}^{B} R^{(b)}.
\label{total_reward}
\end{equation}
The resulting total reward accumulated \eqref{total_reward} values from our experiments are shown by Table \ref{rewardacc}
\begin{table}[H]
	\centering { \color{black} 
	\begin{tabular}{lrrrr}
	\toprule
\thead{Approach}	& \thead{INTC} & \thead{MSFT} & \thead{VOD} & \thead{FB}  \\ 
\midrule
		Q-learning & 4,213,840.38 & 7,696,781.21 & 655,062.69 & 15,035,261.72 \\ 
		DynaQ-ARIMA & 4,213,890.65 & 7,696,920.73 & 655,089.19 & 15,035,559.59 \\ 
		DynaQ-LSTM  & 4,213,264.39 & 7,696,561.65 & 654,962.30 & 15,033,668.45 \\ 
		DynaQ-DMM & 4,213,901.02 & 7,696,949.70 & 655,054.11 & 15,035,270.46 \\ 
		RDMM-NoU & 4,213,749.96 & 7,696,599.59 & 655,076.06 & 15,033,878.16 \\ 
		TWAP3 & 4,213,359.78 & 7,695,776.42 & 655,031.33 & 15,033,006.60 \\ 
		RDMMx & 4,213,965.91 & 7,697,035.90 & 655,108.51 & 15,035,694.30 \\ 
		RDMM & 4,213,982.71 & 7,697,035.90 & 655,108.47 & 15,035,720.19 \\ 
		\bottomrule
\end{tabular} }
     \caption{Total reward accumulated $R_{acc}$ in US dollars for all test sets}
\label{rewardacc}   
\end{table}

{\color{black} 
The total reward accumulated values in Table \ref{rewardacc} naturally follow the same narrative seen in Table \ref{rewardavg} where the RDMM and the RDMMx outperform all other baselines models, and DynaQ-DMM and DynaQ-ARIMA alternate between the  third and fourth places.}

{\color{black}  For every batch we compute the difference between the total reward of regular RDMM and the benchmark,
\begin{equation} \label{differences}
d_b=R^{(b)}_{RDMM}-R^{(b)}_{benchmark}  \qquad \text{for b=1,2,3,...}
\end{equation}
where $R^{(b)}$ is the total reward of the batch $b$ as defined in Equation \ref{acc_rewards}}, and we conduct a paired t-test of the mean difference of total reward between the RDMM approach and every benchmark. The null hypothesis is that the true mean difference is equal to zero, with the alternative that the difference is positive in favour of the RDMM.

The mean differences with the resulting p-values from the statistical test, indicated in parentheses, can be found in Table \ref{pairedtest_real}. 

%

\begin{table}[H]
	\centering  {\color{black} 
	\begin{tabular}{lcccc} 
		\toprule
		\thead{Approach}	& \thead{INTC} & \thead{MSFT} & \thead{VOD} & \thead{FB}  \\ 
		\midrule
		{Q-learning} & 0.3558 & 0.6367 & 0.4578 & 1.1462 \\[-0.03in]
  		& ($3.7 \times 10^{-52}$) & ($2.1 \times 10^{-37}$) & ($1.2 \times 10^{-21}$) & ($2.6 \times 10^{-29}$) \\ [0.1in]
		{DynaQ-ARIMA}& 0.2301 & 0.2879 & 0.1928 & 0.4015 \\[-0.03in]
		& ($1.8 \times 10^{-29}$) & ($8.9 \times 10^{-15}$) & ($1.8 \times 10^{-20}$) & ($5.4 \times 10^{-7}$) \\ [0.1in]
		{DynaQ-LSTM}& 1.7958 & 1.1856 & 1.4617 & 5.1294 \\[-0.03in]
		& ($1.0 \times 10^{-14}$) & ($2.7 \times 10^{-3}$) & ($7.4 \times 10^{-4}$) & ($1.1 \times 10^{-4}$) \\ [0.1in]
		{DynaQ-DMM} & 0.2042 & 0.2155 & 0.5436 & 1.1243 \\ [-0.03in]
		& ($6.5 \times 10^{-29}$) & ($4.1 \times 10^{-15}$) & ($1.0 \times 10^{-16}$) & ($8.4 \times 10^{-32}$) \\ [0.1in]
		{RDMM-noU}  & 0.5819 & 1.0908 & 0.3241 & 4.6051 \\ [-0.03in]
		& ($1.5 \times 10^{-233}$) & ($3.1 \times 10^{-266}$) & ($1.1 \times 10^{-48}$) & ($1.5 \times 10^{-250}$) \\ [0.1in]
		{TWAP3 }& 1.5573 & 3.1487 & 0.7714 & 6.7840 \\ [-0.03in]
		& ($2.0 \times 10^{-299}$) & ($9.1 \times 10^{-315}$) & ($2.8 \times 10^{-66}$) & ($1.7 \times 10^{-311}$) \\[0.1in]
		{RDMMx}  & 0.0420 & 0.0000 & -0.0004 & 0.0647 \\ [-0.03in]
		& ($1.5 \times 10^{-18}$) & ($10.0 \times 10^{-1}$) & ($9.0 \times 10^{-1}$) & ($3.5 \times 10^{-68}$) \\ [0.1in]
		\bottomrule
	\end{tabular}}
	\caption{Mean differences statistic  $\bar{d}=\frac{1}{B}\sum_{b} d_b$ of the total reward  of the paired t-test  with the  associated p-values in parentheses.}
	\label{pairedtest_real}
\end{table}

From Table \ref{pairedtest_real} we see in all cases that the estimated mean difference is positive {\color{black}  in almost all cases except when the RDMM is running against the RDMMx model in MSFT and VOD datasets, where the mean differences are not statistically significant.}

{\color{black}  
For the other baselines models, we have enough evidence  from our experiments to assert that the financial gain of the RDMM approach is statistically significantly higher than the gain achieved with these benchmark models (excluding the RDMMx)  when looking at the financial metrics defined by Equations \eqref{avg_total_reward} and \eqref{total_reward}. }

 As mentioned before, it is a challenge to gauge which part or idea behind the RDMM is more relevant for producing good results. The RDMM-NoU and DynaQ-DMM approaches were an attempt at grasping the benefits of using a filtered process instead of the raw observations, or the use of the state uncertainty as an input for the deterministic policy in the RDMM model. From the results in Tables \ref{rewardavg}, \ref{rewardacc} and \ref{pairedtest_real}, we see that, on some occasions, these approaches perform well, but they are always outperformed by the complete model. 

Despite of our efforts to determine the significance of the individual components of the RDMM via ablation studies\index{ablation studies}, we should also consider that the performance of the DMM in the DynaQ-DMM approach could be partially bounded by a Q-learning type framework. In the RDMM-NoU approach, the challenge lies in defining what is the best practice regarding the sizes of the neural networks involved in the policy search process, once part of the inputs (i.e., the state uncertainty) of the deterministic function has been eliminated.

{\color{black}  
With the RDMMx we find limited evidence that artificially increasing the number of trajectories enhances the policy search. Producing equal and slightly superior mean rewards for the MSFT and VOD datasets, respectively, compared to the standard RDMM.  In the INTC and FB cases, the RDMMx found inferior results.
}

\section{Discussion and Conclusions}
\label{conclusions}

We demonstrated that the RDMM architecture proposed outperforms classical approaches like Q-learning and variations of DynaQ in a simple setting like the mean reversion problem. The performance improvement becomes more pronounced when price dynamics are more complex, and this has been demonstrated using real data sets from the limit order book of Facebook, Intel, Vodafone and Microsoft.   Since our approach requires fewer training examples (as we allowed the benchmarks to have access to the entire dataset to train while the RDMM used just half of it), we can conclude that the method is very data efficient. Additionally, as defined in the method core assumptions, this approach can handle noisy and incomplete observations since the observations are processed in a POMDP framework, where the RL policy search is performed with respect to the filtered process provided by a modified DMM. Another important feature of the RDMM is that it has been formulated while taking into account the impact caused by the agent to the environment.

It should be highlighted that all experiments were conducted using  a small range for inventory and prices. In real life applications, where prices vary considerably more or the agent is required to deal with large amounts of shares to sell, implementing procedures like Q-learning and DynaQ might become infeasible due to an unmanageable number of states to visit.  An attempt to circumvent this issue would be to use larger price intervals on the Q-tables to reduce the number of states (binning) but it is reasonable to conjecture that this might result in loss of information and inferior results.

Although the RDMM model has reasonable assumptions, we need to verify to what extent its architecture helps us understand the environment and generate profits. An experiment to evaluate the importance of direct connections between actions and latent variables representing the environment, and the degree to which the agent's actions can deteriorate the quality of policies obtained by the benchmarks and the RDMM should be conducted. 

A challenge posed by the RDMM, is the visualization of its policies. A heatmap gives an accurate representation for a simplistic approach  based only on value functions like Q-learning. However, it does not provide an adequate representation of the rationale behind the actions taken by the RDMM, given the complexity of the neural networks involved, where an LSTM summarising past information is used.  This was identified in the graphical representation in Figure \ref{fig:arch}, and in more details, in Equations \eqref{eq:pol} and \eqref{eq:approx_dist}. In other words, the policy derived from the RDMM architecture does not depend only on the current price and inventory, it also depend on all historical data such as prices, inventory, rewards and actions.

Finally, the RDMM was formulated and implemented to handle much more information than the prices and inventory, such as high frequency snapshots of the LOB containing the type of orders, prices, depths, as well as auxiliary information coming from different sources not restricted to the LOB. Unfortunately, due to time constraints we could not fully explore the RDMM's ability to address different kinds of inputs, where we believe this method's advantages might become more evident.

\section*{Acknowledgments}
We would like to thank Sebastian Jaimungal and David Duvenaud for their guidance, and other members of the Department of Statistical Sciences of University of Toronto for helpful discussions.

\begin{appendices}

\section{Activation Functions}

In this section we will define the activation functions used in this article. Figure \ref{fig:Activation} presents a plot of the curves described.

\subsection{Rectified linear unit (ReLU)}

\begin{equation}
{\displaystyle f(x)={\begin{cases}0&{\text{for }}x<0\\x&{\text{for }}x\geq 0\end{cases}}}
\end{equation}

\subsection{Hyperbolic tangent (tanh)}
\begin{equation}
{\displaystyle f(x)=\tanh(x)={\frac {(e^{x}-e^{-x})}{(e^{x}+e^{-x})}}}
\end{equation}
\subsection{Softplus}
\begin{equation}
{\displaystyle f(x)=\ln(1+e^{x})}
\end{equation}
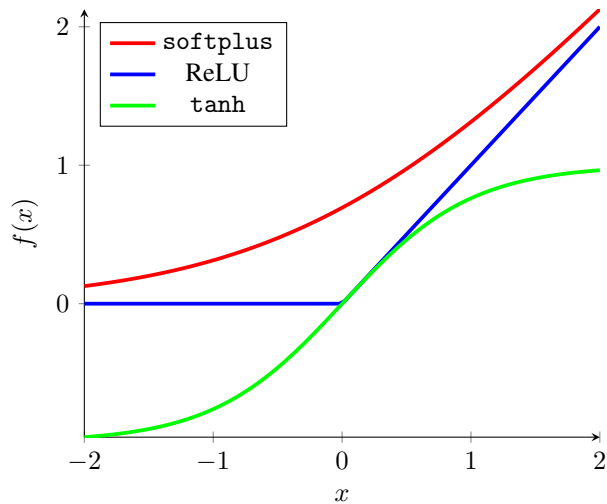
\begin{figure}
\centering
\begin{tikzpicture}[scale=1.0]
\begin{axis}[
    axis lines = left,
    xlabel = $x$,
    ylabel = {$f(x)$},
    legend pos = north west,
]
\addplot [
    domain=-2:2, 
    samples=100, 
    color=red,
    line width=0.5mm,
]
{ln(1+exp(x))};
\addlegendentry{\texttt{softplus}}
\addplot [
    domain=-2:2, 
    samples=100, 
    color=blue,
    line width=0.5mm,
    ]
    {x>0 ? x : 0};  
\addlegendentry{ReLU}
\addplot [
    domain=-2:2, 
    samples=100, 
    color=green,
    line width=0.5mm,
    ]
    {tanh(x)};  
\addlegendentry{\texttt{tanh}}
\end{axis}
\end{tikzpicture}
\caption{Activation Functions} \label{fig:Activation}
\end{figure}

\section{Neural Network Architectures for Sequential Data}
\subsection{LSTM}
\label{LSTM_ex}

A long short-term memory (LSTM) \citep{Hochreiter1997} is an recurrent neural network (RNN) where each building unit is composed of a memory cell and 3 gates; an input gate, an output gate and a forget gate:

\begin{align*}
f_{t}&=\sigma _{g}(W_{f}x_{t}+U_{f}h_{t-1}+b_{f}) \quad \textsf{forget gate}\\
i_{t}&=\sigma _{g}(W_{i}x_{t}+U_{i}h_{t-1}+b_{i}) \quad \textsf{input gate}\\
o_{t}&=\sigma _{g}(W_{o}x_{t}+U_{o}h_{t-1}+b_{o}) \quad \textsf{output gate}\\
c_{t}&=f_{t}\circ c_{t-1}+i_{t}\circ \sigma _{c}(W_{c}x_{t}+U_{c}h_{t-1}+b_{c}) \quad \textsf{memory cell}\\
h_{t}&=o_{t}\circ \sigma _{h}(c_{t})
\end{align*}

Here, $x_t$ is the input to the memory cell layer at time $t$, $W_{\_}$ and $U_{\_}$ are weight matrices, and $b_{\_}$ are bias vectors.  

\begin{figure}[H] 
	\centering
	\includegraphics[width=0.75\textwidth]{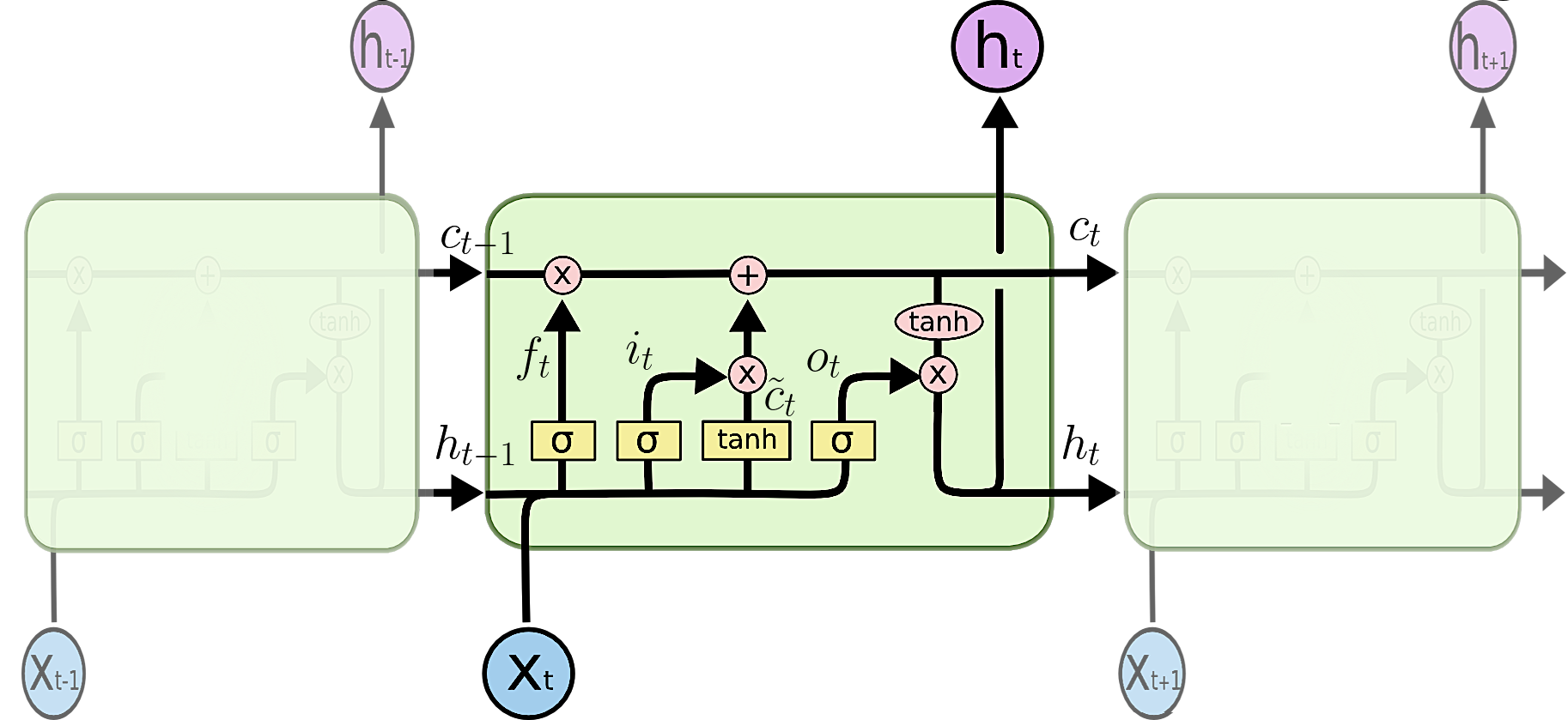}
	\caption{\small Graphical representation of an unrolled LSTM.  Image courtesy of Christopher Olah, used with permission.}
	\label{fig:unlstm}
\end{figure}

A graphical representation of an unrolled LSTM is shown by Figure \ref{fig:unlstm}.

\subsection{GRU}
\label{GRU_ap}

Gated recurrent units (GRU) \citep{Cho2014} are a simplified version of an LSTM with fewer parameters (U and W are smaller) but with comparable performance than LSTMs \citep{Chung2014}.

\begin{align*}
z_{t}&=\sigma _{g}(W_{z}x_{t}+U_{z}h_{t-1}+b_{z}) \quad \textsf{update gate}\\
r_{t}&=\sigma _{g}(W_{r}x_{t}+U_{r}h_{t-1}+b_{r}) \quad \textsf{reset gate}\\
h_{t}&=(1-z_{t})\circ h_{t-1}+z_{t}\circ \sigma _{h}(W_{h}x_{t}+U_{h}(r_{t}\circ h_{t-1})+b_{h})  \quad \textsf{output vector}
\end{align*}

Here, $x_t$ is the input vector, $W_{\_}$ and $U_{\_}$ are weight matrices, and $b_{\_}$ are biases vectors.

\section{Variational Auto-encoder}
\label{s:VAE}
\begin{figure}[H] 
	\centering
		\includegraphics[width=0.40\textwidth]{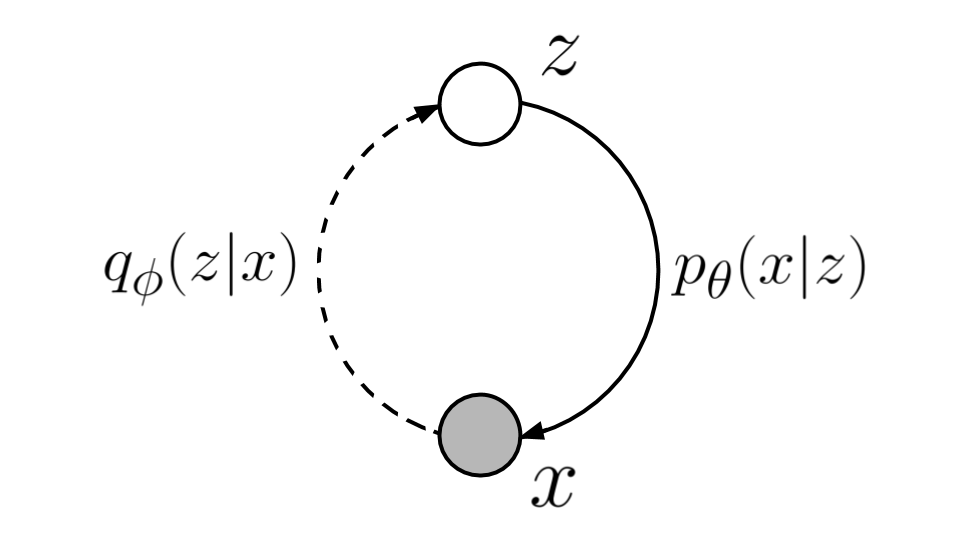}
	\caption{Variational autoencoder diagram}
	\label{fig:VAE}
\end{figure}

\citet{Kingma2014} and \citet{Rezende2014} independently introduced a powerful approach to treat intractable posteriors of direct probabilistic models with continuous latent variables. It is assumed that $X=\left\lbrace x^{(1)},...,x^{(n)} \right\rbrace$ is a random sample generated by a conditional distribution $p_{\theta}(x|z)$, where $z$ is an unobserved random variable, which in turn, is generated by some prior distribution $p_{\theta}(z)$. Briefly we want to write $p_{\theta}(x|z)$ as a normal distribution parametrized by an MLP:

\begin{equation}
\log p_{\theta}(x|z) = \log \mathcal{N} (\mu_{\theta},\sigma_{\theta}^{2} I)
\label{decoder}
\end{equation}
where
\begin{eqnarray*}
h &=& \tanh(W_{1}^{(\theta)} z + b_{1}^{(\theta)})\\
\mu_{\theta}&=&W_{2}^{(\theta)}h +b_{2}^{(\theta)} \\
\log \sigma_{\theta}^{2}&=&W_{3}^{(\theta)} h + b_{3}^{(\theta)},
\end{eqnarray*}
which makes the posterior $p_{\theta}(z|x)$ intractable. In the variational autoencoder approach, \cite{Kingma2014} and \cite{Rezende2014} replace the intractable posterior $p_{\theta}(z|x)$ by an approximation $q_{\phi}(z|x)$ parametrized by neural networks called \emph{recognition models} (Figure \ref{fig:VAE}) and they introduce a method to learn the parameters $\phi$ and $\theta$ (Algorithm \ref{alg:VAE}):

\begin{equation}
\log q_{\phi}(x|z) = \log \mathcal{N} (\mu_{\phi},\sigma_{\phi}^{2} I)
\label{encoder}
\end{equation}
where
\begin{eqnarray*}
h &=& \tanh(W_{1}^{(\phi)} z + b_{1}^{(\phi)})\\
\mu_{\phi}&=&W_{2}^{(\phi)}h +b_{2}^{(\phi)} \\
\log \sigma_{\phi}^{2}&=&W_{3}^{(\phi)} h + b_{3}^{(\phi)}
\end{eqnarray*}

The learning is driven by the maximisation of the variational lower bound:

\begin{equation}
\log p_\theta(x)=KL \left[ q_{\phi}(z|x)||p_{\theta}(z|x) \right] +\mathcal{L}(\theta,\phi,x)
\end{equation}

where
\begin{equation}
\mathcal{L}(\theta,\phi,x)=-KL \left[ q_{\phi}(z|x)||p_{\theta}(z) \right]  + E_{q_{\phi}(z|x)} \left[ \log p_{\theta}(x|z) \right]
\label{lowerbound}
\end{equation}
is maximised making use of the \emph{reparametrization trick}:

\begin{equation}
E_{q_{\phi}(z|x)} \left[ \log p_{\theta}(x|z) \right] \approx \frac{1}{L} \sum_{l=1}^{L}  \log p_{\theta}(x|z^{(l)}) 
\end{equation}
with
\begin{equation}
z^{(l)}=\mu_{\phi} + \sigma_{\phi} \epsilon^{(l)}, \qquad \epsilon^{(l)} \sim N(0,1)
\end{equation}

\begin{algorithm} [H]
 Initialize $\theta$ and $\phi$\;
 \Repeat{convergence of $\theta$ and $\phi$ is reached}{
  Sample a mini-batch $x^{M}$\;
  $\epsilon_{\phi} \sim N(0,1)$, $z=\mu_{\phi}+\sigma_{\phi} \epsilon_{\phi}$ as in Equation \ref{encoder}\;
  $\nabla_{\phi,\theta} \mathcal{L}_{M}(\theta,\phi,x_{M},\epsilon)$ as in Equation \ref{lowerbound} \;
  Update $\theta$ and $\phi$ using stochastic gradient descent
	}
 \caption{VAE - Learning}
 \label{alg:VAE}
\end{algorithm}

\section{Automatic Differentiation}
\label{autodiff}

Algorithmic or automatic differentiation\index{automatic differentiation} are techniques designed to compute derivatives without the need to explicitly compute them by hand. These techniques were developed long before deep learning was established.  There are many variants of AD such as forward-mode \citep{Wengert64}, reverse-mode \citep{Seppo1970}\index{reverse-mode AD}, among many others. The appropriateness of each type will depend on the type of function from which the derivatives are computed. In our application, reverse-mode AD is most appropriate and we provide a brief review limited to the computation of gradients, used extensively by deep learning packages like Theano, PyTorch or Tensorflow. For a general and comprehensive analysis on AD, we recommend the book by \cite{Griewank08}, where the authors do not confine AD to machine learning applications. On the other hand, \citet{JMLR:v18:17-468} provide an up-to-date survey specific to machine learning applications.

Reverse-mode AD is most appeling when the function being differentiated $f: \mathbb{R}^{n} \rightarrow  \mathbb{R}^{m}$ has a codomain dimension  that is significantlly smaller than the domain, ie, $m \ll n$. In this case only $m$ sweeps are required to compute the derivative in contrast to the forward method, which requires $n$ sweeps. For this reason, reverse-mode AD has become a cornerstone in the developement of deep learning models where gradient based optmizations are widespread.

We present reverse-mode AD through an example. Consider the following function, $f: \mathbb{R}^{2} \rightarrow  \mathbb{R}$:
\begin{eqnarray} \label{exAD}
f(x_1,x_2)=\sin(x_1)+x_1 \exp(x_2).
\end{eqnarray}

The first step of reverse-mode AD is to convert the function target into a sequence of primitive operations (add, multiply, log, exp, etc.. ). This sequence is called the \textit{Wengert list},\index{Wengert list} which in this example is
\begin{eqnarray*} \label{wengert}
	w_1 &=& x_1,\\
	w_2 &=& x_2,\\
	w_3 &=& \exp(w_2), \\
	w_4 &=& w_1 w_3,\\
	w_5 &=& \sin(w_1),\\
	w_6 &=& w_4 + w_5.
\end{eqnarray*}

\begin{figure}[H] 
	\centering
	\includegraphics[width=0.3\textwidth]{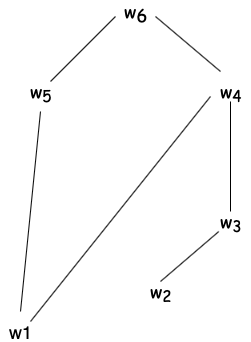}
	\caption{\small Computational graph associated with the Wengert list above}
	\label{fig:compgraph}
\end{figure}

This sequence of operations forms a computational graph representing the inter-relations of the series of operations as displayed  in Figure \ref{fig:compgraph},

The next step is to compute all the derivatives in the reverse order using the following formula, (essentialy an application of the chain rule for a sequence of composite functions)
\begin{eqnarray}
\bar{w_{i}}=\sum_{j \in \pi(i)} \bar{w_{j}} \frac{\partial w_{j}}{\partial w_{i}}  \qquad i=N,N-1,N-2,...    \quad   \text{and}   \quad  \bar{w_{N}}=1,
\end{eqnarray}
where $\bar{w_{i}}$ are called adjoints\index{adjoints}, $\pi(i)$ represents all parent indexes of $i$ in the computational graph in Figure \ref{fig:compgraph}, and $N$ is the total number of elements on the Wengert list. For example,  $\pi(1)=\left\lbrace 4,5 \right\rbrace $ and $\pi(4)=\left\lbrace 6\right\rbrace$. Therefore, the list of adjoints is:
\begin{eqnarray*} \label{adjoints}
	\bar{w_6} &=& 1,\\
	\bar{w_5} &=& 1,\\
	\bar{w_4} &=& 1,\\
	\bar{w_3} &=& w_1,\\
	\bar{w_2} &=& w_1 \exp(w_2),\\
	\bar{w_1} &=& w_3+\cos(w_1).
\end{eqnarray*}

For a real valued function $f: \mathbb{R}^{n} \rightarrow  \mathbb{R}$ ,the full gradient of the function $f$ is given by $\nabla f= (\frac{\partial f}{\partial x_1},..,\frac{\partial f}{\partial x_n})=(\bar{w_1},...,\bar{w_n})$, which can be computed in just one sweep. 

In our particular example, the full gradient is:
\begin{eqnarray*}
	\nabla f &=& (\bar{w_1},\bar{w_2}),\\
	\nabla f &=& (w_3+\cos(w_1),w_1 \exp(w_2)),\\
	\nabla f &=&  (\cos(x_1)+\exp(x_2),x_1 \exp(x_2)).
\end{eqnarray*}

\section{Deep Learning Packages}

Deep learning packages like Theano, PyTorch or Tensorflow use reverse-mode AD extensively. In addition, these libraries have several optimisations including GPU computations, arithmetic simplification, merging of similar subgraphs and improvements to numerical stability to name a few. In our experiments we use Theano \citep{2016arXiv160502688short} to implement the RDMM model.

\section{Q-learning and DynaQ algorithms}
\label{qlearning}
In Algorithm \ref{alg:qlearning} we find the Q-learning used in this article, while the base algorithm for DynaQ used in our experiments is presented in Algorithm \ref{alg:dynaq}. In both cases Q(x,q,a) represents the action-value function, $x \sim U(a,b)$ and $q \sim U(a,b)$ represents the price $x$ and inventory being drawn from a uniform distribution on the interval (a,b), $round(a,b)$ represents rounding off $a$ up to $b$ number of digits.
\begin{algorithm}[H] \small
 Initialise Q(x,q,a)\;
 $\epsilon=0.9,\alpha=0.9$\;
 \For{$i\leftarrow 1$ \KwTo $1,000,000$}{
 	\If{i mod 100,000==0}{
 	$\epsilon=\epsilon \times 0.9$\;
 	$\alpha=\alpha \times 0.9$\;
 	}
 	$x \sim U(9.3,10.7), q \sim U(0,10)$\;
 	$x=round(s,2),q=round(q,0)$\;
 	\While{$q>0$}{
 	\uIf{$U(0,1)<\epsilon$}{
 	$a=\underset{a'}{\mathrm{argmax}} Q(x,q,a')$\;
 	}
 	\uElse{$a \sim U(0,q)$\;
 	$a=round(a,0)$\;}
 	$r=x*a- c_2 a^2 -  c_3 q^2$\;
 	$x'=f(x,a)$\;
 	$q'=q-a$\;
 	$Q(x,q,a) \leftarrow Q(x,q,a)+\alpha \left( r + \max_{a'} Q(x',q',a')-Q(x,q,a) \right)$\;
 	$q=q', x=x'$\;
 	}
 	}
 \caption{Q-Learning - Adapted from Sutton [1998]}
 \label{alg:qlearning}
\end{algorithm}

DynaQ planning is conducted in an online manner, i.e., we combined real experiences sampled from the environment with simulated experiences sampled from the model:

\begin{algorithm} [H]\small
 Initialise Q(x,q,a)\;
 $\epsilon=0.9,\alpha=0.9$\;f
 \For{$i\leftarrow 1$ \KwTo $1,000,000$}{
 	\If{i mod 100,000==0}{
 	$\epsilon=\epsilon \times 0.9$\;
 	$\alpha=\alpha \times 0.9$\;
 	}
 	$x \sim U(9.3,10.7), q \sim U(0,10)$\;
	$x=round(s,2),q=round(q,0)$\;
 	\While{$q>0$}{
 	\tcc{choose action using $\epsilon$-greedy} 
 	\uIf{$U(0,1)<\epsilon$}{
 	$a=\underset{a'}{\mathrm{argmax}} Q(x,q,a')$\;
 	}
 	\uElse{$a \sim U(0,q)$\;
 	$a=round(a,0)$\;}
 	$r=x*a- c_2 a^2 - c_3 q^2$\;
 	$x'=f(x,a)$\;
 	$q'=q-a$\;
 	$Q(x,q,a) \leftarrow Q(x,q,a)+\alpha \left( r + \max_{a'} Q(x',q',a')-Q(x,q,a) \right)$\;
 	$q=q', x=x'$\;
 \HiLi   \tcc{ Simulated experience} 
 	\For{$i\leftarrow 1$ \KwTo $5$}{
 	$x,q \leftarrow$ random previously observed state\;
 	$a \leftarrow$ random action taken on state $x,q$\;
 	$r,x',q'=M(x,q,a)$ (model) \;
 	$Q(x,q,a) \leftarrow Q(x,q,a)+\alpha \left( r + \max_{a'} Q(x',q',a')-Q(x,q,a) \right)$\;
 	}
 	}
 	}
 \caption{DynaQ - Adapted from Sutton, 1998}
 \label{alg:dynaq}
\end{algorithm}

\subsection{DynaQ-ARIMA}

In the DynaQ-ARIMA, for a state $s_t=(x_t,q_t)$ and an action $a_t$ obtained from the simulated experience in Algorithm \ref{alg:dynaq}, the predictions of the model M are:

\begin{equation}
    M(x_{t},q_{t},a_t) 
\begin{dcases}
\hat{x}_{t+1}=&\mu+x_{t}+\phi_{1}(x_{t}-x_{t-1}) \\
\hat{r}_{t+1}=&x_{t} a_t- c_2 a_t^2 - c_3 q_t^2 \\
\hat{q}_{t+1}=&q_{t}-a_{t}
\end{dcases}
\end{equation}
with $c_2=c_3=0.001$

\subsection{DynaQ-LSTM}

We used a cross-validated long short-term memory (LSTM) network (see Appendix \ref{LSTM_ex}) as the model M in Algorithm \ref{alg:dynaq} in the DynaQ-LSTM benchmark.
Therefore, for a state $s_t=(x_t,q_t)$ and an action $a_t$ obtained from the simulated experience in Algorithm \ref{alg:dynaq} the predictions of the model M are:

\begin{equation}
    M(x_{t},q_{t},a_t) 
\begin{dcases}
\hat{x}_{t+1}=&LSTM \\
\hat{r}_{t+1}=&x_{t} a_t- c_2 a_t^2 - c_3 q_t^2 \\
\hat{q}_{t+1}=&q_{t}-a_{t}
\end{dcases}
\end{equation}
with $\phi_{1}=0.02$ and $c_2=c_3=0.001$

\end{appendices}

\bibliographystyle{plainnat}

\end{document}